\documentclass[twocolumn]{aastex631}

\usepackage{appendix}

\definecolor{darkgoldenrod}{rgb}{0.72, 0.53, 0.04}

\usepackage{xcolor}

\begin{document}

\title{Using Gravitational Waves \& Multi-messenger Astronomy to reverse-engineer the properties of galactic nuclei}

\author[0000-0002-5956-851X]{K.E. Saavik Ford}
\affiliation{Center for Computational Astrophysics, Flatiron Institute, 
162 5th Ave, New York, NY 10010, USA}
\affiliation{Department of Astrophysics, American Museum of Natural History, New York, NY 10024, USA}
\affiliation{Department of Science, BMCC, City University of New York, New York, NY 10007, USA}
\affiliation{Graduate Center, City University of New York, 365 5th Avenue, New York, NY 10016, USA}

\author[0000-0002-0786-7307]{Barry McKernan}
\affiliation{Center for Computational Astrophysics, Flatiron Institute, 
162 5th Ave, New York, NY 10010, USA}
\affiliation{Department of Astrophysics, American Museum of Natural History, New York, NY 10024, USA}
\affiliation{Department of Science, BMCC, City University of New York, New York, NY 10007, USA}
\affiliation{Graduate Center, City University of New York, 365 5th Avenue, New York, NY 10016, USA}



\begin{abstract}
Active galactic nuclei (AGN) are powered by accretion disks onto supermassive black holes in the the centers of galaxies. AGN are believed to play important roles in the evolution of both supermassive black holes and their host galaxies over cosmic time. AGN and the nuclear star clusters (NSCs) that interact with them remain unresolved with present and planned telescopes. As a result, the properties of AGN and NSCs are highly uncertain. Here we review how binary black hole (BBH) mergers can occur in AGN disks and how both the gravitational wave (GW) and electromagnetic wave (EM) properties of such mergers allow us to reverse-engineer the properties of AGN disks and NSCs over cosmic time. We point out that the feature in the BBH mass spectrum around $\sim 35M_{\odot}$ is an excellent probe of hierarchical merger models. Likewise constraints on the spins of upper-mass gap BH ($\gtrsim 50M_{\odot}$) test the AGN channel. The effective spin ($\chi_{\rm eff}$) distribution, including asymmetry, islands of structure and magnitudes are excellent tests of AGN model predictions. We also argue, that the rate of AGN-driven BBH mergers as a function of redshift should scale slightly shallower than the AGN number density, at least out to redshifts of $\sim 2$, and should turnover at the same redshift as the AGN number density. Finally, we emphasize a determination of an AGN fraction of observed BBH mergers ($f_{\rm BBH,AGN}$), \emph{regardless of the actual value}, allows us to infer the average properties of AGN disks and NSCs out to high redshift. 

\end{abstract}



\section{Introduction}
\label{sec:intro}

Nearly all galactic nuclei in massive galaxies are thought to harbor a supermassive black hole (SMBH; $\sim 10^{6}-10^{10}M_{\odot}$) \citep{KormendyHo13}. At any given time, however, most nuclei are quiescent or `inactive': their SMBH is not accreting much mass, and our only evidence of its existence comes from the orbits of stars within its sphere of influence. Fortunately, the vast majority of galactic nuclei also simultaneously host nuclear star clusters (NSCs) within the central few $\rm{pc}^{3}$, often comparable in mass to their SMBH ($\sim 10^{6-8}M_{\odot}$) \citep{Neumayer20}. These NSCs enable us to detect SMBHs in nearby quiescent galactic nuclei.

A small fraction of galaxies $\sim 1-10\%$ (depending on classification \citep[e.g.][]{Ho08}) are classified as `active', hence, Active Galactic Nuclei (AGN). As these were first identified observationally, there is a zoo of AGN classifications and subclasses, hosting (or not) varied physical features \citep{Padovani17}. The most critical structure which features in many classifications of AGN is a relatively geometrically thin, optically thick, thermally emitting, accretion disk of dense gas which is falling onto the central SMBH\footnote{Not all nuclei classified as AGN may host such an accretion disk; here we discuss AGN which do host such a disk.}. The conversion of gravitational potential energy to heat, through processes often modeled as a viscosity, are ultimately responsible for enormous electromagnetic (EM) luminosities, which can span $\sim 10^{9-11}L_{\odot}$ for Seyfert AGN, up to $\sim 10^{12-15}L_{\odot}$ for quasars, where $L_{\odot}$ is the solar luminosity. Such luminosities, prevent us from probing the stellar population in the NSC directly, as the accretion disk can be comparable to the \textit{stellar luminosity of the entire galaxy}. 

Despite their relative rarity, AGN disks and their associated outflows are believed to play two very important roles in our Universe. First, large gas disks can exchange orbital angular momentum with SMBH binaries (SMBHBs) at small ($\sim \rm{pc}$) separations. We expect such binaries to form due to galaxy mergers; however it is unclear what processes drive SMBHBs to merger---dynamical friction and interactions with a NSC can be inefficient in bringing the SMBHB into the efficiently gravitational wave (GW) emitting regime, the so-called 
``final parsec'' problem \citep{finalpc03}. However, angular momentum exchange with a gas disk can drive SMBHB to merger very efficiently, accelerating the mass growth of SMBH, including at early times \citep{BertiVolonteri08}. Second, vigorous outflows from AGN can quench star-formation in the host galaxy \citep{DiMatteo05,Tremmel17,Koudmani24}. For example, powerful radio jets heat up gas in the intergalactic medium, inhibiting inflow \citep{Sijacki07,Byrne24}. Thus, AGN are believed to be important in evolving \emph{both} SMBH \emph{and} their host galaxies over cosmic time .

NSCs are also important structures in our Universe. They represent a direct record of the history of mergers and star formation/AGN in their host galaxy. Around $\sim 1/2$ of the Globular Clusters that form in/around a Galaxy are believed to gravitationally decay into the NSC in a Hubble time \citep{Generozov18}, with a higher average infall rate expected in tri-axial potentials \citep{Capuzzo05}. NSCs can also play a role in driving SMBH binaries towards merger by acting as a source of dynamical friction in the nucleus \citep{BinneyTremaine87}. 
NSCs are widely believed to segregate by mass over time, with more massive objects sinking towards the center and less massive objects diffusing outwards \citep{BahcallWolf76,Antonini12}. As a result, the highest density of stellar mass black holes in the Universe may be in the region surrounding SMBH in galactic nuclei \citep{Morris93,Miralda00,Antonini15,Hailey18}.

Despite their importance, AGN are poorly constrained in their basic properties. We do not know: how long AGN live, how big and dense their disks are, how outflows occur, all to orders of magnitude uncertainty. Current and future telescopes cannot resolve AGN ($0.01^{"}$ corresponds to $5$pc linear size at $100$Mpc distance), so our only constraints on their basic properties have so far come from spectral and variability inferences and simple disk modelling. Likewise, NSCs are also poorly constrained. We only resolve two NSCs in our Universe; those in the Milky Way and Andromeda(M31). So, despite their apparent importance there is large uncertainty in the role AGN and NSCs can play in our Universe.  

In this review, we discuss the role GW from BBH mergers can play in establishing the properties of AGN and NSC across cosmic time. We will begin by reviewing what is broadly known about AGN and NSCs and their evolution. We will then outline the key physics which underlies the efficient merger of BBH in AGN. This will be followed by a review of the broad predictions of the AGN channel for BBH mergers, divided into GW predictions and EM predictions, highlighting robust versus more speculative or context-dependent predictions. We will then highlight  the usefulness of the $\sim 35M_{\odot}$ primary BH mass peak in the BBH mass function observed by LVK as a probe of hierarchical merger channels in general and AGN in particular. We will then discuss the predicted redshift distribution of GW detectable BBH mergers (out to redshift $z\sim 2$). We will review what we can already conclude about AGN, NSCs, and their contribution to GW observations to date. Finally we will highlight the most important areas of future work, and summarize our conclusions.

\section{Why is there an AGN channel for BBH mergers?}
Swarms of stars, but also stellar-mass black holes, neutron stars and white dwarfs (collectively: NSC objects) are expected to live in NSCs \citep{Morris93,Generozov18, Neumayer20}. For those galactic nuclei hosting an AGN disk, some fraction of NSC objects will have orbits approximately coincident with the AGN disk plane. Another fraction of NSC objects will end up dynamically cooled and captured by interacting with the AGN disk over its lifetime \citep{Ostriker83,Syer91,Artymowicz93,MacLeod20,Fabj20,Nasim22,Wang24}. Thus, a population of stars and stellar remnants are expected to live in AGN disks. 

Embedded objects in AGN disks will experience gas torques and migrate within the disk \citep{Levin07,McK12}. The result is close-encounters, sometimes at low relative velocities, allowing binaries to form \citep[e.g.][]{Stan23,Rowan23,Rozner23,Qian24,DodiciTremaine24,Whitehead24}.  If two stellar origin black holes (BH) form a binary (BBH), the interplay of gas hardening and dynamics can drive the BBH to the gravitational wave (GW) dominated regime, rendering their merger detectable with LVK  \citep{McK14,Bartos17,Stone17,ArcaSedda23}.

This AGN channel for BBH mergers has been substantially fleshed out over the last decade and broad predictions for AGN channel mergers include: 

\begin{itemize}

\item Mergers in the (upper) pair instability supernova mass gap (PISN) \citep[see][for PISN]{Farmer19} \citep[see e.g.][for mergers in the gap]{Bellovary16,Yang19,Secunda19,Hiromichi20,Gayathri21}; 
\item A higher IMBH formation efficiency compared to all other channels \citep[e.g.][]{McK12,Bellovary16,Yang19,Secunda19,Hiromichi20}; 
\item Asymmetric mass ratio mergers \citep{Secunda19,Yang19, McK20b,Hiromichi20}; 
\item A bias towards positive effective spin $\chi_{\rm eff}$ while retaining characteristics of dynamical assembly \citep[i.e. non-negligible $\chi_{\rm eff}<0$][]{qX22,Santini23,McKF24,AlexD24,Cook24};  
\item Potentially residual non-zero eccentricity \citep{RomeroShaw20, Hiromichi21ecc, Gayathri22, Samsing22,Isobel22}.
\end{itemize}

Furthermore, since the merged BBH forms a new, rapidly spinning BH and likely experiences a kick through hot gas, a simultaneous EM flare is \emph{guaranteed} \citep{Bartos17, McK19, Hiromichi23}.  However, it is very difficult for an accreting, newly merged BH, to outshine an AGN, even if it is kicked out of the AGN on the side facing the observer. 
Only if the newly merged BH launches a jet that persists long enough, can enough luminosity be generated so that observers could detect an EM counterpart \citep{Graham20,Cabrera24,Hiromichi24}.

Since the details of BBH mergers in the AGN channel must depend on basic AGN disk properties and the NSC population, by observing the GW/EM properties of BBH mergers occuring in AGN, we can finally constrain the basic properties of both AGN and NSC \citep[e.g.][]{McK18, Grobner20, Vajpeyi22,Ford22}. With LVK at present, and future instruments including LISA \citep{LISA23} and next generation GW detectors such as Cosmic Explorer \citep{CE19} and the Einstein Telescope \citep{ET25}, we can test the importance of the role that we think AGN play in our Universe over cosmic time. 

\section{What do we know about AGN and NSCs?}

\subsection{AGN}
Approximately one-third of all galaxies in the nearby Universe display some low-level of nuclear activity probably associated with the SMBH \citep{Ho08}, but about $5-10\%$ of galactic nuclei display Seyfert-like activity (bolometric luminosity ($L_{\rm bol}$) estimated in the approximate range $L_{\rm bol} \sim \mathcal{O}(10^{42-45}{\rm erg/s})$) and $\mathcal{O}(1\%)$ display quasar-like activity ($L_{\rm bol} \geq 10^{45}{\rm erg/s}$) \citep{Padovani17}. 

Most of this luminosity is believed to be driven by gas accreting onto the SMBH from a disky flow, generating a bolometric luminosity ($L_{\rm bol}= \eta \dot{M} c^{2}$) or
\begin{equation}
L_{\rm bol} \sim  10^{46}{\rm erg/s}\left(\frac{\eta}{0.1} \right)\left(\frac{\dot{M}}{\dot{M}_{\rm Edd}} \right) \left( \frac{M_{\rm SMBH}}{10^{8}M_{\odot}}\right)
\end{equation}
where $\dot{M}_{\rm Edd}$ is the Eddington accretion rate onto an SMBH of mass $M_{\rm SMBH}$ and $\eta$ is the efficiency of conversion of potential energy into radiation.
If there is a high accretion rate ($\dot{M}$) onto the SMBH, conservation of angular momentum implies vigorous outflows are expected. Indeed outflows are commonly observed across AGN, detectable as blue-shifted absorption features across optical/UV/X-ray bands \citep[e.g.][]{Yamada24} and occasionally as prominent jets \citep[e.g.][]{Silpa22}.

The most numerous type of AGN, low luminosity AGN (LLAGN), including objects classified as low ionization nuclear emission region (LINER) galaxies, do not possess a `big blue bump' (generally thought to be evidence of a thermally emitting, geometrically thin, but optically thick, accretion disk). This absence makes them an inherently different kind of object from Seyfert and quasar AGN. The evidence of high-energy emission processes associated with the SMBH in these nuclei \citep{Gonzalez09} leads to two possible explanations for their low luminosities: 1) they are weakly accreting (very small $\dot{M}$), possibly also in a radiatively inefficient fashion, or 2) they are strongly accreting, but in a different mode than the Seyferts and quasars (and in this case \textit{must} be accreting radiatively inefficiently, and do not possess a geometrically thin/optically thick accretion disk). We will return to this in our discussions of the GW predictions of the AGN channel (\S\ref{sec:gwpred}).

The classic So\l tan argument holds in part that if $\mathcal{O}(1\%)$ of galactic nuclei are quasars, and assuming we do not live at a special time in the Universe, then the fraction of the Hubble time ($t_{\rm H}$) that a galactic nucleus spends active is $\mathcal{O}(1\%)t_{\rm H} \sim 100{\rm Myr}$ \citep{Soltan82}. The division of the $\mathcal{O}(100{\rm Myr})$ of activity per nucleus is unclear and probably varies significantly between galactic nuclei. On one hand, the fraction of quasars observed with no narrow $[\rm{O}_{\rm{III}} ]$ lines ($\sim 1\%$) \citep{ChenChen24} implies that low-density gas at $\mathcal{O}(1-10{\rm kpc})$ from the AGN has not yet been ionized by the AGN radiation field, and so $\sim 1\%$ of quasars are $<10^{4}{\rm yr}$ old. On the other hand, the prominence of $\sim{\rm Mpc}$-scale jets around a small fraction of AGN, implies that these AGN have been active for several Myrs at least.

All galactic nuclei hosting an SMBH are thought to go through at least one (more likely multiple) periods of activity over cosmic history. The uncertainty on the typical length of an AGN phase is compounded by our uncertainties on the causes of AGN onset and cessation. We can imagine a model in which the ultimate source of the accretion disk gas is a giant molecular cloud (GMC) which arrives in a galactic nucleus due to a random walk; in this case, the distribution of cloud sizes might follow a powerlaw, with more small clouds and fewer large ones, leading to a powerlaw lifetime distribution (assuming a roughly consistent mass accretion rate, with the lifetime then set by the available gas mass). However, structures like bars, which supply gas at a roughly constant rate due to instabilities,  might lead to a lifetime distribution more like a Gaussian around some well-defined average.

Most SMBH in the nearby Universe that we can test with e.g. X-ray reflection methods, have high spins ($a>0.9$) \citep{Reynolds21}. Such high spins imply  accretion onto the SMBH consistent with repeated episodes in a single plane, rather than chaotic accretion which should spin down the central SMBH on average \citep{BertiVolonteri08}. Thus, it is possible that AGN in the nearby Universe consist of relatively short-lived episodes of SMBH fueling primed by pulses of low angular momentum gas from a fuel reservoir in the nuclear star cluster. Again, a powerlaw distribution of infalling masses due to instabilities in the fuel reservoir, would set a powerlaw distribution of lifetimes for the resulting co-planar AGN episodes.

Crucially, we wish to emphasize the growing evidence that AGN activity (especially among Seyferts and quasars) is \textit{NOT} triggered by interactions between galaxies. This was a reasonable hypothesis several decades ago, motivated by the need for gas to lose angular momentum in order to be delivered to a galactic nucleus and form an accretion disk \citep{Sanders88}. However, at this point there is a wealth of observational evidence \citep{Malkan99,Gabor09,McK10,Storchi19} that purely internal secular processes are likely the ultimate cause of the onset of most AGN, particularly at $z<1$.

What then is the cause of the observed variation of AGN fraction (or galaxy activity) with redshift (or cosmic time)? We observe a peak in nuclear activity around $z\sim2$ \citep{Ueda14} as a function of AGN luminosity, and this variation is not small---the number density of the most luminous AGN has fallen by nearly 3 orders of magnitude since their peak. This is not unlike the variation in star formation rate density (SFRD), though the peak in redshift of AGN activity and SFRD is slightly, but measureably offset, and the decrease in SFR is notably less steep. We suggest \citep[as have e.g.][]{Yang23MIRI} that the variation in AGN density with redshift is, like SFRD, primarily driven by a larger gas supply at earlier epochs.

\subsection{NSCs}
\label{sec:nscs}
NSCs are only directly observable in the very local universe \citep[see ][for an excellent review]{Neumayer20}. The NSC we know best is in our own galactic nucleus, Sgr A* \citep[see e.g. ][]{Genzel03,Ghez04}; however we must bear in mind that it may be atypical due to the possibility it has been through a recent episode of AGN activity, or for other reasons of historical accident \citep[e.g.][]{Levin07,Akiba24}. With that caveat in mind, we can consider the high density of stars of varying ages, and the fact that the stars we see most prominently (the S-stars) are B-type main sequence stars, and reasonably conclude that there should be many stellar mass black holes within a very small distance around our own SMBH. Indeed, 
\citet{Morris93,Antonini15,Generozov18} find $\mathcal{O}(10^4){\rm pc^{-3}}$ stellar mass BH predicted around Sgr A*, due to both in situ stellar evolution \citep{Bartko10}, and the infall of globular clusters over the history of the Milky Way \citep{Antonini12,Arca14,Gnedin14}. This is in good agreement with the BH population inferred from observations \citep{Hailey18,Mori21,Chen23}.

Looking at slightly more distant nuclei, we see that the presence or absence of a detectable NSC is dependent on both the galaxy type (bulge- or disk-dominated) and galaxy stellar mass. The probability of a galaxy hosting an NSC is large and fairly independent of galaxy type, for galaxies  with stellar masses between $10^{8}~M_{\odot}$ and $10^{10}~M_{\odot}$. For larger mass bulge-dominated galaxies, the probability of finding a detectable NSC decreases to near zero for galaxies with stellar masses of $10^{12}~M_{\odot}$, but remains significant for disk-dominated galaxies even at high mass \citep[see][for extensive discussion]{Neumayer20}.
However, we also note the detectability of an NSC is dependent on fitting isophotes to the radial light profile of a galaxy, and for larger mass galactic bulges, a substantial NSC could be present but undetectable. In addition, the processes which lead to NSCs (star formation, dynamical friction delivering star clusters to the nucleus) are likely to be ubiquitous and ongoing throughout cosmic time, thus, nearly every SMBH is likely to host at least a modest ($M\sim 10^7~{\rm M_{\odot}}$) NSC at $z\sim0$. In our own Galaxy, a disky distribution of stars with a top-heavy stellar mass function is observed, suggestive of star formation in a dense gaseous disk \citep{Bartko10}. The processes which may deplete NSCs are of interest to us as they include both episodes of AGN activity and SMBHB hardening and merger; however, under no plausible circumstances should either process cause an NSC to entirely vanish \citep[see e.g.][]{KritosRapster24}.

However, NSCs on their own (without the addition of a gas disk) produce extremely low rates of merging BBH \citep[e.g.][]{Fragione19, CMC22,Ford22}. This is due to the large velocity dispersion expected (and observed) in NSCs containing an SMBH. Left to its own secular processes, equipartition drives each stellar mass object in the nucleus to prefer to form a binary with the SMBH rather than with each other \citep{Hills88}; hence few BBH form, and fewer still persist until merger, if the nucleus remains unperturbed and allowed to fully relax.

The timescale for a cluster of stars to relax via two-body energy exchange is \citep{BinneyTremaine87}
\begin{equation}
t_{\rm relax} = 0.3 \frac{\sigma^{3}}{G^{2}m_{\ast}\rho_{\ast}\ln{\Lambda}}
\end{equation}
where $\sigma$ is the 1-d velocity dispersion of a population $N$ objects of mass $m_{\ast}$ and volume density $\rho_{\ast}$ and $\ln \Lambda \sim 15$ is the Coulomb logarithm. For orbiters around a supermassive black hole of mass $M_{\rm SMBH}$, $t_{\rm relax}$ can be parameterized as \citep{RauchTremaine96}
\begin{eqnarray}
    t_{\rm relax} &\approx& \frac{40{\rm Gyr}}{\ln N_\ast}\left(\frac{M_{\rm SMBH}}{N_{\ast}m_{\ast}} \right)\left(\frac{M_{\rm SMBH}}{10^{8}M_{\odot}} \right)^{1/2} \nonumber \\
    &\times& \left(\frac{m
    \ast}{1M_{\odot}} \right)^{-1}\left(\frac{r}{1{\rm pc}} \right)^{3/2}
\end{eqnarray}
so that in our own Galactic nucleus (where $M_{\rm SMBH}=4 \times 10^{6}M_{\odot}, N_{\ast}\sim 10^{6}$), $t_{\rm relax} \sim {\rm Gyr}$. This process is mass-dependent, with rates of BH delivery to central regions $\mathcal{O}(150(450))\rm{Gyr^{-1}}$ for $10(20)M_{\odot}$ BH expected in NSCs \citep{Hopman06a}.

In potentials dominated by the central SMBH, orbits are approximately Keplerian, so precession is small and eccentric orbits can persist in the same plane over multiple orbits. The resulting coherent torques between orbiters in such `resonant' potentials, lead to a timescale ($t_{\rm res, relax}$) on which angular momentum rather than energy is relaxed . $t_{\rm res, relax}$ can be significantly faster than $t_{\rm relax}$ by approximately\citep{RauchTremaine96}
\begin{equation}
 t_{\rm res, relax} \sim 7 \left( \frac{N_{\ast}m_\ast}{M_{\rm SMBH}}\right) t_{\rm relax}.
\end{equation}
As a result, the rate of delivery of stellar mass black holes (BH) to the central regions of a galactic nucleus (as EMRIs in the absence of gas) can therefore be strongly enhanced by up to an order of magnitude over energy exchange by (scalar) resonant relaxation \citep{RauchTremaine96,Hopman06}.  

We have few observational constraints on the evolution of NSCs with redshift. Due to the difficulty of observing NSC beyond $z\sim0$, we are forced to rely on models of galactic evolution to infer the likely redshift evolution of NSCs \citep{Antonini15,Generozov18}. Naturally, such models may be highly degenerate, as they are anchored only by observational tests at $z\sim0$; worse, such models often derive from cosmological simulations with scales insufficient to resolve the actual NSC, or processes that inform their structure. Nevertheless, during quiescent periods, we can use our robust understanding of N-body dynamics to model the development of NSCs \citep{RauchTremaine96}, with appropriate caveats regarding uncertainties on our boundary conditions, and come to some reasonably robust conclusions about the development of a typical NSC over cosmic time. In particular, we expect NSCs to have accumulated $\mathcal{O}(10-50\%)$ of their total stellar mass by $z\sim2$ \citep{Antonini15,Generozov18}.

Thus, in the relatively nearby universe ($z<2$), the variation in AGN fraction is far larger than the variation in NSC mass and occupation fraction.

\section{What are the key drivers of BBH mergers in AGN?}
\label{sec:key}
The crucial insight of the AGN Channel is that NSCs must \textit{be part of an} AGN. 
To a first approximation, the rate of BBH mergers in the AGN channel depends on the initial number density of BH in NSCs and the density of AGN disk gas. The initial BBH fraction is also likely an important factor, although this factor has been investigated less than the first two. 

The number of BH embedded in an AGN disk depends on both the initial NSC population density, as well as fundamental properties of the AGN disk including how large and dense the disk is and how long it persists. The initial NSC population interacting with the AGN disk in turn depends on how much (expected) mass segregation has occurred in the galactic nucleus, and the evolution history of the NSC. The process proceeds as follows: 

First, for a sufficiently dense gas disk, initially eccentric prograde embedded orbits should tend to circularize due to gas damping (equivalently: orbital dynamical cooling). During this process, any initially retrograde embedded orbiters (particularly in a dense inner disk) will tend to flip to prograde quite quickly \citep{AlexD24}. An additional population of low-inclination orbiters can also be captured relatively quickly \citep[e.g.][]{MacLeod20,Fabj20,WZL24,Rowan25,Whitehead25}.

Second, once a population of circularized orbiters builds up, a population of migrators will change semi-major axes due to (mass-dependent, generally inwards) gas torques given by
\citep{Paaardekooper10,JM17,Grishin23}
\begin{equation}
 \Gamma_{\rm mig} = C \left(\frac{H}{r} \right) \Gamma_{0}    
\end{equation}
where 
\begin{equation}
    \Gamma_{0}= q^{2} \Sigma r^{4} \Omega^{2} \left( \frac{H}{r}\right)^{-3}
\end{equation}
where $q$ is the mass ratio $q=m_{\rm bh}/M_{\rm SMBH}$ of a BH of mass $m_{\rm BH}$, located in the disk at distance $r$ from the SMBH, with Keplerian orbital frequency $\Omega$, local disk surface density ($\Sigma$) and local disk aspect ratio ($H/r$) and $C$ is the coefficient of the migration torque prescription depending on disk model and feedback conditions \citep{Grishin23}. As a result of the slow inspiral of migration, close encounters at low relative velocity can occur and BBH can form efficiently \citep{Stan23,Whitehead24,Rowan25a}, particularly compared to NSCs without AGN \citep{Ford22}. 

Third, new (and pre-existing) BBH will migrate within the AGN disk and undergo gas hardening \citep[e.g.][]{Baruteau11, Rixin22,Calcino23} and (typically, but not always) dynamical encounters. Fourth, dynamical encounters with other embedded objects can ionize, soften or harden the BBH \citep{DeLaurentiis23,Qian24,DodiciTremaine24,Whitehead24}. Fifth, once a BBH is `hard' (small enough semi-major axis around its own center of mass) with respect to the velocity dispersion of the embedded population, any other dynamical encounters will drive it towards the strongly GW-emitting regime and merger \citep[e.g.][]{Leigh18, Yihan21, Jairu22, Rowan23, Wang24}. We elaborate on some of the crucial details of migration torques, and binary hardening mechanisms below.

Once BBH form, gas effects tend to drive BBH to smaller binary separations, whether through direct gravitational drag effects \citep{Baruteau11,Rixin22} or eccentricity pumping \citep{Calcino23}. Gas drag effects are expected to become less efficient as the BBH orbital velocity becomes substantially supersonic w.r.t. disk gas \citep{Ostriker99,Sanchez14}. As these gas effects operate, the BBH itself continues to migrate in the disk, allowing for dynamical encounters with other single or binary embedded objects \citep{Leigh18, Yihan21, Jairu22, Wang24, Whitehead24}. Such dynamical encounters can soften (widen) or harden (shrink) the BBH. As long as the BBH binding energy is greater than the typical tertiary encounter energy , the BBH is considered `hard' relative to typical dynamical encounters or 
\begin{equation}
\frac{GM_{1}M_{2}}{2a_{\rm b}} > \frac{1}{2}m_{3}\sigma^{2} 
\end{equation}
for a binary ($M_{1},M_{2}$) of separation ($a_{\rm b}$) encountering mass $m_{3}$ with typical velocity dispersion $\sigma$. A sufficiently hard BBH can be driven to merger via some combination of gas and dynamical hardening. 
 The AGN lifetime over which BBH formation and hardening processes can act, before the NSC returns to quiescence and mutual interactions between objects lead to dynamical reheating, is also therefore a crucial parameter in determining the rate of BBH mergers. 
 
The most robust predictions of the AGN Channel are thus driven by the disk gas density, size, the initial population (especially the masses) of the NSC, and the disk lifetime. Everything else is first-order or higher corrections.





Of course, the processes described above depend on the \textit{existence} of a nuclear star cluster, however we believe the approximate content of the NSC has far less variation than the AGN parameters at $z<2$. To understand the AGN channel at $z>2$, we need better models of NSC formation and more observational and theoretical understanding of AGN at high redshift. Neverthless AGN-driven BBH mergers should be expected to decline at higher redshift due to both the decline in AGN activity and the expected incomplete formation of NSCs.

\section{What are the GW predictions for the AGN channel?}
\label{sec:gwpred}
Broadly testable consequences from the AGN channel can be grouped into predictions for: masses, spins, correlations between mass and spin  as well as the change in the expected merger rate as a function of cosmic time (redshift).

\subsection{Masses}
\subsubsection{Robust production of mass-gap BH and IMBH}

BBH can have strong kicks at merger depending on relative masses, spins and spin orientations \citep{Campanelli07, Lousto12}. However, BBH mergers in AGN disks occur in the deepest gravitational potentials in the Universe. As a result, unlike all other channels, even large merger kicks are mere  orbital perturbations \footnote{Keplerian orbital velocity at $a=10^{3}r_{g}$ is $\mathcal{O}(10^{4})$km/s, larger than \emph{any} possible merger kick.}, rather than escape opportunities \citep{Gerosa19}. As a result, the hierarchical merger chain in the AGN channel, can proceed farther, building more massive IMBH, more efficiently, than any other channel.

\begin{figure}
    \centering
    \includegraphics[width=1\columnwidth]{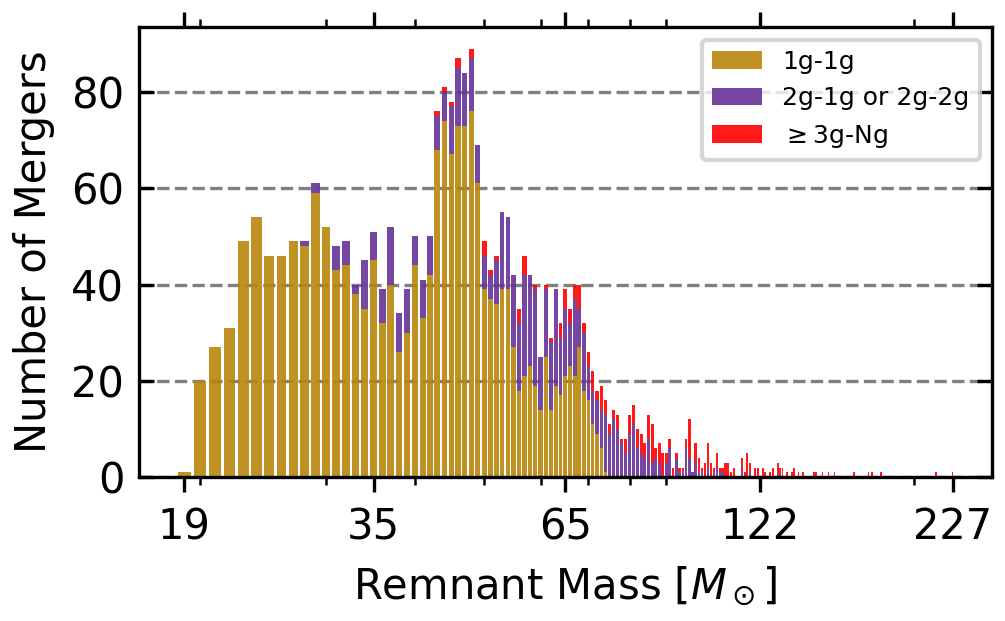}
    \caption{From \citet{McFACTS}. Number of BBH mergers per generation of BH as a function of merged mass ($M_\odot$) for a BH initial mass function (IMF) of $M^{-1}$ spanning $M_{\rm BH}=[10,40]\,M_{\odot}$ and simple Gaussian pile-up centered on $35\,M_{\odot}$ for a 0.5Myr, \citet{SG03} disk model around a $M_{\rm SMBH}=10^{8}\,M_{\odot}$. Gold denotes 1g-1g mergers, purple denotes 2g-mg ($m\leq 2$)and red denotes 3g-ng ($n \leq 3$) and higher, where 1g is first generation BH (not previously involved in a merger), 2g is second generation (the result of 1 merger) etc. The code producing this result and plot is open-source, reproducible and publically available at www.github.com/McFACTS/McFACTS.
   }
    \label{fig:remnantmass}
\end{figure}

\begin{figure}
    \centering
    \includegraphics[width=1\columnwidth]{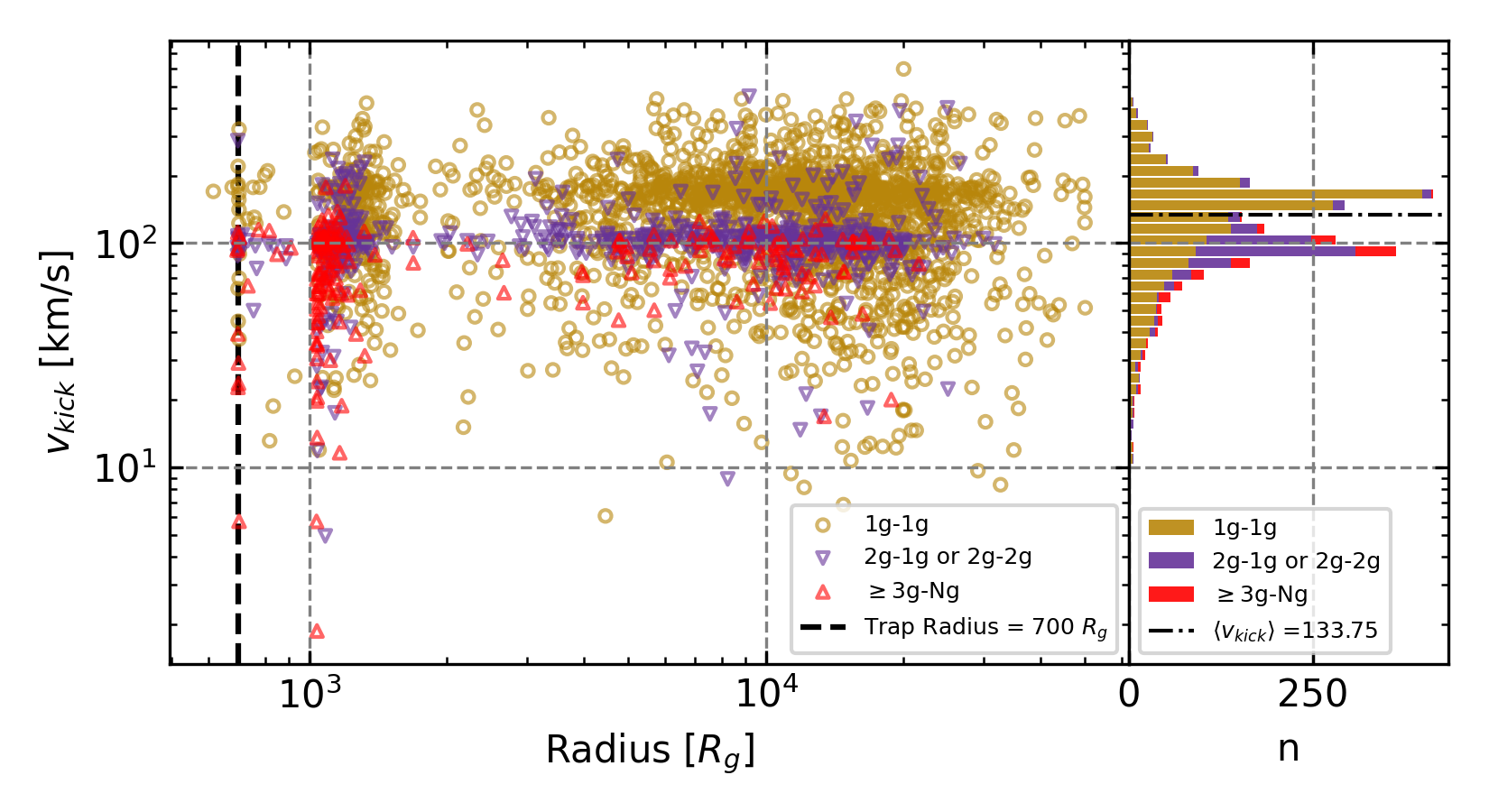}
    \caption{Kick velocity associated with each BBH merger as a function of disk radius and generation for the BBH mergers in Fig.~\ref{fig:remnantmass}. Right hand panel is a histogram of the merger number and generation as a function of velocity. The code producing this result and plot is open-source, reproducible and publically available at www.github.com/McFACTS/McFACTS.
   }
    \label{fig:vkick}
\end{figure}

The main, robust, universal prediction of the AGN channel is that AGN should host IMBH creation events, due to hierarchical mergers \citep{McK12,Bellovary16,Bartos17,Yang19,Hiromichi20,Vaccaro24}. Fig.~\ref{fig:remnantmass} shows an illustrative mass merger hierarchy resulting from a Monte Carlo simulation (\texttt{McFACTS}\footnote{link here?}) of 100 AGN with a Milky Way-like NSC interacting over 0.5Myr with a \citet{SG03} AGN disk accreting onto a SMBH of mass $M_{\rm SMBH}=10^{8}M_{\odot}$ (see \citet{McFACTS} for details). Black holes with masses $> 100~M_{\rm \odot}$ plainly form.

In this particular simulation, the BH initial mass function takes the form of a $M^{-1}$ powerlaw between [10,40]$M_{\odot}$ with a Gaussian excess centered on $35M_{\odot}$. 
The reason IMBH can form so efficiently in this example is apparent in Fig.~\ref{fig:vkick}. Merger kicks are assumed to depend on the relative masses, spins, and spin orientations and follow the prescription given in \citet{Akiba24}. Most 1g-1g mergers have kicks around $\sim 200{\rm km/s}$, and higher generation mergers have kicks around $\sim 100{\rm km/s}$ but the strongest kicks under this prescription are around $\sim 600{\rm km/s}$. \emph{Not a single kicked BH in Fig.~\ref{fig:vkick}  will escape these AGN.} Note, the approximation used here is likely an underestimation of the kick magnitudes experienced by the higher generation mergers. When we feed the merging BBH above through  \texttt{surfInBH} \citep{SurfinBH}, which includes a full G.R. evolution of the BBH, the higher generation mergers (2g-1g, 2g-2g, 3g-ng) can experience much larger kicks up to $\sim 2000{\rm km/s}$ (see Ray et al. 2025 in prep.). \emph{Even including higher kicks up to $\sim 2000{\rm km/s}$, virtually all the BH $<5 \times 10^{4}r_{g}$ will be retained by the AGN}. This is why the mass merger hierarchy in AGN is so efficient (and the IMBH fraction is so high with respect to other channels).

It may be hard to distinguish the 1g-1g mergers in AGN from other dynamical channels (e.g. Globular Clusters or Nuclear Star Clusters \citep{Antonini15,Rodriguez19}). This is due to the  merger timescale for BBH embedded in AGN disks being less than the expected accretion timescale onto the BH. Thus, spins of 1g-1g mergers don't have much preference for alignment with the gas disk, even if the orbital inclinations of the binaries are well-aligned. Thus, 1g-1g mergers look very much like gas-free NSC mergers---i.e. nearly random pairing of masses from the black hole IMF and nearly isotropically paired spins. There is a slightly larger probability of unequal mass mergers in AGN versus in gas-poor dynamical channels, which we explore in \ref{sec:massratio} below.

\begin{figure}
    \centering
    \includegraphics[width=1\columnwidth]{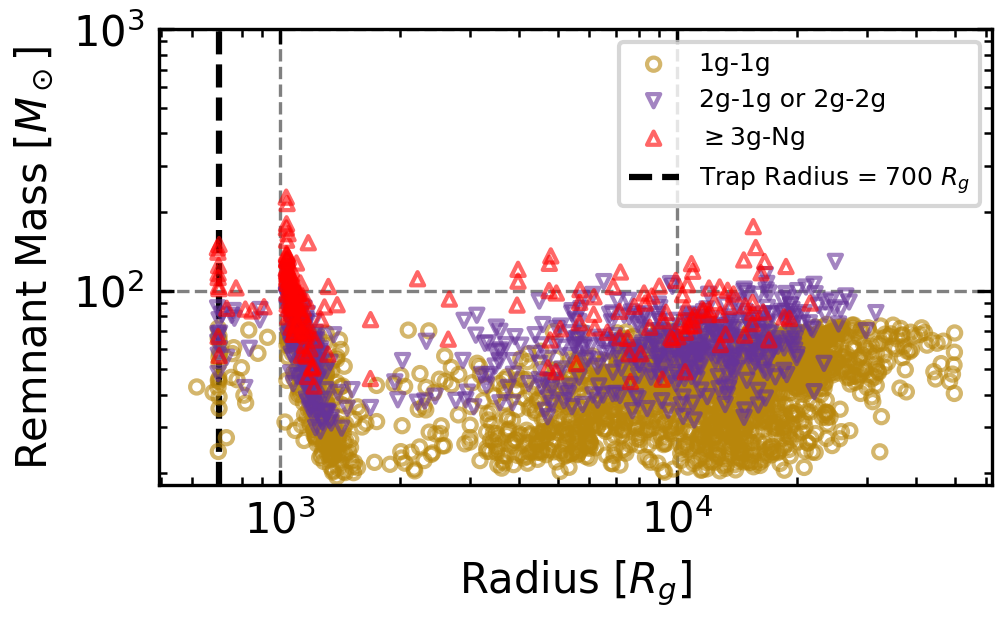}
    \caption{Mass and generation of BH merger as a function of disk radius for the BBH mergers in Fig.~\ref{fig:remnantmass}. The code producing this result and plot is open-source, reproducible and publically available at www.github.com/McFACTS/McFACTS.
   }
    \label{fig:radiusmass}
\end{figure}

However, the fraction and mass function of higher generation BH mergers may allow us to distinguish the AGN channel from other channels. For example, how far up the mass-merger hierarchy AGN-driven mergers can go depends on the average AGN disk lifetime (only 0.5Myrs in Fig.~\ref{fig:remnantmass}). Another factor that determines the fraction of higher generation mergers in AGN is the presence or absence of a migration trap/swamp. Migration is the competition among gas torques on an embedded object between inward and outward. Typically the resultant torque is inward. A migration trap occurs when the migration torques experienced by an embedded object cancel \citep{Bellovary16,Grishin23}. A migration swamp occurs when the net torque experienced by an embedded object decreases significantly (often in the inner disk as the surface density drops), leading to a pile-up as the in-migration rate changes. Fig.~\ref{fig:radiusmass} shows \textit{where} in the disk the mergers from Fig.~\ref{fig:remnantmass} occur. We see clearly the location of a merger trap around $\sim 700r_{g}$, where all migration ceases, and a migration  swamp around $1000r_{g}$ where migration torques in the inner disk drop and a `traffic jam' of BH arises, leading to close encounters, BBH formation and merger. The locations (and even the existence) of traps are highly disk model dependent \citep{Grishin23} but it is probable that  migration swamps are ubiquitous in areas of AGN disks where the surface density changes rapidly (usually near the hot inner disk).

Fig.~\ref{fig:radiusmass} also shows that most mergers clearly occur in the bulk outer disk ($>10^{4}r_{g}$), but higher generation mergers (and IMBH formation) are preferred at two regions:  the migration trap (in this model highlighted by the vertical dashed line at $700r_{g}$), and the migration swamp around $\sim 10^{3}r_{g}$, where the inward migration torque drops by $\sim 30\%$ approaching the trap (see e.g. Fig.~2 in \citet{Grishin23} for an illustration) and a pile-up occurs among the in-migrating BH before they reach the trap. If we assume $10^{8}$ galaxies per $\rm Gpc^{-3}$ of which $\sim 1\%$ are active, then we find a convenient merger rate scaling of $1  {\rm  merger/AGN/Myr}=1{\rm Gpc^{-3} yr^{-1}}$. The equivalent rate in Fig.~\ref{fig:radiusmass} is then  $\mathcal{R}=40{\rm Gpc^{-3} yr^{-1}}$.

\begin{figure}
    \centering
    \includegraphics[width=1\columnwidth]{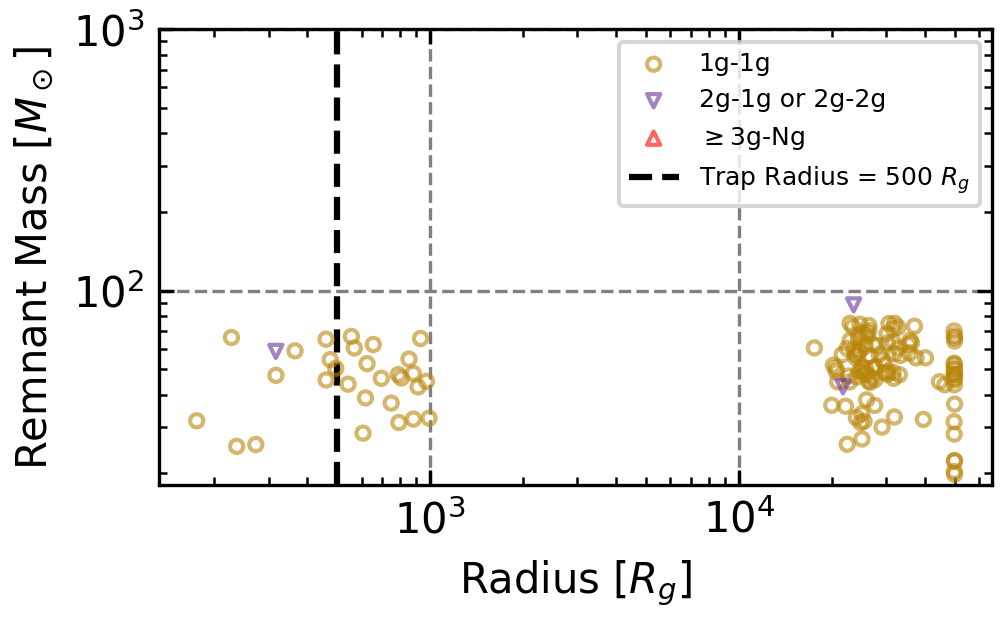}
    \caption{As Fig:~\ref{fig:radiusmass} except using the lower density \citet{TQM05} disk model run for 3Myr.
   }
    \label{fig:tqm}
\end{figure}

In contrast to models such as \citep{SG03}, lower density disk models such as \citet{TQM05} and \citet{Hopkins24} will yield drastically lower rates of BBH formation. We show the equivalent plot to Fig.~\ref{fig:radiusmass} for a \citet{TQM05} model disk in Fig.~\ref{fig:tqm} that has been run for 3Myr. We see a far lower overall merger rate ($\mathcal{R}\sim 0.3 {\rm Gpc^{-3} yr^{-1}}$) in part because we assume an initial binary fraction ($f_{\rm bin,0} = 0$). For these low-density disk models, $f_{\rm bin,0}$ will be a very important driver of the overall BBH merger rate. We also note the pile-up of mergers in the inner disk where there is a rapid change in surface density around $\sim 10^{3}r_{g}$. Likewise, out-migration torques due to thermal feedback in low optical depth regions of the disk tends to drive mergers in the outer parts of the disk. 

The efficiency of converting lower generation mergers into high generation mergers (i.e. the ratio of 1g-1g to ng-mg mergers) depend on the average of these factors across all AGN probed by GW detectors. The duration of a hierarchical merger chain in AGN will also depend on how rapidly IMBH can decay into the SMBH over potentially multiple cycles of activity and quiescence. An IMBH of $\sim 10^{2}M_{\odot}$ that builds up at a migration trap at $\sim 700r_{g}$ from a $10^{7}M_{\odot}$ SMBH will decay on a timescale given by \citep{Peters64}
\begin{equation}
    \tau_{\rm GW} = \frac{5}{128}\frac{c^{5}}{G^{3}}\frac{a^{4}}{M_{\rm b}^{2}\mu_{b}} (1-e_{b}^{2})^{7/2}
\end{equation}
which can be parameterized as
\begin{equation}
    \tau_{\rm GW} \sim 1.5{\rm Gyr}\left(\frac{M_{1}}{10^{7}M_{\odot}}\right)^{2}\left(\frac{M_{2}}{10^{2}M_{\odot}}\right)^{-1} \left(\frac{a}{700r_{g,M_{1}}} \right)^{4}
\end{equation}
where $r_{g,M_{1}}=GM_{1}/c^{2}$ and we assume $e_{b}\sim 0$. So, while IMBH can be produced efficiently during a brief AGN episode, they will eventually be lost to the SMBH, but potentially detectable with LISA as an IMRI \citep{LISA23}.

\subsubsection{Mass ratios}
\label{sec:massratio}
The relative occurrence of high mass ratio mergers ($q=M_{2}/M_{1} < 0.3$) is another useful test of the AGN channel. High mass ratio mergers are expected in an environment where the BH IMF is relatively broad, IMBH formation is efficient and migration drives a high rate of high mass ($M_{1}$) encounters with lower mass ($M_{2}$), slower migrators. However, most mergers are still $q \sim 1$ \citep[e.g.][and many others since]{Yang19ApJ, Secunda20, Hiromichi20}. Enhanced rates of unequal mass ratio mergers (due to differential migration rates by mass) are a slightly less robust prediction of the AGN channel than the formation of IMBH. This is because dynamical ($2+1$) exchange interactions for migrating BBH should suppress highly asymmetric mass ratios. However this depends on the number density of embedded objects, the details of gas hardening, the presence of migration swamps, disk turbulence, and the details of dynamical interactions. The interplay between these factors will determine the frequency of $q < 0.3$ mergers in the AGN Channel.
Unlike in the bulk AGN disk, a relatively high rate of high mass ratio mergers is also to be expected around migration traps in AGN disks \citep{Bellovary16,Secunda19,Yang19,Vaccaro24}, due to smaller black holes merging with an IMBH (but these are not the majority of the $q < 0.3$ mergers seen by most modelers).

\subsubsection{Tests of hierarchical mergers using features in the BH IMF}

\label{sec:35msun}

Any `features' in the BH initial mass function (IMF) will be amplified and echoed in hierarchical merger scenarios. Interestingly, among BH observed merging in GW, an excess in primary mass around $M_{1}\sim 35M_{\odot}$ is observed over a powerlaw fit to the mass function \citep{o3b}. The astrophysical origin of this feature is uncertain, but if this excess feature appears in the IMF of BH in any hierarchical merger channel, such as AGN, globular clusters or nuclear star clusters, then high mass `echoes' of such features must appear in the mass spectrum of those channels at a relative strength that depends on the channel merger efficiency. Importantly, hints of a $\sim 70M_{\odot}$ peak are also observed in GWTC-3 \citep{Ignacio24}. Such features represent an excellent test case for hierarchical merger channels. 

The number and relative strength of high mass `echoes' of the $\sim 35M_{\odot}$ peak can test the channel of BBH origin. For example, if a $70M_{\odot}$ peak is confirmed in the mass spectrum, but no further peaks in multiples of $35M_{\odot}$ or $70M_{\odot}$ are observed, this implies mergers in a relatively shallow potential and AGN are \emph{not responsible} for these mergers. This is because AGN are the most efficient merger channel at retaining merger products (merger kicks are generally irrelevant, see above). Conversely, multiples of the $35M_{\odot}$ peak should be observed (such as at $70M_{\odot}+70M_{\odot}$), and the relative strengths of those peaks should indicate the efficiency of retaining merger products (equivalently the depth of the potential well in which the mergers occur). 

The prominence of such peaks also tests the BH IMF in the hierarchical merger environment. For example, in AGN, we might expect a top-heavy IMF due to mass segregation within galactic nuclei. Thus, the relative rates of merger between $M_{1}\sim 35M_{\odot}$ BH and lower mass $M_{2}$ BH can constrain the local BH IMF.
As an example, in Fig.~\ref{fig:remnantmass} we observe remnant peaks due to $\sim 10M_{\odot}+10M_{\odot}$, $\sim 35M_{\odot}+10M_{\odot}$ and $\sim 35M_{\odot}+35M_{\odot}$ mergers among the first generation (1g-1g) of BBH mergers (in gold). These specific peak combinations arise because the $\sim 10M_{\odot}$ BH are the most numerous in these disks, but the $\sim 35M_{\odot}$ BH are the fastest  migrators. For this input IMF, among higher generation mergers (purple and red) a weaker peak is seen around $\sim 70M_{\odot}+20M_{_\odot}$ and (very weakly) at $70M_{\odot}+35M_{\odot}$ and $70M_{\odot}+70M_{\odot}$ for similar reasons. A BH IMF that is biased towards high masses, due to mass segregation and/or high mass star formation would suppress the lower mass population ($10M_{\odot}$ in this example) and drive the relative magnitude of the $35M_{\odot}+35M_{\odot}$ peak. If this model runs for a longer time ($>0.5$Myr) we should expect to observe peaks at $35M_{\odot}+70M_{\odot}$, $70M_{\odot}+70M_{\odot}$ and $140M_{\odot} +70M_{\odot}$. \emph{The relative strengths of these peaks tests the depth of the potential well the mergers occur in, as well as segregation within and between AGN phases}. For example, the absence of specific mass peaks along a hierarchical merger chain tests the effect of mass segregation within that merger chain. Such mass segregation  can occur within an individual AGN phase due to rapid radial migration or dynamical partner exchange in $2+1$ encounters (driving $q=M_{2}/M_{1} \rightarrow 1$) as outlined above. Alternatively,  segregation between AGN phases can occur due to sufficiently strong kicks that remove more massive merger products from a relatively short-lived AGN phase. Once a new AGN phase begins, mass-dependent recapture will segregate amongst the NSC population.


\subsection{Spins}
\label{sec:spins}
BH embedded in an AGN disk are interacting with a mass reservoir with a strong angular momentum preference. Therefore we expect signatures of this bias to be imprinted on the AGN channel population. Binaries are also expected to have their orbital angular momentum, $\vec{L}_{b}$, aligned or anti-aligned to the disk angular momentum $\vec{L}_{d}$, since binaries are expected to form most efficiently at the midplane, and gas torques tend to align their inclination angle with that of the disk \citep{Bardeen75,NatarajanArmitage99,King05,Whitehead25}.

LVK measures the effective spin parameter ($\chi_{\rm eff}$) of a BBH merger, i.e. the mass weighted projection of the spins ($\vec{a}_{1},\vec{a}_{2}$) of the binary masses ($M_{1},M_{2}$) onto the binary orbital angular momentum ($\vec{L}_{b}$) given by
\begin{equation}
    \chi_{\rm eff} = \frac{(M_{1}\vec{a}_{1} + M_{2}\vec{a}_{2})}{M_{1}+M_{2}}\cdot \vec{L}_{b}.
\end{equation}
Disky accretion by embedded BH might generally be expected to spin-up those BH and torque the spin angle of the BH into alignment with the disk.

However, the details of how embedded BH spins evolve depend on the orbital parameters of the BH. Interestingly, embedded BH on sufficiently eccentric orbits will tend to accrete \emph{retrograde} from the AGN disk \citep{YaPing2022,Chen22} and \emph{spin down over time}.

\begin{figure}
\begin{center}
\includegraphics[width=0.85\linewidth,angle=0]{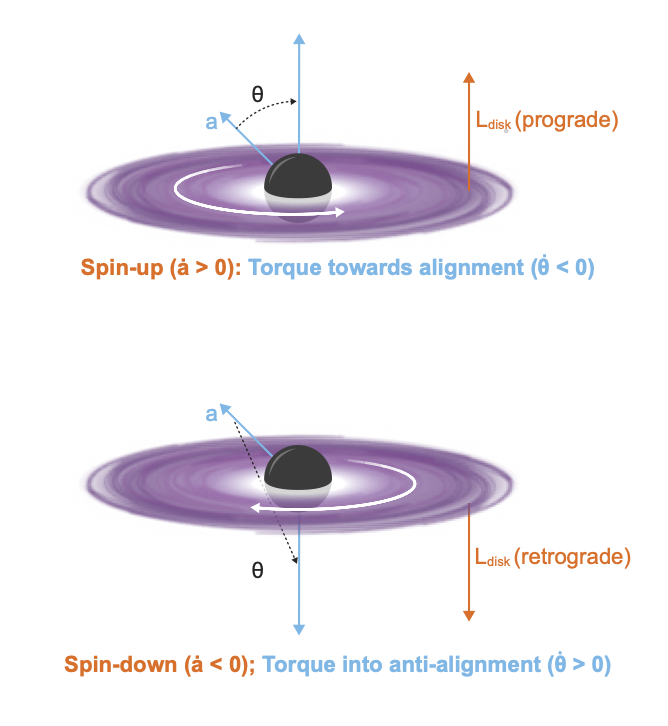}
\end{center}
\caption{From \citet{McKF24}. Cartoon illustrating prograde and retrograde accretion onto BH embedded in an AGN. Top panel shows a gas minidisk (with prograde orbital angular momentum $L_{\rm disk}$) accreting onto a BH. Blue vector labelled $a$ corresponds to an initial BH spin vector mis-aligned with the accretion flow. Dashed line shows the direction of torque of the BH spin over time, through decreasing angle $\theta$ towards alignment with mini-disk, but also increasing spin magnitude (longer final blue vector parallel to $L_{\rm disk}$). Bottom panel is similar except the accretion minidisk has retrograde orbital angular momentum. BH spin at first \emph{decreases} in magnitude towards $a=0$ (vanishing vector) and then grows increasingly negative ($a<0$) over time approaching full anti-alignment with the greater AGN disk (unlabelled downward pointing final spin vector).
\label{fig:acc_mode}}
\end{figure}

Once circularized however, BH will accrete prograde and tend to both spin up and torque towards alignment with the AGN disk \citep{AlexD24,Calcino23}. So, predictions for spin evolution of embedded objects depends on the timescale on which disk damps the orbital eccentricity, $t_{\rm damp}$.

\begin{figure*}
\begin{center}
\includegraphics[width=0.85\linewidth,angle=0]{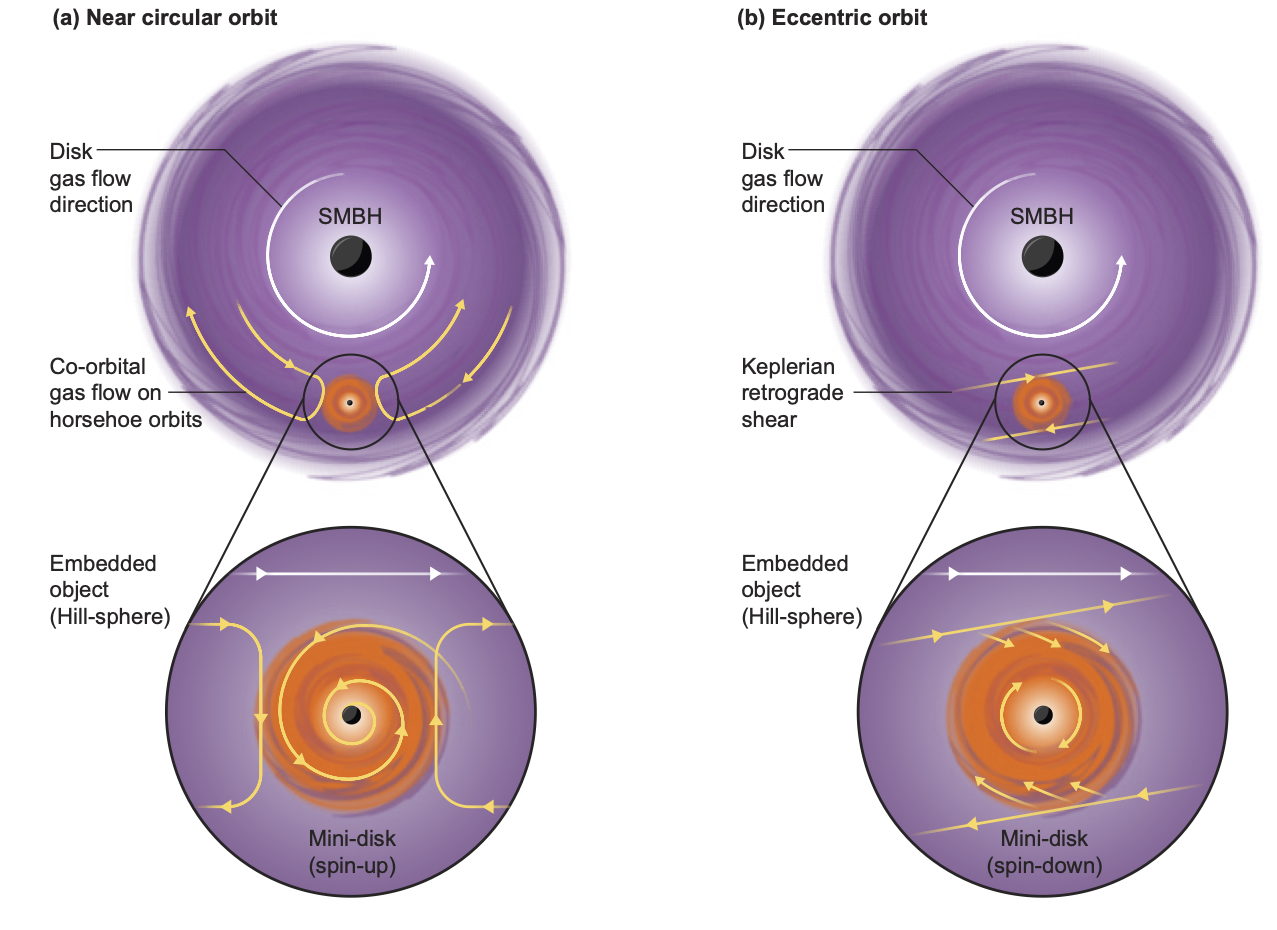}
\end{center}
\caption{From \citet{McKF24}. Cartoon of accretion onto a BH embedded in an AGN disk on a circularized orbit (left hand panel) and an eccentric orbit (right hand panel). White arrow indicates AGN gas flow direction and direction of orbit of the embedded BH. Yellow arrows indicate the relative flow of gas in the frame of the embedded BH. Bottom panels show  zoom in to Hill sphere of embedded BH.   Note: arrows do not indicate velocity magnitude, only the expected direction of the gas flow.
\label{fig:circ_ecc}}
\end{figure*}

Gas damping should drive BH orbits towards circularization, and the characteristic orbital damping timescale for a prograde orbit $t_{\rm damp}$ is \citep{Tanaka04}: 
\begin{equation}
    t_{\rm damp}=\frac{M_{\rm SMBH}^{2}(h/r)^{4}}{m_{\rm BH}\Sigma \,a^{2} \Omega}
\end{equation}
where $M_{\rm SMBH}$ is the supermassive black hole (SMBH) mass, $h/r$ is the disk scale height, $m_{\rm BH}$ is the embedded black hole mass, $\Sigma$ is the disk surface density, $a$ is now the orbital semi-major axis and $\Omega$ the Keplerian orbital frequency of embedded BH of mass $m_{\rm BH}$. The strong dependence of $t_{\rm damp}$ on the disk aspect ratio ($(h/r)^{4}$) shows that rapid circularization is most dependent on average AGN disk thickness.
Embedded BH with modest initial eccentricity ($e_{0}<2h$) should damp quickly $e(t)=e_{0}\exp{-t/t_{\rm damp}}$ \citep{Papaloizou00}, while those with larger $e_{0}$ tend to circularize more slowly \citep{Bitsch10}.

If merger timescales are short compared to gas accretion timescales, the spins and $\chi_{\rm eff}$ parameters of 1g-1g mergers do not show notable signatures of the AGN disk. However, 2g and higher generation mergers should, especially because the spin of merger remnants is dominated by the $\vec{L}_{b}$ of the binary that created the remnant. For 1g-1g mergers, these should be strongly biased towards alignment (or occasionally, anti-alignment) with the AGN disk.

\begin{figure} \includegraphics[width=1\columnwidth]{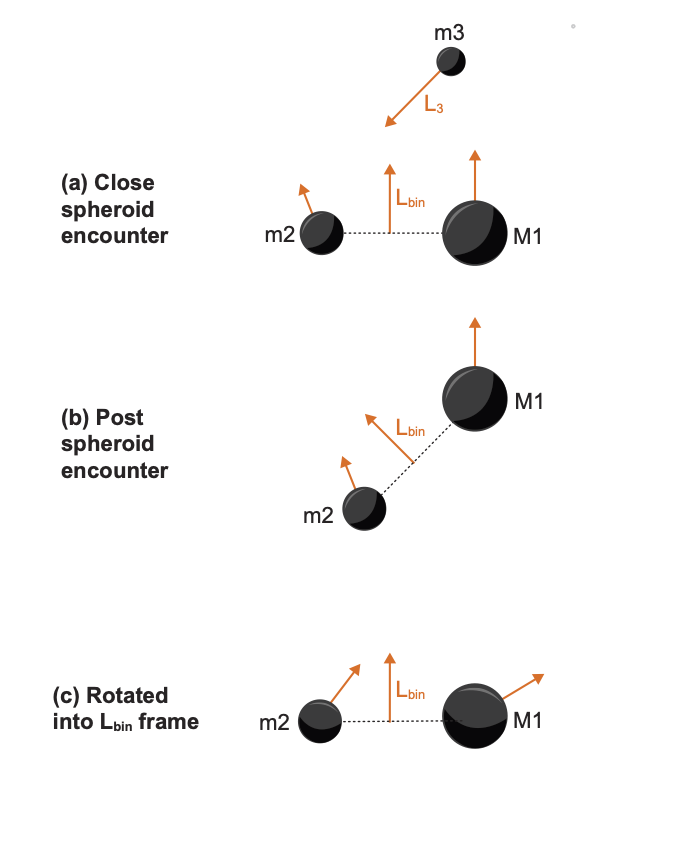}
    \caption{From \citet{McKF24}, a cartoon illustrating a spheroid encounter with an embedded BBH that drives the BBH out of the disk. The resulting merger (if it happens before disk-recapture) will have a significant spin component in the plane of the BBH. 
   }
    \label{fig:chi_p_cartoon}
\end{figure}

Embedded BBH in AGN face a further complication due to possible encounters with spheroid NSC orbiters. Conservation of angular momentum implies that the BBH will tend to be perturbed and kicked out of the disk. BH spins are unaltered in this encounter, but the \textit{projection} of those spins onto the BBH orbital angular momentum vector ($\chi_{\rm eff}$) could change substantially \citep{Hiromichi20spin,Samsing22,McKF24}. Because higher mass binaries have a larger cross-section to encounter, this effect is more likely to occur for higher generation mergers. We illustrate the geometry of such an encounter in Fig.~\ref{fig:chi_p_cartoon}. 

\begin{figure} \includegraphics[width=1\columnwidth]{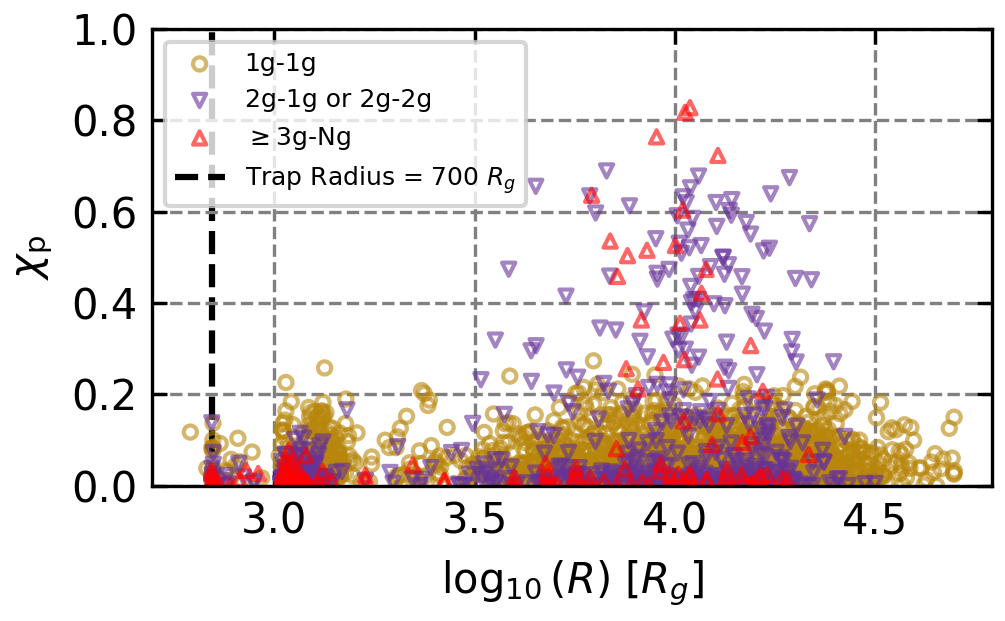}
      \includegraphics[width=1\columnwidth]{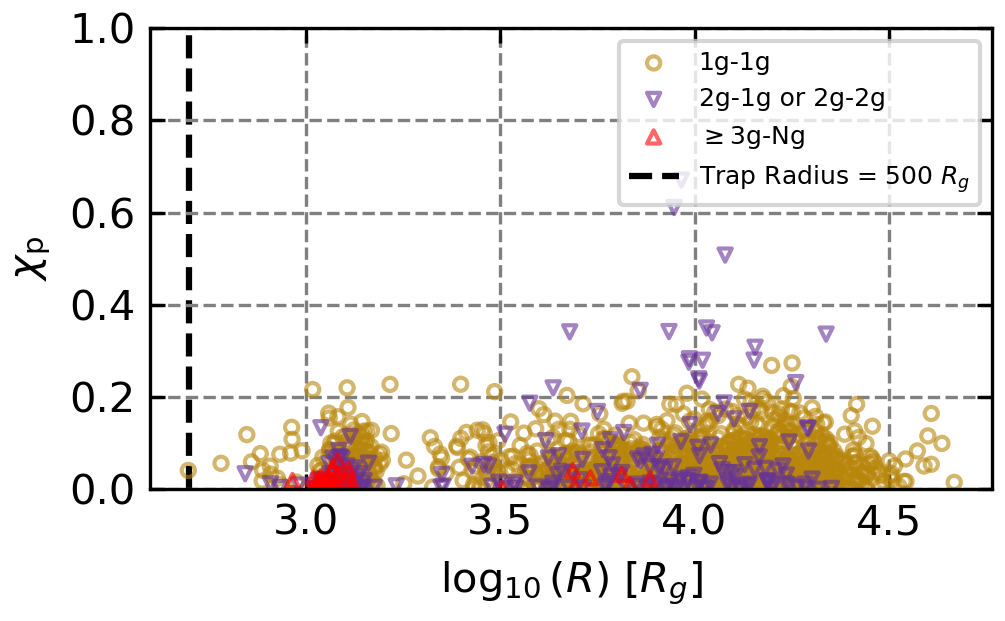}
    \caption{$\chi_{\rm p}$ for the BBH mergers in Fig.~1 as a function of disk radius where spheroid encounters are on (top panel) and spheroid encounters are off (bottom panel). 
   }
    \label{fig:chi_p}
\end{figure}

If the binary merges before its inclination is damped, this results in a smaller $\chi_{\rm eff}$ but a larger in-plane spin component ($\chi_{\rm p}$). This result and the mass (and generation) dependence of it is visible in Fig~\ref{fig:chi_p}

 Fig.~\ref{fig:chi_p} shows $\chi_{\rm p}$ as a function of BBH merger location in the disk, for the mergers in Fig.~1, with spheroid orbiter encounters on/off (top/bottom panel respectively). The highest values of $\chi_{\rm p}$ are overwhelmingly driven by higher generation (higher mass) mergers (purple and red points in upper panel). Higher mass BBH have a larger cross-section for spheroid encounters (larger $R_\mathrm{H}$), but higher mass makes them harder to ionize via spheroid stellar encounters. This feature of the AGN channel should be directly testable in O4 by LVK.

While the details of how many AGN-driven mergers will have what distribution of $\chi_{\rm eff}$ and $\chi_{\rm p}$ remains flexible due to uncertainties in input parameters and the efficiency of gas hardening, \textit{non-isotropic} spin distributions for higher generation (and thus, higher mass), clearly hierarchical mergers are a smoking gun signature of AGN-driven dynamical mergers. In addition, high values of the spin parameter in merging BBH systems can \textit{only} come from gas accretion, and not from dynamics alone \citep{Kritos24}. 

The spin distribution of the $\sim 35M_{\odot}$ BHs and associated hierarchical echo peaks at $70(140)M_{\odot}$ (see \S\ref{sec:35msun} above) is another test of channel origin. If the possible $70M_{\odot}$ feature is due to mergers among the $35M_{\odot}$ population then the merged products ($\sim 70M_{\odot}$) and subsequent merger products (e.g. $\sim 140M_{\odot}$) are likely to have high spin parameters ($|a| \sim 0.5-0.8$, depending on the initial BH spins). The ratio of negative to positive spins among hierarchical merger peaks tests the BBH formation mechanism (retrograde vs prograde).

The distribution of spin magnitudes among hierarchical merger products constrains the role of gas in the channel. Very high spins ($a>0.9$) among upper mass gap BH ($\gtrsim 50M_{\odot}$) are a \emph{smoking gun for AGN channel origin}. This is because such high spin can \emph{only} result from persistent prograde gas accretion or a series of repeated prograde mergers, both of which should be expected in an AGN environment. 

Going forward,  parameter estimation ($\chi_{\rm eff}$ and $\chi_{p}$) for BBH with progenitors in the suspected PISN mass gap will be key discriminators between `pure' dynamics channels such as GCs and NSCs, and those dynamics channels which include the influence of gas such as AGN \citep{Vajpeyi22,McKF24}.

\subsection{Correlations in mass-spin parameter space}

One intriguing feature from GWTC-3 was the apparent anti-correlation between BBH mass ratio ($q=m_{2}/M_{1} \leq 1$) and the effective spin parameter ($\chi_{\rm eff}$) \citep{Callister21}. The astrophysical origin of the anti-correlation in \citep{Callister21} is unclear, but possibilities include mass-transfer in stellar evolution \citep{Olejak24} as well as mergers in the AGN channel \citep{qX22,Santini23,Cook24}. 

\begin{figure}
    \centering
    \includegraphics[width=1\columnwidth]{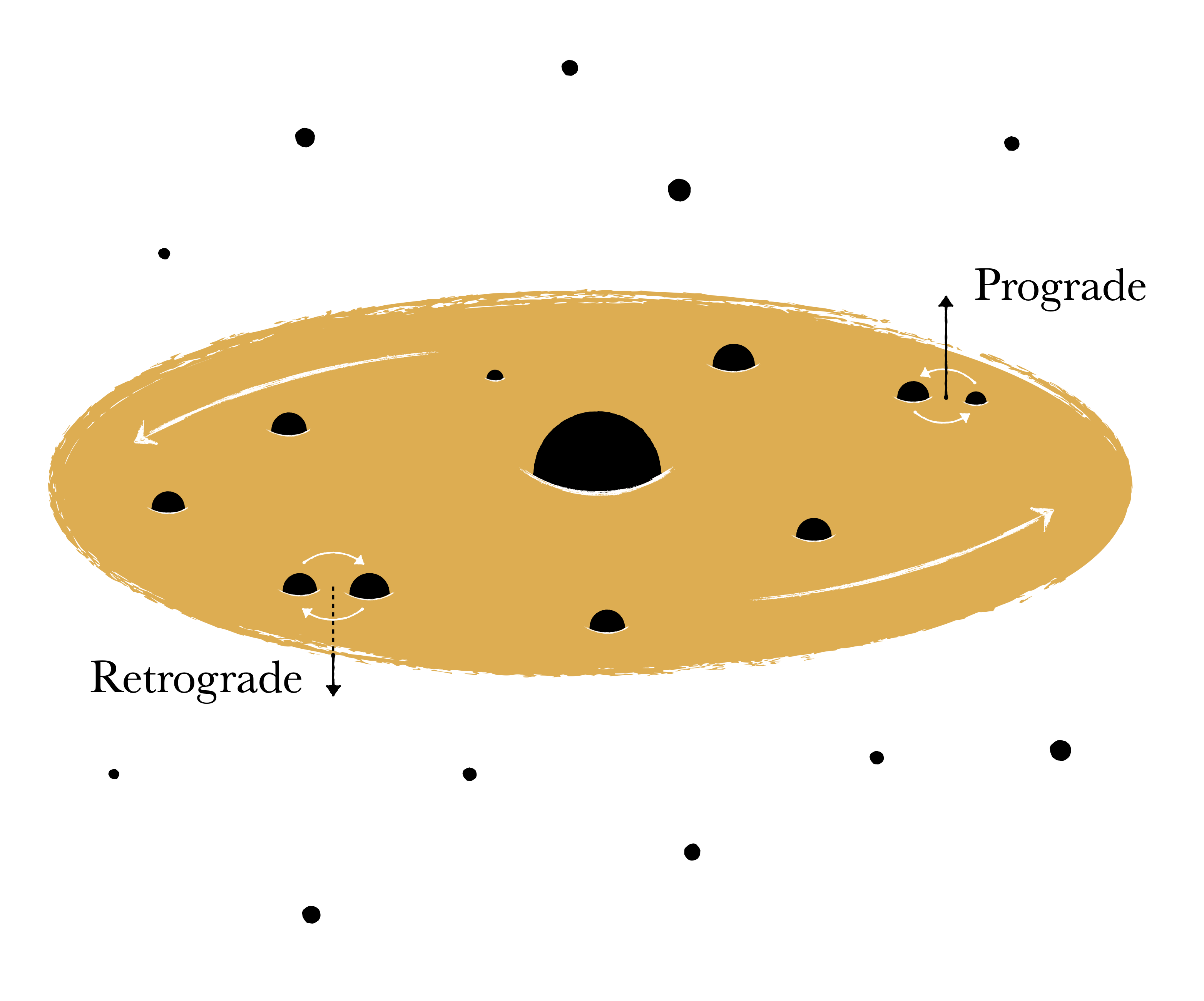}
    \caption{From \citet{qX22}, cartoon showing prograde and retrograde binary black holes (BBH) embedded in an AGN disk. Also illustrated are other (single) BH embedded in the disk and BH on disk-crossing orbits. There are several reasons to expect the fraction of retrograde BBH in AGN is supressed (see text). \emph{Credit: T.Callister}
   }
    \label{fig:retro_bbh}
\end{figure}

The anti-correlation in the AGN channel depends on three factors: 1) substantial suppression of mergers of retrograde binaries\footnote{A retrograde BBH is one that orbits its own center of mass in the opposite sense to the gas flow in the disk} (see Fig.~\ref{fig:retro_bbh}), 2) disk capture of objects being mass dependent, 3) some level of disruption of the orderly inward migration of large mass (IMBH) objects, which might otherwise `sit' at a migration trap or migration swamp and produce an overabundance of equal mass, large $\chi_{\rm eff}$ mergers. 

Suppression of mergers of retrograde  binaries is likely for two reasons. First, a retrograde BBH  that is not identically anti-aligned with the AGN disk experiences a gas torque (dependent on the binary accretion rate) that tends to flip the BBH to prograde \citep{AlexD24}. Note that interactions between embedded objects and the spheroid NSC, or gas turbulence, will tend to move retrograde BBH away from perfect anti-alignment. Second, dynamical hardening of prograde BBH is preferred over retrograde BBH \citep{Yihan21}.

Mass-dependent capture of objects by the disk is expected from considerations of dynamical friction \citep[e.g.][]{MacLeod20,Fabj20, Rowan25}. Since the dynamical friction force depends on the mass of the orbiter, more massive black holes will be captured into disk-embedded orbits faster. Note this is less true for stars, which experience aerodynamic drag in addition to dynamical friction \citep[see][for a thorough treatment of disk capture]{WZL24}.

The disruption of standard inwards migration is likely for several reasons including thermal feedback driving out-migration \citep{Hankla20,Grishin23}, stronger than expected merger kicks segregating the merger population from the disk BH propulation (Ray et al. 2025 in prep.), or turbulent perturbations in the viscous gas. Note that these effects are all more important for higher mass black holes, so this preferentially causes the higher generation merger remnants to have more chaotic paths through the disk and neatly prevents high rates of equal mass/high spin merger events.

\begin{figure}
    \centering
    \includegraphics[width=1\columnwidth]{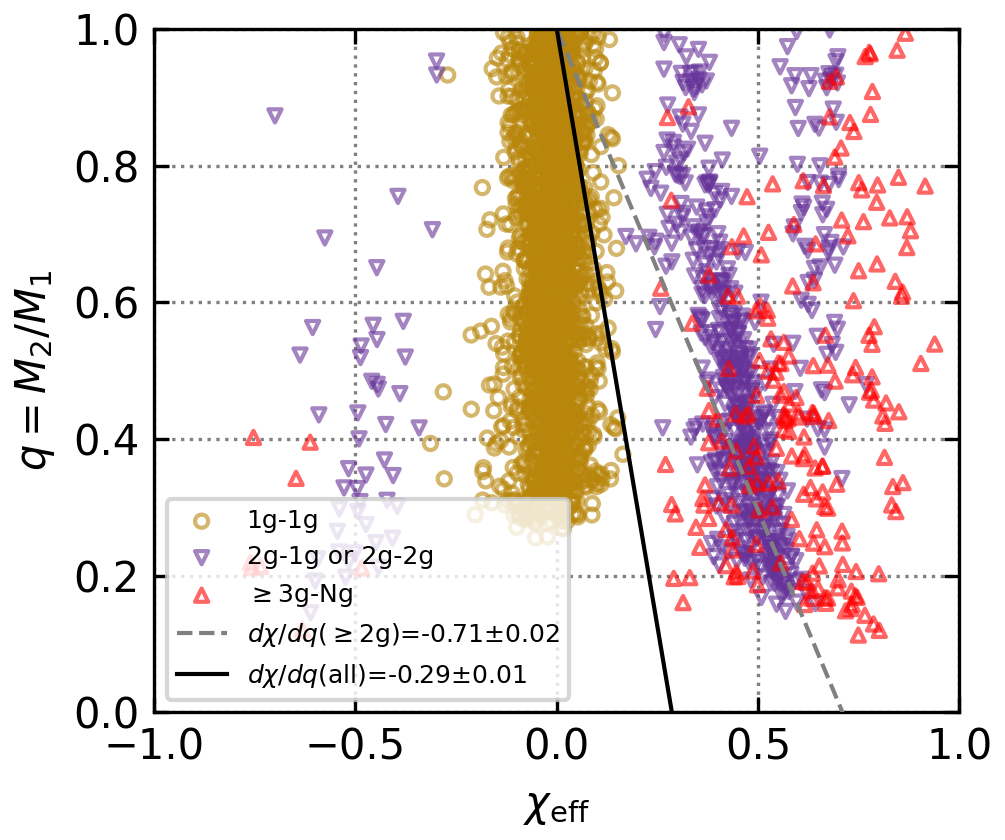}
    \caption{Effective spin parameter ($\chi_{\rm eff}$) as a function of mass ratio ($q=M_{2}/M_{1} \leq 1$) for the BBH mergers in Fig.~\ref{fig:remnantmass}, where the fraction of retrograde BBH is assumed to be $0.1$. The code producing this result and plot is open-source, reproducible and publically available at www.github.com/McFACTS/McFACTS \citep{McFACTS}.
   }
    \label{fig:qchi}
\end{figure}

Fig.~\ref{fig:qchi} shows the distribution of mass ratio ($q=M_{2}/M_{1} \leq 1$) as a function of BBH generation and $\chi_{\rm eff}$.
Evidently, most 1g-1g mergers are centered around $\chi_{\rm eff} \sim 0$, with the width depending on the variance of the initial spin distribution, and the extent in $q$ depending on the ratio of the minimum to maximum initial BH masses for the 1g BH \citep{Cook24}. This is because mergers happen rapidly enough that mass accretion is negligible prior to most 1g-1g mergers. Higher generation BH form with high spin ($|a| \sim 0.5-0.8$ typically) derived from the orbital angular momentum of the merging BBH and we assume that these BH merge in the plane of the AGN disk, so their new, rapid spin vector is aligned (or anti-aligned) with the disk \citep{Vajpeyi22, Santini23}. 

In Fig.~\ref{fig:qchi} we have assumed that $\sim 10\%$ of BBH form and merge retrograde (anti-aligned) w.r.t. the AGN disk. The ratio of prograde to retrograde BBH (or $\chi_{\rm eff}>0/\chi_{\rm eff}<0$) is a direct test of the efficiency of the gas  torque that flips retrograde BBH to prograde in the AGN channel.

In the AGN channel, predicted features in ($q,\chi_{\rm eff}$) space are a direct test of the NSC IMF and the average properties of AGN disks \citep{Cook24}. 
For hierarchical mergers, we also expect spin signatures. High spins ($\chi_{\rm eff} >0.9$) among BH with masses in the mass gap, imply gas and/or mergers have been involved in building up the spins of these BH, in alignment with the BBH orbital angular momentum.



\subsection{Merger rate as function of redshift}
A new result we discuss in this paper is the expected merger rate in the AGN channel, as a function of redshift, out to redshift $\sim2$. As we have noted earlier, the AGN fraction is changing by orders of magnitude \citep[][]{Ueda14, Yang21a, Yang23MIRI}, while the NSC mass and content likely changes by factors of $\sim$few to at most an order of magnitude \citep[][]{Antonini15, Generozov18}. We also note that the nuclei expected to dominate the AGN Channel rate contribution are those with $M_{\rm SMBH} \sim 10^{8}~M_{\odot}$ \citep{Delfavero24}. Thus the functional form of the merger rate should closely follow the AGN fraction as a function of redshift, and in particular the AGN fraction around $M_{\rm SMBH} \sim 10^{8}~M_{\odot}$, \textit{modified} by the NSC growth rate over the same span of time.

\begin{figure}
    \centering
    \includegraphics[width=1\columnwidth]{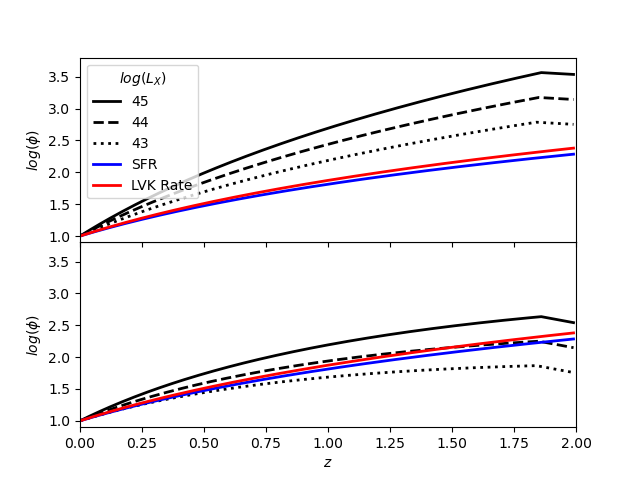}
    \caption{Variation of factors affecting BBH merger rate with redshift. \textit{Top panel:} we show the number density of AGN (arbitrary normalization) as a function of redshift for AGN at 3 X-ray luminosities: $L_x=10^{43}, 10^{44}, 10^{45}~{\rm ergs~s^{-1}}$ in black dotted, dashed, and solid lines, respectively. We also show the star formation rate variation in blue and the current LVK measured BBH merger rate variation in red. We drew our powerlaw indices for the AGN redshift variation from \citet{Ueda14}. Assuming the the typical X-ray luminosity is 10\% of the total AGN luminosity, these X-ray luminosities should correspond to $M_{\rm SMBH}=10^8~M_{\odot}$ accreting at $1, 10, 100\%$ of the Eddington rate. $10^8~M_{\odot}$ black holes hosting NSC should dominate the AGN-driven BBH merger rate, and considering that rate alone would suggest the measured BBH merger rate rises too slowly with redshift for AGN to represent an important contributor.
    \textit{Bottom panel:} We show the same AGN density, adjusted for NSC growth. Here we have assumed the NSCs will grow linearly by a factor of 10 from $z=2$ to $z=0$, consistent with \citet{Antonini15}. The expected AGN-driven BBH merger rate among $L_x=10^{44}~{\rm ergs~ s^{-1}}$ galaxies (which should closely track the $M_{\rm SMBH}=10^8~M_{\odot}$ accreting at $10\%$ of the Eddington accretion rate---i.e. bog-standard Seyfert galaxies) closely follows both the SFR and observed BBH merger rate (which is only well constrained to $z\sim1$). If NSCs instead grow more slowly \citep[consistent with][]{Generozov18}, somewhat lower X-ray luminosities would need to dominate the AGN-driven rate to be consistent with current observations (but $L_x=10^{43}~{\rm ergs~ s^{-1}}$ or $1\%$ Eddington accretors would still be a reasonable match).
   }
    \label{fig:ratesz}
\end{figure}

There is broad agreement that the functional form of the AGN number density rises sharply from $z=0$ to $\sim 2$, proportional to $(1+z)^{\gamma}$, where $\gamma$, depends on the X-ray (or bolometric) luminosity of the AGN (with larger values corresponding to larger luminosities), at least out to some cutoff redshift, typically $z\sim1.86$ for higher luminosity AGN \citep[but the cutoff redshift is also luminosity dependent---see e.g.][for details]{Ueda14}. We show in the top panel of Fig. \ref{fig:ratesz} the shape of the AGN number density as a function of redshift for AGN with X-ray luminosities, $L_x$, of $10^{43}, 10^{44}, 10^{45}~{\rm ergs~s^{-1}}$ drawn from \citet{Ueda14} (where $\gamma=3.94, 4.78, 5.62$ respectively, to a cutoff redshift of $z=1.84, 1.85, 1.86$). Assuming a $10\%$ bolometric correction, these correspond to $M_{\rm SMBH} =10^{8}~M_{\odot}$ accreting at $1-100\%$ of the Eddington luminosity. We also plot the \citet{madaudickinson} rate of star formation with redshift ($\propto (1+z)^{2.7}$), and the most recent LVK determination of the BBH merger rate variation with redshift \citep[$\propto (1+z)^{2.9}$][]{o3b} for comparison. All are arbitrarily normalized to intersect at $z=0$. The AGN rate varies far more strongly than the star formation rate or the BBH merger rate.

However, if the typical NSC is adding black holes between $z=2$ and now, this will soften the slope of the AGN-driven BBH merger rate, and we show an example of this in the lower panel of Fig. \ref{fig:ratesz}. Here we have assumed the NSC black hole population grew linearly and by a factor of 10 since $z=2$, consistent with models from \citet{Antonini15}. Given the uncertainty in the measured slope of the variation in rate of BBH mergers (and since current constraints only go out to $z\sim 1$), the expected AGN driven BBH merger rate is certainly consistent with current observations, and would still be consistent, even if NSCs grow somewhat less than we have supposed here \citep[e.g.][]{Generozov18}.

It is notable that the expected dominant contributors to the AGN-driven BBH merger rate are nuclei containing SMBH of $10^8~M_{\odot}$, along with an NSC. It is therefore quite interesting that the redshift trend which most closely matches the BBH merger observations is for AGN with $L_x=10^{44}~{\rm ergs~s^{-1}}$. These are disproportionately likely to be $M_{\rm SMBH}=10^8~M_{\odot}$ black holes accreting at $\sim10\%$ of the Eddington rate. Those would be extremely typical Seyfert galaxies, i.e. the most numerous type of AGN with a large, dense accretion disk.

At redshifts near 2, the AGN number density does appear to decrease, though recent mid-infrared observations \citep{Yang23MIRI} suggest the decrease is not nearly as steep as earlier X-ray observations implied \citep{Ueda14}. While we would also expect to see a decrease in the number of black holes hosted in NSCs at higher redshift, this decrease should be relatively monotonic; we should therefore expect a turnover in the AGN-driven rate of BBH mergers at approximately the redshift of the turnover of the AGN density. It would be interesting to investigate how many GW-detected BBH merger events are required to measure the turnover redshift to high enough precision to distinguish between the multiple channels with somewhat different expected turnover redshifts.

\subsection{Observable Effects of the SMBH}
\label{sec:smbh}
In the AGN channel, individual BH and BBH may end up on orbits close to the SMBH. In this case, multiple environmental effects may be testable using GW waveforms \citep{Barausse15}. Individual BH can form a binary with the SMBH and generate an extreme mass-ratio inspiral (EMRI), or intermediate mass-ratio inspiral (IMRI), detectable with LISA \citep{Amaro_Seoane_2007,LISA23}. Binary BH can be lensed by the SMBH during their orbit \citep{Kocsis11,Dorazio20,BenceGW22,Postiglione25}. In such cases the BBH may be eccentric \citep{Samsing22,Fabj24,Knee24,Stegmann25} and the GW waveform must climb out of the the potential well, leading to a possible over-estimation of the BBH mass \citep{XianChen19,Peng21}. The acceleration and jerk of the BBH center of mass in its orbit may also be detectable in the GW waveform \citep{Inayoshi17,Aditya23,Han24}. However, these are difficult measurements to carry out at present; the LISA and ET/CE era are when it will likely become essential to account for these effects.

\section{What are the EM predictions for the AGN channel?}
\label{sec:empred}
\subsection{Direct counterparts}
For any BBH merger occuring within an AGN disk, there is an interaction with the disk gas, which can emit light, so EM counterparts are more or less guaranteed to happen. But BBH mergers in AGN occur  in an AGN disk, which is already extremely luminous, so detectability is a problem.  The detectability of any EM counterpart therefore depends on parameters of both the AGN disk and the emission mechanism in some detail, in addition to the natural dependence on the parameters of the BBH merger event itself.

The physical and optical thickness of the AGN disk are key obstacles to detection---an event of almost any energy that occurs at the midplane of a typical optically thick AGN disk and that does not physically break out of the disk will always be undetectable. This is because  emergent luminosity $L_{\rm EM}=E_{\rm EM}/t_{\rm diff}$ from an event of energy $E_{\rm EM}$ is washed out over diffusion timescale $t_{\rm diff}$. $t_{\rm diff}$ is very long for high optical depth disk midplanes, and therefore $L_{\rm EM} \ll L_{\rm AGN}$.

\begin{figure*}
    \centering
    \gridline{
        \fig{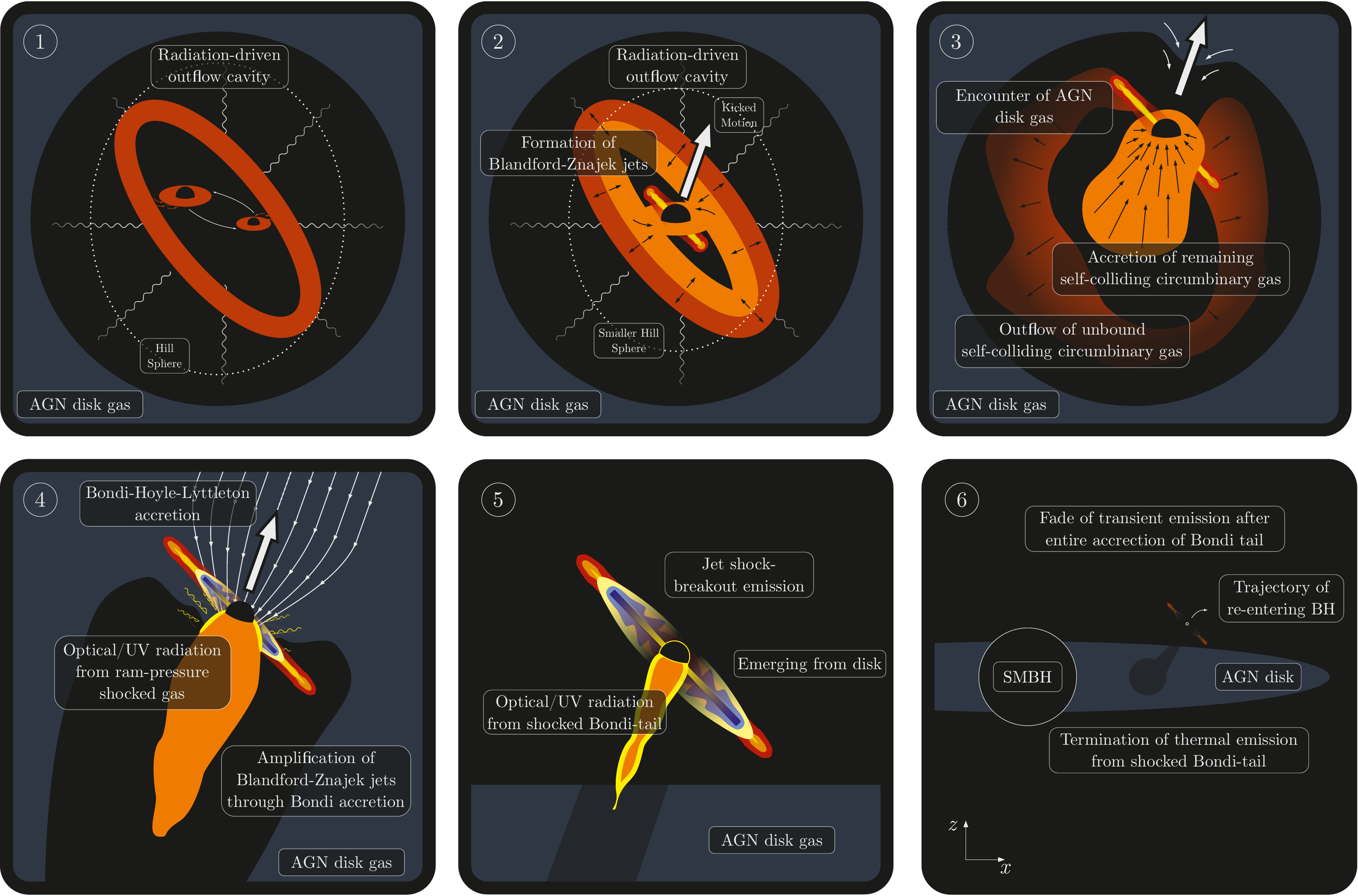}{0.95\textwidth}{}
    }
    \caption{
         Adapted from \citet{Cabrera24}, multi-panel schematic showing a representation of the mechanism believed to underpin luminous EM counterparts to BBH mergers in AGN disks.
         In panel 1, the pre-merger BBH  accretes from mini-disks within its Hill sphere in the AGN disk midplane and blows a cocoon within the disk via feedback.
        In panel 2, the merger happens, forming a highly spinning BH (dimensionless spin parameter $a \sim 0.7$ typically).
        A jet is presumed to form at this stage (although it has yet to be established whether such a jet can persist for long, or whether it is choked off by high mass accretion).
        Mass and spin asymmetries in the progenitor black holes lead to a kick at merger (depicted by the arrow in panel 2).
        Panels 3 and 4 show the development of BHL accretion as the newly merged BH exits its original Hill sphere into the rest of the AGN disk, powering a luminous transient.
        In panel 5 the BH emerges from the AGN disk, dragging disk gas with it in its Bondi tail. In panel 6 the EM emission fades as the disk material is consumed and the BH continues on an inclined orbit around the SMBH, and will re-enter the AGN disk on half the orbital timescale.
    }
    \label{fig:model}
\end{figure*}

From \citet{Cabrera24}, Fig.~\ref{fig:model} shows a cartoon of the process required to generate a luminous EM counterpart to a BBH merger in an AGN disk. In panel 1, a BBH embedded in the midplane of the disk is hard enough that it has decoupled from much of the gas (orange) elsewhere in its Hill sphere. Feedback from mini-disks around each BH blows away some of the gas in and around the Hill sphere, generating a cavity. Gas outside the Hill sphere is depicted in dark gray. In panel 2, the BBH has merged, creating a kicked, rapidly spinning remnant BH moving through hot plasma. As a result a jet tries to form \citep[e.g.][]{KimMost24}, tapping the spin energy of the BH. In panel 3, the BH undergoes Bondi accretion from the remaining gas within the original BBH Hill sphere and cavity as that gas attempts to follow the kicked BH. In panel 4 the BH has left the original cavity and travels within the bulk disk, accreting at an extremely high (Bondi) rate and the jet is assumed to persist. In panel 5, the kicked BH leaves the AGN disk and emerges on the side facing the observer. Jetted emission if directed at the observer should yield X-ray emission \citep{Hiromichi24}. Otherwise, optical/UV emission emerges from the interaction between the jet and surrounding/accreting  gas \citep{Graham20}. Emission lasts as long as the jet persists and as long as the BH still accretes from material dragged with it from the midplane. In panel 6, the emission has shut off. Either the jet has been choked off, or the accretion onto the kicked BH has shut off, or both. The kicked BH is now on an inclined orbit w.r.t. the disk and will impact the disk on a timescale depending on the new orbital inclination angle, $M_{\rm SMBH}$ and the semi-major axis.

The details of what sort of EM counterpart we should expect are not yet clear. It seems that in order to outshine relatively bright AGN, the formation and persistence of a jet that taps the spin energy of the newly merged BH is essential. Such a jet may drive an initial prompt, GRB-like X-ray flare. Intruigingly, some models suggest possible associated neutrino emission \citep{Hiromichineutrinos23}. In O4, a GRB was detected by Swift within 11 seconds of BBH merger S241125n. The odds of a coincidence within this time over a 10yr mission are $\sim 20\%$, but it has been argued that the GRB and BBH are associated \citep{241125}. The details of the optical/UV emission from jetted kicked BH are less clear, but luminous, short-lived flaring is expected \citep[e.g.][]{Lazzati22,Hiromichi23,Nemmen25,Ma25,ChenEM25}, with several candidate counterparts for O3 \citep{Graham23}, although see \citep{Palmese190521}. Only a few candidate counterparts have been identified for O4 so far \citep{Cabrera24,241125}.

The Hubble parameter $H_{0}$ can be independently measured from GW detected from BBH mergers, the so-called `dark sirens' \citep{LVKH0, PalmeseH0}.
If direct EM counterparts to BBH mergers in AGN can be confirmed, the combination of GW and EM distance estimates from AGN redshifts can significantly improve the accuracy of estimates of the Hubble parameter $H_{0}$ \citep{GayathriH0}.

\subsection{Indirect counterparts}
\label{sec:indirect}

AGN are intrinsically EM-bright sources, which are also relatively rare on the sky (compared to quiescent galactic nuclei). This means we can potentially detect an overdensity of AGN in GW-detected BBH error volumes, and use that association to measure to $f_{\rm BBH,AGN}$, even in the absence of a direct EM counterpart \citep{Bartos17stats}. The fundamental principle is easily illustrated by considering a situation where GW-detected BBH error volumes are so small, the expected rate of AGN detection in the volume is a (very small) background rate, perhaps 0.1 AGN per error volume. After a few detections with such a small error volume, if each one had an AGN in the middle of the error volume, one would reasonably conclude the AGN are associated with the BBH mergers. Unfortunately, we cannot yet achieve such small error volumes; instead the expected rate of AGN per error volume is 10s to 100s. Nevertheless, if a BBH merger was AGN-driven, there should be an `extra' AGN in each error volume \textit{above the background rate}.

This strategy clearly relies on having reasonably complete catalogs of AGN, such that we can independently measure their background rate; we also require such catalogs to reliably count the number of AGN in an error volume. With current error volume sizes, we unfortunately require far more than `a few' detections. Depending on the completeness and possible contamination of our catalogs, as well as the underlying background rate of the relevant AGN, this experiment requires 1-several 100 BBH merger detections with current GW localizations \citep{Bartos17stats,FordWP}.

Since we are approaching 100 public catalogued detections with full parameter estimation available (and nearing 300 if we include O4 alerts), \citet{Veronesi22, Veronesi23, Veronesi25a, Zhu25} have actually tried to determine $f_{BBH,AGN}$ using this indirect strategy. Unfortunately, the most stringent results can be obtained with the most complete catalogs, which means focusing on the brightest AGN---which (as discussed elsewhere) may not be the dominant contributor to the AGN Channel rate. There is also added risk to drawing conclusions based on the low-latency alert stream maps, as \citet{Veronesi25a} have done, since both the catalog of events and the maps themselves are subject to change (compare, for example, the public alerts to the O3 published catalog, where roughly half of the alerts are not included in the catalog, and half the catalog had no associated low-latency alert).

The current statistics from \citet{Veronesi25a} suggest that the brightest AGN, with bolometric luminosities of $10^{44.5}$ ($10^{45}$) ${\rm ergs ~s^{-1}}$ (or X-ray luminosities of $10^{43.5}$ ($10^{44}$) ${\rm ergs ~s^{-1}}$ do not contribute more than 21 (11) percent of the GW detected BBH merger population. We strongly suggest this analysis should be redone when the O4 catalog is released, as results could change substantially. In addition, it would be useful to determine which GW events provide the most power on this measurement, and confirm the spectroscopic redshifts of the AGN in or near the GW localization volume, as photometric redshifts can become quite unreliable at low redshift, where the volumes will be smallest. 

If this result is repeated, the remarkably small AGN contribution inferred may be in tension with the AGN contribution one would infer from the rate of IMBH formation events. Such a finding would be extremely constraining on the types of AGN disks and NSCs permitted in our universe, and might require a wholly different mechanism to produce some of the large mass mergers we have see to date. Nevertheless, it is interesting to note that alternative approaches suggest that the more numerous, less luminous AGN may in fact contribute most significantly to the inferred AGN channel rate \citep{Zhu25}. 

\section{Where are we now, and where are we going?}

\subsection{Observations: Where Are We Now?}
At time of writing, populations of GW events that are public correspond to those from LVK O1-O3. We anticipate the release of O4a later this year (2025), which may include surprises, but is likely to continue the broad-brush picture so far, namely: a sizeable fraction ($\sim 5-10\%$) of detected GW events are IMBH-formation events in the upper mass gap, suggestive of hierarchical merger environments. Effective spins appear biased to small positive values with a broad ($q,\chi_{\rm eff}$) anti-correlation expected.
Some BBH mergers have occurred with $q<<1$, e.g. \citet{GW190814}. 
From \citet{o3b} there is support for $\chi_{\rm eff}<0$ in the population of merging BBH, at the $\sim 25\%$ level. This implies a moderate contribution to the merger rate from a dynamical channel, although no single BBH merger event has possessed a clearly negative measured value for $\chi_{\rm eff}$.
The merger rate rises with redshift out to $z\sim 1$ at a rate consistent with the SFR (but with large uncertainty).

\subsection{Implications for the AGN Channel}
\label{sec:imp_agn}
Clues that might allow us to begin to constrain the AGN channel in the upcoming O4 data releases can appear in mass, spin and correlation spaces. In mass, the fraction of upper mass gap events puts a lower limit on the contribution of all hierarchical merger models (including AGN). 
From discussions in previous sections, BH signatures that would strongly suggest an AGN origin in O4:
\begin{itemize}
\item{Very high BH spin ($a>0.9$) and BH mass in the upper mass gap. High mass signifies an origin not from isolated binaries and very high spin can only arise from consistent disk-plane accretion (i.e. AGN).}
\item{Mergers in a chain from the $35M_{\odot}$ peak (including e.g. $70M_{\odot}+70M_{\odot}$). The relative strength of the peaks (e.g.: $(70+70)/(35+35)=f_{70}/f_{35}$ is a measure of the efficiency of retention of merger products. A high fraction $f_{70}/f_{35}$ implies a deep gravitational well, capable of holding onto kicked merger products. A fraction $f_{140}/f_{35}$ would be a clear signature of 3rd generation mergers in this context.}
\item{Very small mass ratio mergers ($q\ll 0.1$). Very low $q$ in AGN channel mergers could correspond to mergers between IMBH and moderate mass BH, or mergers between moderate mass BH and the results of e.g. NS-NS mergers. The former implies a migration trap environment \citep{Yang19,Secunda19,Vaccaro24}. The latter implies a merger in the bulk disk due to migration \citep{McK20b}. }
\item{If mergers in the upper mass gap do not look isotropic in $\chi_{\rm eff}$, this is a signature of an additional bias in a hierarchical channel (i.e. AGN).}
\item{Parallel, in-plane, spin components ($\chi_{\rm p}$ large) are strongly suggestive of a BBH that previously was aligned with $L_{\rm disk}$ and then experienced a kick out of the disk from a close encounter with a spheroid orbiter.}
\end{itemize}

From such events, the task can then begin of estimating the overall fraction of observed events that come from AGN ($f_{\rm AGN}$) by estimating the fraction of events above that appear in AGN channel distributions. 

\subsection{Missing Links in Modelling: Where should we go?}
\subsubsection{Disks}
The AGN channel is a fascinating interplay between gas and dynamics, with large questions still outstanding. As a result, much early work in the AGN channel involved simplifying assumptions. We require broad testing of each of the model components in order to understand which variables are dominant population drivers. 

For example, all AGN disk models so far are flawed. \citet{SG03} is a thin-disk model that is moderately plausible at small disk radii but is inconsistent at large radii. \citet{TQM05} is relatively plausible at large disk radii, but is not well modelled at small disk radii. \citet{Hopkins24} seems very promising as a descriptor of outer stable disks (Toomre's $Q>1$) due to magnetic pressure support, but it is still unclear whether such highly magnetized, high accretion rate models could produce the observed big blue bump characteristic of quasars. Note also that the implied very high throughput ($\times 10^{2}\dot{M}_{\rm Edd}$) in the $\sim 80-800$AU regions of the disk in the \citet{Hopkins24} model implies a too-fast rate of growth for SMBH in the local Universe (although maybe not in the early Universe). 

All AGN disk models are wrong in some respect, but all AGN disk models are useful. While all AGN disk models have issues, it seems clear that AGN disks are not well modelled by the classic \citep{SS73} razor-thin disk. This is because, in very thin disks, the timescales of change due to viscous effects ($t_{\nu}$) , i.e. the timescale of changes in the mass flow, are far too long for observed rapid variability in the inner AGN disk in many/most AGN, e.g. from \citep{Stern18}
\begin{equation}
t_{\nu} \sim 400{\rm yr} \left(\frac{h}{0.05}\right)^{-2} \left( \frac{\alpha}{0.03}\right)^{-1} \left( \frac{M_{\rm SMBH}}{10^{8}M_{\odot}} \right) \left(\frac{R}{150r_{g}} \right)^{3/2}
\end{equation}
where $h$ is the disk aspect ratio, $\alpha$ is the thin-disk viscosity parameter and $R$ is the disk radius. Since $t_{\nu} \propto h^{-2}$, a razor-thin disk $h \sim 10^{-3}$ with the same properties as above would have $t_{\nu} \sim 1$Myr and large-scale (order of magnitude) changes in AGN disks should not be possible on human timescales. Probably some combination of all these models can give us something approaching a 'plausible' model for AGN disks, until better models are developed.

The effects of embedded objects in and around AGN disks can be detectable electromagnetically if they perturb the inner disk significantly. Quantifying the rate of strongly variable AGN on short (dynamical) timescales as well as investigating details of their behaviour allows us to probe the dynamics of embedded objects in and around AGN disks. For example, tidal disruption events are expected to occur in newly-forming at a significantly higher rate than in gas-free galactic nuclei \citep{Starfall,Prasad24,Ryu24}. The class of changing-look AGN (CLAGN), change optically on timescales on the order of orbital timescales \citep{Graham20a,Ross20} and are likely due to inner disk perturbations, including TDEs and EMRIs \citep{Speri23}.

\subsubsection{NSCs}
As pointed out in \S\ref{sec:35msun}, features in the mass spectrum of BH are a powerful probe of the merger environment. The relative strength of the $35M_{\odot}$ peak compared to higher mass multiple peaks (e.g. at $70M_{\odot}$ (or $140M_{\odot}$)) is a direct probe of the efficiency of retention of kicked, merged BH. High efficiency of merger retention is a clear signpost to a deep potential well and suggestive of AGN. Of course, then the question becomes, how do the $35M_{\odot}$ BH arrive in AGN? There are only two possible answers: either these BH arrived from elsewhere, or they formed in the galactic nucleus.

Assuming $\mathcal{O}(1/2)$ the globular clusters (GCs) in a galaxy decay into the NSC over cosmic time \citep{Antonini12,Generozov18}, then it is possible that a feature produced due to stellar evolution at low metallicity ($Z\sim 0.1$) in GCs would appear in the NSC population. Thus, a $35M_{\odot}$ feature in the BH mass function that efficiently generates a chain of subsequent hierarchical mergers (e.g. $70M_{\odot}+70M_{\odot})$ could be a direct probe of the decay of GCs into galactic nuclei over cosmic time. We can test this scenario if the relative strength of the $35M_{\odot}$ peak in the BH mass function decreases at higher redshift (when fewer GCs have decayed into NSCs).

Alternatively, stellar evolution in AGN, which is believed to be very unlike that in vacuo \citep{Goodman04,DaviesLin20,Cantiello21,Jermyn22,Chen25}, might drive stars to sufficiently high mass just below the pair-instability threshold such that they collapse yielding BH with masses $\sim 35M_{\odot}$ \citep{Renzo24}. It would be very helpful if future studies of stellar evolution in AGN disks could test this latter hypothesis.

In any case, prominent echo peaks of the $35M_{\odot}$ feature (at say $70M_{\odot}$ and $140M_{\odot}$) would suggest in the AGN channel that the $35M_{\odot}$ BH are finding each other efficiently. In particular, narrow mass features at $70M_{\odot}$  \citep{Ignacio24} but not elsewhere, imply that the $35M_{\odot}$ are not preferentially encountering e.g. $10,20M_{\odot}$ BH. In the AGN channel there are only three ways of doing this: 
\begin{enumerate}
    \item heavy mass segregation within the NSC itself, such that the BH IMF mass peak in AGN is close to $\sim 35M_{\odot}$,
    \item mass segregation within or between individual AGN phases or
    \item dynamical partner exchanges in multiple close encounters  driving the BBH mass ratio $q \rightarrow 1$. 
\end{enumerate} 

First, efficient mass segregation within an NSC implies short timescales of relaxation ($t_{\rm relax}, t_{\rm res, relax}$ in \S\ref{sec:nscs} above). Since $t_{\rm relax} \propto M_{\rm SMBH}^{3/2}$,  strong mass segregation might imply a bias towards lower mass $M_{\rm SMBH} \sim 10^{6-7}M_{\odot}$ as sites for AGN channel mergers. Since $t_{\rm res, relax} \propto (M_{\rm NSC}/M_{\rm SMBH})$, further mass segregation enhancement might arise in NSCs that are underweight relative to their central SMBH. Additional  mass segregation can arise when AGN disks arrive in galactic nuclei. The drag force due to Bondi accretion experienced by an object of mass $m$ embedded in a gas flow is \citep{Ostriker99,Fabj20}
\begin{equation}
F_{\rm BHL} = \frac{4\pi G^{2}m^{2}\rho_{\rm disk}}{{\rm v}_{\rm rel}^{2}}
\end{equation}
where $\rho_{\rm disk}$ is the disk gas density and ${\rm v}_{\rm rel}$ is the relative velocity between the embedded object and the gas flow. $F_{\rm BHL}$ acts to slow the embedded object and causes a loss of angular momentum, dropping it inwards through the disk towards the SMBH. Since $F_{\rm BHL} \propto m^{2}$, Bondi drag will preferentially add heavier embedded masses to the inner AGN disk as it forms.

Second, mass segregation \emph{within} a single AGN phase can occur due to mass-dependent radial migration. Gas torques ($\Gamma_{0}$ in \S\ref{sec:key}) depend on mass ($\Gamma_{0} \propto q^{2}=(m/M_{\rm SMBH})^{2}$), disk surface density ($\Gamma_{0} \propto \Sigma$) and location in the disk ($\Gamma_{0} \propto r^{4}\Omega^{2}$) as well as the disk aspect ratio (height/distance, $\Gamma_{0} \propto (H/r)^{-3}$). So efficient mass segregation due to migration should imply relatively small AGN disks (so the population is not too large) that are also thin and dense. 

Mass segregation \emph{between} different AGN phases is a function of the kicks experienced by BBH at mergers and their location at merger. Efficient mass segregation between AGN phases implies that heavy merger products are kicked out of the disk and not recaptured before the AGN disk ends. This is strongly suggestive of relatively short-lived AGN disks as well as mergers that tend to occur further out in the in disk. The odds of escape from the disk are effectively $\mathcal{O}(\rm v_{\rm kick}/\rm v_{\rm orb})$ and are larger for large $\rm v_{\rm orb}$.

Third, dynamic encounters between a binary  ($M_{1},m_{3}$) and singleton $M_{2}$ where $M_{1},M_{2} > m_{3}$ have long been known to have a significant probability of partner exchange as
\begin{equation}
(M_{1},m_{3}) + M_{2} \rightarrow (M_{1},M_{2}) + m_{3}
\end{equation}
depending on the details of the encounters  \citep{Hills75,Heggie96}. Inertia means that the least massive party to the chaotic encounter often (but not always) ends up with the largest relative velocity and ejected. Thus, the mass ratio of binaries in a dense dynamical environment tends towards the ratio of the most massive components. In the AGN channel, we can test for this mechanism by searching for residual eccentricity by the time the BBH enters the LVK band \citep{Samsing22,Isobel22}.

\subsubsection{Gas \& Dynamics}
We are still in the early stages of understanding exactly how (M)HD gas disks interact with embedded objects. One thing is clear: we expect AGN disks to be turbulent. Such turbulence will drive a stochastic jitter on top of any net migration torque in an AGN disk \citep[e.g.][]{AdamsBloch09,Nelson10,Parker13,Trani25}. The net effect of such jitter is to smear out the radial locations of mergers in the disk. If we add a phenomenological jitter drawn from a Gaussian with variance on the order of $\Gamma_{0}$, the mergers in Fig.~\ref{fig:radiusmass} occur across a wider range of radii and the mass build up due to mergers at the trap(swamp) at $\sim 700(1000)r_{g}$ drops from $\sim 150(250)M_{\odot}$ to $\sim 100(150)M_{\odot}$. Jitter from turbulence will also tends to drive a random component to binary formation and should lead to a higher rate of asymmetric (lower $q$) mass mergers than otherwise. Thus, LVK upper limits on IMBH masses and lower limits on $q$ are a useful constraint on models of turbulence in AGN disks (as well as the presence of migration traps in the latter case).

\subsection{GW: The future}
Constraints from LVK observations on $f_{\rm BBH,AGN}$ can be cross-checked by LISA, via the rate and population properties of IMRI, BBH and EMRI mergers \citep{LISA23}. Doppler shifts of BBH close to the SMBH might be detectable (or constrainable) with LVK, but only if the BBH is very close to the SMBH. Lensing signatures may also arise in GW signals in this case \citep{XianChen19,Peng21,BenceGW22,Postiglione25}. Lower mass BBH or NS-NS could be detectable at larger radii. Such accelerations among BBH around SMBH should be more detectable with LISA \citep{Kaze19,Aditya23} or DECIGO  \citep{Decigo06}. Gas effects on the GW waveform may also be detectable with LISA \citep[e.g.][]{Mudit22}. Future observatories (Einstein Observatory \citep{ET25} and Cosmic Explorer \citep{CE19}) are expected to detect \emph{all} BBH mergers out to $z\sim 20$ and can therefore find $f_{\rm AGN}$ as a function of cosmic time. Finding $f_{\rm AGN} (z)$ will allow us to map the changing properties of AGN, NSCs and SMBH throughout cosmic time, allowing us detailed tests of our models of galaxy formation and evolution ($\Lambda$CDM) as well as models of SMBH growth and merger.

\subsection{EM: The future}
Early simulations of kicked, merged BH in AGN disks suggest jets are commonly attempting to form \citep{KimMost24}. The important next step lies in understanding whether and how long these jets may persist. Once a Blandford-Znajek type jet persists, it taps the spin energy of the rotating merged black hole, or $\mathcal{O}(10^{55}{\rm erg})(M_{\rm BH}/50M_{\odot})$. A small fraction of this energy can emerge as an X-ray burst, or optical/UV emission from shocks and reprocessing. Understanding the rate of occurrence of such EM counterparts might allow for a better constraint on the rate of EM counterparts to BBH mergers in the AGN channel \citep{Hiromichi23} and therefore better joint GW and EM constraints on the Hubble parameter $H_{0}$.

In the absence of confident, direct EM counterparts, we also look forward to the increasing power of the indirect counterpart `association' strategy we described in \S\ref{sec:indirect}. Current catalogs may not have sufficient accuracy in redshift (especially at low redshift) to ensure we are able to confidently localize the AGN of interest---efforts to improve the completeness of AGN catalogs, including with spectroscopic redshifts may be an important limiting factor in making this measurement. However, if we can make a reliable, indirect inference of $f_{\rm BBH, AGN}$ \citep{Gayathri23,Zhu25}, we can count on using a `hybrid siren' method to measure $H_{0}$ faster than via the `dark siren' method \citep{PalmeseH0}.

A complementary approach involves testing the presence of an embedded, scattered or disk-crossing population in AGN by studying short-timescale, high amplitude variability in AGN \citep{Graham20a}. Ongoing study of the perturbation of the AGN disk by embedded objects and disk crossers, such as in AGN-TDEs \citep{Ryu24}, changing-look AGN \citep{Graham20a}, and turning-on AGN \citep{Sanchez24} as well as associated QPEs \citep{Hernandez25} allows us to independently constrain both the embedded and disk-crossing populations in AGN.

\subsection{Fitting everything, everywhere, all at once}
Finally determing the AGN fraction ($f_{\rm BBH,AGN}$) of observed BBH mergers will allow us to infer a great deal about the average properties of AGN disks and NSCs out to the detection horizon of current and future GW detectors. However,  at present many channels could in principle account for the rate of BBH mergers observed \citep{IlyaFloor22}. Clearly, attempting to model the rate of BBH mergers alone is insufficient to distinguish between models. Instead, the best approach appears to be to generate distributions for \emph{everything} from the GW populations (rates, masses, spins, mass ratios) \emph{all together}, including apparently odd-ball or otherwise `rare' events such as GW190814 \citep{GW190814}. The best way of modeling all of these parameters remains population synthesis codes, and we are fortunate that the AGN Channel now has at least 3 of those: \texttt{McFACTS} \citep[][]{McFACTS}, \citet{Xue25}, and an adaptation of \texttt{FastCluster} \citep[as described in][]{Vaccaro24}, each with their own formulations of the appropriate physics, but all producing fast output that allows exporation of which input parameters and physics drives which population features in the AGN Channel.

We live in one Universe, the probability of generating rare events should allow us to discriminate between channels while fitting all available data, and allow us to converge more rapidly on channel fractions. With the release of O4 data, the presence or absence of AGN-like events (\S\ref{sec:imp_agn}) should allow us to begin to constrain $f_{\rm BBH,AGN}$ from above and below.
 
\section{Conclusions} 
 A fraction of mergers observed by LVK come from the AGN channel ($f_{\rm BBH,AGN}$). The value of $f_{\rm BBH,AGN}$ will reveal important facts about our Universe over cosmic time, \emph{regardless of what that value turns out to be}. 
 
 A small value of $f_{\rm BBH,AGN}$ likely implies that either AGN disks are small and short-lived ($\lesssim 1$Myr), or very long-lived ($\sim 100$Myr), or very low-density. If AGN disks are generally small and short-lived, they cannot generate long-lived feedback as apparently required for $\Lambda$CDM and so our present model of AGN feedback sculpting galaxy evolution requires revision. Small AGN disks are also unlikely to exchange much orbital angular momentum between merging SMBH so the rate of SMBH binaries may be lower than expected for LISA/PTA.

A high value of $f_{\rm BBH,AGN}$ likely implies that AGN out to $z \sim 1$ consist of repeat relatively short-lived (few Myr) events in a similar plane, powered by moderate-sized, thin disks, embedded in NSCs with substantial mass segregation. Such AGN may be efficient at solving the final pc problem and may be sufficient to drive extensive AGN feedback required for $\Lambda$CDM.

\begin{acknowledgments}
KESF \& BM are supported by NSF AST-2206096, AST-1831415 and Simons Foundation Grant 533845 as well as Simons Foundation support to release the public, open-source code used in this paper. The Flatiron Institute is supported by the Simons Foundation. The \texttt{McFACTS} code was developed and is maintained and updated by BM, KESF, Harry Cook, Vera Delfavero, Kaila Nathaniel, Jake Postiglione, Shawn Ray, Emily McPike, and Richard O'Shaughnessy and we are intensely grateful for their work and collaboration as colleagues. We also acknowledge the wider AGN Channel community, which has done an excellent job remaining collegial while also carefully scrutinizing results from varied and diverse groups. This spirit has been especially apparent at recent meetings such as `New Ideas on the Origin of Black Hole Mergers' (at NBIA) and `Dynamix' (at the IoA), ably organized by (for NBIA): Johan Samsing, Dan D'Orazio, Lorenz Zwick, Alessandro Trani and Christopher Tiede \& (for IoA): Mor Rozner, Barry Ginat, Johan Samsing, Bence Kocsis, and Cathie J. Clarke, respectively.
\end{acknowledgments}

%

\vspace{5mm}


\software{Astropy, \citep{2013A&A...558A..33A,2018AJ....156..123A}, pAGN \citep{pAGN24}, NumPy \citep{harris2020array}, SciPy \citep{2020SciPy-NMeth}, Matplotlib \citep{Hunter_2007}, McFACTS \citep{McFACTS}.
          }





\bibliography{refs1}{}

\begin{thebibliography}{}
\expandafter\ifx\csname natexlab\endcsname\relax\def\natexlab#1{#1}\fi
\providecommand{\url}[1]{\href{#1}{#1}}
\providecommand{\dodoi}[1]{doi:~\href{http://doi.org/#1}{\nolinkurl{#1}}}
\providecommand{\doeprint}[1]{\href{http://ascl.net/#1}{\nolinkurl{http://ascl.net/#1}}}
\providecommand{\doarXiv}[1]{\href{https://arxiv.org/abs/#1}{\nolinkurl{https://arxiv.org/abs/#1}}}

\bibitem[{{Abac} {et~al.}(2025){Abac}, {Abramo}, {Albanesi}, {Albertini}, {Agapito}, {Agathos}, {Albertus}, {Andersson}, {Andrade}, {Andreoni}, {Angeloni}, {Antonelli}, {Antoniadis}, {Antonini}, {Arca Sedda}, {Artale}, {Ascenzi}, {Auclair}, {Bachetti}, {Badger}, {Banerjee}, {Barba-Gonz{\'a}lez}, {Barta}, {Bartolo}, {Bauswein}, {Begnoni}, {Beirnaert}, {Bejger}, {Belgacem}, {Bellomo}, {Bernard}, {Grazia Bernardini}, {Bernuzzi}, {Berry}, {Berti}, {Bertone}, {Bettoni}, {Bezares}, {Bhagwat}, {Bisero}, {Bizouard}, {Blanco-Pillado}, {Blasi}, {Bonino}, {Borghese}, {Borghi}, {Borhanian}, {Bortolas}, {Botticella}, {Branchesi}, {Breschi}, {Brito}, {Brocato}, {Broekgaarden}, {Bulik}, {Buonanno}, {Burgio}, {Burrows}, {Calcagni}, {Canevarolo}, {Cappellaro}, {Capurri}, {Carbone}, {Casadio}, {Cayuso}, {Cerd{\'a}-Dur{\'a}n}, {Char}, {Chaty}, {Chiarusi}, {Chruslinska}, {Cireddu}, {Cole}, {Colombo}, {Colpi}, {Comp{\`e}re}, {Contaldi}, {Corman}, {Crescimbeni}, {Cristallo}, {Cuoco}, {Cusin}, {Dal Canton}, {D{\'a}lya}, {D'Avanzo},
  {Davari}, {De Luca}, {De Renzis}, {Della Valle}, {Del Pozzo}, {De Santi}, {Ludovico De Santis}, {Dietrich}, {Dimastrogiovanni}, {Domenech}, {Doneva}, {Drago}, {Dupletsa}, {Duval}, {Dvorkin}, {Elias-Rosa}, {Fairhurst}, {Fantina}, {Fasiello}, {Fays}, {Fender}, {Fischer}, {Foucart}, {Fragos}, {Foffa}, {Franciolini}, {Gair}, {Gamba}, {Garcia-Bellido}, {Garc{\'\i}a-Quir{\'o}s}, {{\'A}rp{\'a}d Gergely}, {Ghirlanda}, {Ghosh}, {Giacomazzo}, {Gittins}, {Giudice}, {Goncharov}, {Gonzalez}, {Gori{\'e}ly}, {Graziani}, {Greco}, {Gualtieri}, {Guidi}, {Gupta}, {Haney}, {Hannam}, {Harms}, {Harutyunyan}, {Haskell}, {Haungs}, {Hazra}, {Hemming}, {Heng}, {Hinderer}, {van der Horst}, {Hu}, {Husa}, {Iacovelli}, {Illuminati}, {Inguglia}, {Izquierdo Villalba}, {Janquart}, {Janssens}, {Jenkins}, {Jones}, {Kacskovics}, {Klessen}, {Kokkotas}, {Kuan}, {Kumar}, {Kuroyanagi}, {Laghi}, {Lamberts}, {Lambiase}, {Larrouturou}, {Leaci}, {Lenzi}, {Levan}, {Li}, {Li}, {Liang}, {Limongi}, {Liu}, {Llanes-Estrada}, {Loffredo}, {Long},
  {Lope-Oter}, {Lukes-Gerakopoulos}, {Maggio}, {Maggiore}, {Mancarella}, {Mapelli}, {Marchant}, {Margiotta}, {Mariotti}, {Marriott-Best}, {Marsat}, {Mart{\'\i}nez-Pinedo}, {Maselli}, {Mastrogiovanni}, {Matos}, {Melandri}, {Mendes}, {Mendon{\c{c}}a Soares de Souza}, {Mentasti}, {Mezcua}, {M{\"o}sta}, {Mondal}, {Moresco}, {Mukherjee}, {Muttoni}, {Nagar}, {Narola}, {Nava}, {Navarro Moreno}, \& {Nelemans}}]{ET25}
{Abac}, A., {Abramo}, R., {Albanesi}, S., {et~al.} 2025, arXiv e-prints, arXiv:2503.12263, \dodoi{10.48550/arXiv.2503.12263}

\bibitem[{{Abbott} {et~al.}(2021){Abbott}, {Abbott}, {Abbott}, {Abraham}, {Acernese}, {Ackley}, {Adams}, {Adhikari}, {Adya}, {Affeldt}, {Agathos}, {Agatsuma}, {Aggarwal}, {Aguiar}, {Aiello}, {Ain}, {Ajith}, {Allen}, {Allocca}, {Aloy}, {Altin}, {Amato}, {Anand}, {Ananyeva}, {Anderson}, {Anderson}, {Angelova}, {Antier}, {Appert}, {Arai}, {Araya}, {Areeda}, {Ar{\`e}ne}, {Arnaud}, {Aronson}, {Arun}, {Ascenzi}, {Ashton}, {Aston}, {Astone}, {Aubin}, {Aufmuth}, {AultONeal}, {Austin}, {Avendano}, {Avila-Alvarez}, {Babak}, {Bacon}, {Badaracco}, {Bader}, {Bae}, {Baird}, {Baker}, {Baldaccini}, {Ballardin}, {Ballmer}, {Bals}, {Banagiri}, {Barayoga}, {Barbieri}, {Barclay}, {Barish}, {Barker}, {Barkett}, {Barnum}, {Barone}, {Barr}, {Barsotti}, {Barsuglia}, {Barta}, {Bartlett}, {Bartos}, {Bassiri}, {Basti}, {Bawaj}, {Bayley}, {Bazzan}, {B{\'e}csy}, {Bejger}, {Belahcene}, {Bell}, {Beniwal}, {Benjamin}, {Berger}, {Bergmann}, {Bernuzzi}, {Berry}, {Bersanetti}, {Bertolini}, {Betzwieser}, {Bhandare}, {Bidler}, {Biggs},
  {Bilenko}, {Bilgili}, {Billingsley}, {Birney}, {Birnholtz}, {Biscans}, {Bischi}, {Biscoveanu}, {Bisht}, {Bitossi}, {Bizouard}, {Blackburn}, {Blackman}, {Blair}, {Blair}, {Blair}, {Bloemen}, {Bobba}, {Bode}, {Boer}, {Boetzel}, {Bogaert}, {Bondu}, {Bonnand}, {Booker}, {Boom}, {Bork}, {Boschi}, {Bose}, {Bossilkov}, {Bosveld}, {Bouffanais}, {Bozzi}, {Bradaschia}, {Brady}, {Bramley}, {Branchesi}, {Brau}, {Breschi}, {Briant}, {Briggs}, {Brighenti}, {Brillet}, {Brinkmann}, {Brockill}, {Brooks}, {Brooks}, {Brown}, {Brunett}, {Buikema}, {Bulik}, {Bulten}, {Buonanno}, {Buskulic}, {Buy}, {Byer}, {Cabero}, {Cadonati}, {Cagnoli}, {Cahillane}, {Calder{\'o}n Bustillo}, {Callister}, {Calloni}, {Camp}, {Campbell}, {Canepa}, {Cannon}, {Cao}, {Cao}, {Carapella}, {Carbognani}, {Caride}, {Carney}, {Carullo}, {Casanueva Diaz}, {Casentini}, {Caudill}, {Cavagli{\`a}}, {Cavalier}, {Cavalieri}, {Cella}, {Cerd{\'a}-Dur{\'a}n}, {Cesarini}, {Chaibi}, {Chakravarti}, {Chamberlin}, {Chan}, {Chao}, {Charlton}, {Chase}, {Chassande-Mottin},
  {Chatterjee}, {Chaturvedi}, {Cheeseboro}, {Chen}, {Chen}, {Chen}, {Cheng}, {Cheong}, {Chia}, {Chiadini}, {Chincarini}, {Chiummo}, {Cho}, {Cho}, {Cho}, \& {Christensen}}]{LVKH0}
{Abbott}, B.~P., {Abbott}, R., {Abbott}, T.~D., {et~al.} 2021, \apj, 909, 218, \dodoi{10.3847/1538-4357/abdcb7}

\bibitem[{{Abbott} {et~al.}(2020){Abbott}, {Abbott}, {Abraham}, {Acernese}, {Ackley}, {Adams}, {Adhikari}, {Adya}, {Affeldt}, {Agathos}, {Agatsuma}, {Aggarwal}, {Aguiar}, {Aich}, {Aiello}, {Ain}, {Ajith}, {Akcay}, {Allen}, {Allocca}, {Altin}, {Amato}, {Anand}, {Ananyeva}, {Anderson}, {Anderson}, {Angelova}, {Ansoldi}, {Antier}, {Appert}, {Arai}, {Araya}, {Areeda}, {Ar{\`e}ne}, {Arnaud}, {Aronson}, {Arun}, {Asali}, {Ascenzi}, {Ashton}, {Aston}, {Astone}, {Aubin}, {Aufmuth}, {AultONeal}, {Austin}, {Avendano}, {Babak}, {Bacon}, {Badaracco}, {Bader}, {Bae}, {Baer}, {Baird}, {Baldaccini}, {Ballardin}, {Ballmer}, {Bals}, {Balsamo}, {Baltus}, {Banagiri}, {Bankar}, {Bankar}, {Barayoga}, {Barbieri}, {Barish}, {Barker}, {Barkett}, {Barneo}, {Barone}, {Barr}, {Barsotti}, {Barsuglia}, {Barta}, {Bartlett}, {Bartos}, {Bassiri}, {Basti}, {Bawaj}, {Bayley}, {Bazzan}, {B{\'e}csy}, {Bejger}, {Belahcene}, {Bell}, {Beniwal}, {Benjamin}, {Benkel}, {Bentley}, {Bergamin}, {Berger}, {Bergmann}, {Bernuzzi}, {Berry}, {Bersanetti},
  {Bertolini}, {Betzwieser}, {Bhandare}, {Bhandari}, {Bidler}, {Biggs}, {Bilenko}, {Billingsley}, {Birney}, {Birnholtz}, {Biscans}, {Bischi}, {Biscoveanu}, {Bisht}, {Bissenbayeva}, {Bitossi}, {Bizouard}, {Blackburn}, {Blackman}, {Blair}, {Blair}, {Blair}, {Bobba}, {Bode}, {Boer}, {Boetzel}, {Bogaert}, {Bondu}, {Bonilla}, {Bonnand}, {Booker}, {Boom}, {Bork}, {Boschi}, {Bose}, {Bossilkov}, {Bosveld}, {Bouffanais}, {Bozzi}, {Bradaschia}, {Brady}, {Bramley}, {Branchesi}, {Brau}, {Breschi}, {Briant}, {Briggs}, {Brighenti}, {Brillet}, {Brinkmann}, {Brito}, {Brockill}, {Brooks}, {Brooks}, {Brown}, {Brunett}, {Bruno}, {Bruntz}, {Buikema}, {Bulik}, {Bulten}, {Buonanno}, {Buskulic}, {Byer}, {Cabero}, {Cadonati}, {Cagnoli}, {Cahillane}, {Bustillo}, {Callaghan}, {Callister}, {Calloni}, {Camp}, {Canepa}, {Cannon}, {Cao}, {Cao}, {Carapella}, {Carbognani}, {Caride}, {Carney}, {Carullo}, {Diaz}, {Casentini}, {Casta{\~n}eda}, {Caudill}, {Cavagli{\`a}}, {Cavalier}, {Cavalieri}, {Cella}, {Cerd{\'a}-Dur{\'a}n}, {Cesarini},
  {Chaibi}, {Chakravarti}, {Chan}, {Chan}, {Chao}, {Charlton}, {Chase}, {Chassande-Mottin}, {Chatterjee}, {Chaturvedi}, {Chatziioannou}, {Chen}, \& {Chen}}]{GW190814}
{Abbott}, R., {Abbott}, T.~D., {Abraham}, S., {et~al.} 2020, \apjl, 896, L44, \dodoi{10.3847/2041-8213/ab960f}

\bibitem[{{Adams} \& {Bloch}(2009)}]{AdamsBloch09}
{Adams}, F.~C., \& {Bloch}, A.~M. 2009, \apj, 701, 1381, \dodoi{10.1088/0004-637X/701/2/1381}

\bibitem[{{Akiba} {et~al.}(2024){Akiba}, {Naoz}, \& {Madigan}}]{Akiba24}
{Akiba}, T., {Naoz}, S., \& {Madigan}, A.-M. 2024, arXiv e-prints, arXiv:2410.19881, \dodoi{10.48550/arXiv.2410.19881}

\bibitem[{Amaro-Seoane {et~al.}(2007)Amaro-Seoane, Gair, Freitag, ller, Mandel, Cutler, \& Babak}]{Amaro_Seoane_2007}
Amaro-Seoane, P., Gair, J.~R., Freitag, M., {et~al.} 2007, Classical and Quantum Gravity, 24, R113, \dodoi{10.1088/0264-9381/24/17/R01}

\bibitem[{{Amaro-Seoane} {et~al.}(2023){Amaro-Seoane}, {Andrews}, {Arca Sedda}, {Askar}, {Baghi}, {Balasov}, {Bartos}, {Bavera}, {Bellovary}, {Berry}, {Berti}, {Bianchi}, {Blecha}, {Blondin}, {Bogdanovi{\'c}}, {Boissier}, {Bonetti}, {Bonoli}, {Bortolas}, {Breivik}, {Capelo}, {Caramete}, {Cattorini}, {Charisi}, {Chaty}, {Chen}, {Chru{\'s}li{\'n}ska}, {Chua}, {Church}, {Colpi}, {D'Orazio}, {Danielski}, {Davies}, {Dayal}, {De Rosa}, {Derdzinski}, {Destounis}, {Dotti}, {Du{\c{t}}an}, {Dvorkin}, {Fabj}, {Foglizzo}, {Ford}, {Fouvry}, {Franchini}, {Fragos}, {Fryer}, {Gaspari}, {Gerosa}, {Graziani}, {Groot}, {Habouzit}, {Haggard}, {Haiman}, {Han}, {Istrate}, {Johansson}, {Khan}, {Kimpson}, {Kokkotas}, {Kong}, {Korol}, {Kremer}, {Kupfer}, {Lamberts}, {Larson}, {Lau}, {Liu}, {Lloyd-Ronning}, {Lodato}, {Lupi}, {Ma}, {Maccarone}, {Mandel}, {Mangiagli}, {Mapelli}, {Mathis}, {Mayer}, {McGee}, {McKernan}, {Miller}, {Mota}, {Mumpower}, {Nasim}, {Nelemans}, {Noble}, {Pacucci}, {Panessa}, {Paschalidis}, {Pfister}, {Porquet},
  {Quenby}, {Ricarte}, {R{\"o}pke}, {Regan}, {Rosswog}, {Ruiter}, {Ruiz}, {Runnoe}, {Schneider}, {Schnittman}, {Secunda}, {Sesana}, {Seto}, {Shao}, {Shapiro}, {Sopuerta}, {Stone}, {Suvorov}, {Tamanini}, {Tamfal}, {Tauris}, {Temmink}, {Tomsick}, {Toonen}, {Torres-Orjuela}, {Toscani}, {Tsokaros}, {Unal}, {V{\'a}zquez-Aceves}, {Valiante}, {van Putten}, {van Roestel}, {Vignali}, {Volonteri}, {Wu}, {Younsi}, {Yu}, {Zane}, {Zwick}, {Antonini}, {Baibhav}, {Barausse}, {Bonilla Rivera}, {Branchesi}, {Branduardi-Raymont}, {Burdge}, {Chakraborty}, {Cuadra}, {Dage}, {Davis}, {de Mink}, {Decarli}, {Doneva}, {Escoffier}, {Gandhi}, {Haardt}, {Lousto}, {Nissanke}, {Nordhaus}, {O'Shaughnessy}, {Portegies Zwart}, {Pound}, {Schussler}, {Sergijenko}, {Spallicci}, {Vernieri}, \& {Vigna-G{\'o}mez}}]{LISA23}
{Amaro-Seoane}, P., {Andrews}, J., {Arca Sedda}, M., {et~al.} 2023, Living Reviews in Relativity, 26, 2, \dodoi{10.1007/s41114-022-00041-y}

\bibitem[{{Antonini} {et~al.}(2015){Antonini}, {Barausse}, \& {Silk}}]{Antonini15}
{Antonini}, F., {Barausse}, E., \& {Silk}, J. 2015, \apj, 812, 72, \dodoi{10.1088/0004-637X/812/1/72}

\bibitem[{{Antonini} {et~al.}(2012){Antonini}, {Capuzzo-Dolcetta}, {Mastrobuono-Battisti}, \& {Merritt}}]{Antonini12}
{Antonini}, F., {Capuzzo-Dolcetta}, R., {Mastrobuono-Battisti}, A., \& {Merritt}, D. 2012, \apj, 750, 111, \dodoi{10.1088/0004-637X/750/2/111}

\bibitem[{{Arca-Sedda} \& {Capuzzo-Dolcetta}(2014)}]{Arca14}
{Arca-Sedda}, M., \& {Capuzzo-Dolcetta}, R. 2014, \mnras, 444, 3738, \dodoi{10.1093/mnras/stu1683}

\bibitem[{{Arca Sedda} {et~al.}(2023){Arca Sedda}, {Naoz}, \& {Kocsis}}]{ArcaSedda23}
{Arca Sedda}, M., {Naoz}, S., \& {Kocsis}, B. 2023, Universe, 9, 138, \dodoi{10.3390/universe9030138}

\bibitem[{{Artymowicz} {et~al.}(1993){Artymowicz}, {Lin}, \& {Wampler}}]{Artymowicz93}
{Artymowicz}, P., {Lin}, D.~N.~C., \& {Wampler}, E.~J. 1993, \apj, 409, 592, \dodoi{10.1086/172690}

\bibitem[{{Astropy Collaboration} {et~al.}(2013){Astropy Collaboration}, {Robitaille}, {Tollerud}, {Greenfield}, {Droettboom}, {Bray}, {Aldcroft}, {Davis}, {Ginsburg}, {Price-Whelan}, {Kerzendorf}, {Conley}, {Crighton}, {Barbary}, {Muna}, {Ferguson}, {Grollier}, {Parikh}, {Nair}, {Unther}, {Deil}, {Woillez}, {Conseil}, {Kramer}, {Turner}, {Singer}, {Fox}, {Weaver}, {Zabalza}, {Edwards}, {Azalee Bostroem}, {Burke}, {Casey}, {Crawford}, {Dencheva}, {Ely}, {Jenness}, {Labrie}, {Lim}, {Pierfederici}, {Pontzen}, {Ptak}, {Refsdal}, {Servillat}, \& {Streicher}}]{2013A&A...558A..33A}
{Astropy Collaboration}, {Robitaille}, T.~P., {Tollerud}, E.~J., {et~al.} 2013, \aap, 558, A33, \dodoi{10.1051/0004-6361/201322068}

\bibitem[{{Astropy Collaboration} {et~al.}(2018){Astropy Collaboration}, {Price-Whelan}, {Sip{\H{o}}cz}, {G{\"u}nther}, {Lim}, {Crawford}, {Conseil}, {Shupe}, {Craig}, {Dencheva}, {Ginsburg}, {VanderPlas}, {Bradley}, {P{\'e}rez-Su{\'a}rez}, {de Val-Borro}, {Aldcroft}, {Cruz}, {Robitaille}, {Tollerud}, {Ardelean}, {Babej}, {Bach}, {Bachetti}, {Bakanov}, {Bamford}, {Barentsen}, {Barmby}, {Baumbach}, {Berry}, {Biscani}, {Boquien}, {Bostroem}, {Bouma}, {Brammer}, {Bray}, {Breytenbach}, {Buddelmeijer}, {Burke}, {Calderone}, {Cano Rodr{\'\i}guez}, {Cara}, {Cardoso}, {Cheedella}, {Copin}, {Corrales}, {Crichton}, {D'Avella}, {Deil}, {Depagne}, {Dietrich}, {Donath}, {Droettboom}, {Earl}, {Erben}, {Fabbro}, {Ferreira}, {Finethy}, {Fox}, {Garrison}, {Gibbons}, {Goldstein}, {Gommers}, {Greco}, {Greenfield}, {Groener}, {Grollier}, {Hagen}, {Hirst}, {Homeier}, {Horton}, {Hosseinzadeh}, {Hu}, {Hunkeler}, {Ivezi{\'c}}, {Jain}, {Jenness}, {Kanarek}, {Kendrew}, {Kern}, {Kerzendorf}, {Khvalko}, {King}, {Kirkby}, {Kulkarni},
  {Kumar}, {Lee}, {Lenz}, {Littlefair}, {Ma}, {Macleod}, {Mastropietro}, {McCully}, {Montagnac}, {Morris}, {Mueller}, {Mumford}, {Muna}, {Murphy}, {Nelson}, {Nguyen}, {Ninan}, {N{\"o}the}, {Ogaz}, {Oh}, {Parejko}, {Parley}, {Pascual}, {Patil}, {Patil}, {Plunkett}, {Prochaska}, {Rastogi}, {Reddy Janga}, {Sabater}, {Sakurikar}, {Seifert}, {Sherbert}, {Sherwood-Taylor}, {Shih}, {Sick}, {Silbiger}, {Singanamalla}, {Singer}, {Sladen}, {Sooley}, {Sornarajah}, {Streicher}, {Teuben}, {Thomas}, {Tremblay}, {Turner}, {Terr{\'o}n}, {van Kerkwijk}, {de la Vega}, {Watkins}, {Weaver}, {Whitmore}, {Woillez}, {Zabalza}, \& {Astropy Contributors}}]{2018AJ....156..123A}
{Astropy Collaboration}, {Price-Whelan}, A.~M., {Sip{\H{o}}cz}, B.~M., {et~al.} 2018, \aj, 156, 123, \dodoi{10.3847/1538-3881/aabc4f}

\bibitem[{{Bahcall} \& {Wolf}(1976)}]{BahcallWolf76}
{Bahcall}, J.~N., \& {Wolf}, R.~A. 1976, \apj, 209, 214, \dodoi{10.1086/154711}

\bibitem[{{Barausse} {et~al.}(2015){Barausse}, {Cardoso}, \& {Pani}}]{Barausse15}
{Barausse}, E., {Cardoso}, V., \& {Pani}, P. 2015, in Journal of Physics Conference Series, Vol. 610, Journal of Physics Conference Series (IOP), 012044, \dodoi{10.1088/1742-6596/610/1/012044}

\bibitem[{{Bardeen} \& {Petterson}(1975)}]{Bardeen75}
{Bardeen}, J.~M., \& {Petterson}, J.~A. 1975, \apjl, 195, L65, \dodoi{10.1086/181711}

\bibitem[{{Bartko} {et~al.}(2010){Bartko}, {Martins}, {Trippe}, {Fritz}, {Genzel}, {Ott}, {Eisenhauer}, {Gillessen}, {Paumard}, {Alexander}, {Dodds-Eden}, {Gerhard}, {Levin}, {Mascetti}, {Nayakshin}, {Perets}, {Perrin}, {Pfuhl}, {Reid}, {Rouan}, {Zilka}, \& {Sternberg}}]{Bartko10}
{Bartko}, H., {Martins}, F., {Trippe}, S., {et~al.} 2010, \apj, 708, 834, \dodoi{10.1088/0004-637X/708/1/834}

\bibitem[{{Bartos} {et~al.}(2017{\natexlab{a}}){Bartos}, {Haiman}, {Marka}, {Metzger}, {Stone}, \& {Marka}}]{Bartos17stats}
{Bartos}, I., {Haiman}, Z., {Marka}, Z., {et~al.} 2017{\natexlab{a}}, Nature Communications, 8, 831, \dodoi{10.1038/s41467-017-00851-7}

\bibitem[{{Bartos} {et~al.}(2017{\natexlab{b}}){Bartos}, {Kocsis}, {Haiman}, \& {M{\'a}rka}}]{Bartos17}
{Bartos}, I., {Kocsis}, B., {Haiman}, Z., \& {M{\'a}rka}, S. 2017{\natexlab{b}}, \apj, 835, 165, \dodoi{10.3847/1538-4357/835/2/165}

\bibitem[{{Baruteau} {et~al.}(2011){Baruteau}, {Cuadra}, \& {Lin}}]{Baruteau11}
{Baruteau}, C., {Cuadra}, J., \& {Lin}, D.~N.~C. 2011, \apj, 726, 28, \dodoi{10.1088/0004-637X/726/1/28}

\bibitem[{{Bellovary} {et~al.}(2016){Bellovary}, {Mac Low}, {McKernan}, \& {Ford}}]{Bellovary16}
{Bellovary}, J.~M., {Mac Low}, M.-M., {McKernan}, B., \& {Ford}, K.~E.~S. 2016, \apjl, 819, L17, \dodoi{10.3847/2041-8205/819/2/L17}

\bibitem[{{Berti} \& {Volonteri}(2008)}]{BertiVolonteri08}
{Berti}, E., \& {Volonteri}, M. 2008, \apj, 684, 822, \dodoi{10.1086/590379}

\bibitem[{{Binney} \& {Tremaine}(1987)}]{BinneyTremaine87}
{Binney}, J., \& {Tremaine}, S. 1987, {Galactic dynamics}

\bibitem[{{Bitsch} \& {Kley}(2010)}]{Bitsch10}
{Bitsch}, B., \& {Kley}, W. 2010, \aap, 523, A30, \dodoi{10.1051/0004-6361/201014414}

\bibitem[{{Byrne} {et~al.}(2024){Byrne}, {Faucher-Gigu{\`e}re}, {Wellons}, {Hopkins}, {Angl{\'e}s-Alc{\'a}zar}, {Sultan}, {Wijers}, {Moreno}, \& {Ponnada}}]{Byrne24}
{Byrne}, L., {Faucher-Gigu{\`e}re}, C.-A., {Wellons}, S., {et~al.} 2024, \apj, 973, 149, \dodoi{10.3847/1538-4357/ad67ca}

\bibitem[{{Cabrera} {et~al.}(2024){Cabrera}, {Palmese}, {Hu}, {O'Connor}, {Ford}, {McKernan}, {Andreoni}, {Ahumada}, {Amsellem}, {Busmann}, {Clark}, {Coughlin}, {Dadiani}, {Diaz}, {Graham}, {Gruen}, {Kunnumkai}, {Postiglione}, {Sommer}, \& {Valdes}}]{Cabrera24}
{Cabrera}, T., {Palmese}, A., {Hu}, L., {et~al.} 2024, arXiv e-prints, arXiv:2407.10698, \dodoi{10.48550/arXiv.2407.10698}

\bibitem[{{Calcino} {et~al.}(2023){Calcino}, {Dempsey}, {Dittmann}, \& {Li}}]{Calcino23}
{Calcino}, J., {Dempsey}, A.~M., {Dittmann}, A.~J., \& {Li}, H. 2023, arXiv e-prints, arXiv:2311.13727, \dodoi{10.48550/arXiv.2311.13727}

\bibitem[{{Callister} {et~al.}(2021){Callister}, {Haster}, {Ng}, {Vitale}, \& {Farr}}]{Callister21}
{Callister}, T.~A., {Haster}, C.-J., {Ng}, K. K.~Y., {Vitale}, S., \& {Farr}, W.~M. 2021, \apjl, 922, L5, \dodoi{10.3847/2041-8213/ac2ccc}

\bibitem[{{Campanelli} {et~al.}(2007){Campanelli}, {Lousto}, {Zlochower}, \& {Merritt}}]{Campanelli07}
{Campanelli}, M., {Lousto}, C.~O., {Zlochower}, Y., \& {Merritt}, D. 2007, \prl, 98, 231102, \dodoi{10.1103/PhysRevLett.98.231102}

\bibitem[{{Cantiello} {et~al.}(2021){Cantiello}, {Jermyn}, \& {Lin}}]{Cantiello21}
{Cantiello}, M., {Jermyn}, A.~S., \& {Lin}, D. N.~C. 2021, \apj, 910, 94, \dodoi{10.3847/1538-4357/abdf4f}

\bibitem[{{Capuzzo-Dolcetta} \& {Vicari}(2005)}]{Capuzzo05}
{Capuzzo-Dolcetta}, R., \& {Vicari}, A. 2005, \mnras, 356, 899, \dodoi{10.1111/j.1365-2966.2004.08433.x}

\bibitem[{{Chen} \& {Dai}(2025)}]{ChenEM25}
{Chen}, K., \& {Dai}, Z.-G. 2025, arXiv e-prints, arXiv:2505.16390, \dodoi{10.48550/arXiv.2505.16390}

\bibitem[{{Chen} {et~al.}(2019){Chen}, {Li}, \& {Cao}}]{XianChen19}
{Chen}, X., {Li}, S., \& {Cao}, Z. 2019, \mnras, 485, L141, \dodoi{10.1093/mnrasl/slz046}

\bibitem[{{Chen} {et~al.}(2022){Chen}, {Bailey}, {Stone}, \& {Zhu}}]{Chen22}
{Chen}, Y.-X., {Bailey}, A., {Stone}, J., \& {Zhu}, Z. 2022, \apjl, 939, L23, \dodoi{10.3847/2041-8213/ac9b3e}

\bibitem[{{Chen} {et~al.}(2025){Chen}, {Jiang}, \& {Goodman}}]{Chen25}
{Chen}, Y.-X., {Jiang}, Y.-F., \& {Goodman}, J. 2025, arXiv e-prints, arXiv:2505.13951, \dodoi{10.48550/arXiv.2505.13951}

\bibitem[{{Chen} {et~al.}(2023){Chen}, {Do}, {Ghez}, {Hosek}, {Feldmeier-Krause}, {Chu}, {Bentley}, {Lu}, \& {Morris}}]{Chen23}
{Chen}, Z., {Do}, T., {Ghez}, A.~M., {et~al.} 2023, \apj, 944, 79, \dodoi{10.3847/1538-4357/aca8ad}

\bibitem[{{Chen} {et~al.}(2024){Chen}, {Chen}, {Peng}, \& {Huang}}]{ChenChen24}
{Chen}, Z.-f., {Chen}, Z.-G., {Peng}, X.-l., \& {Huang}, W.-r. 2024, \apj, 974, 277, \dodoi{10.3847/1538-4357/ad79ff}

\bibitem[{{Cook} {et~al.}(2024){Cook}, {McKernan}, {Ford}, {Delfavero}, {Nathaniel}, {Postiglione}, {Ray}, \& {O'Shaughnessy}}]{Cook24}
{Cook}, H.~E., {McKernan}, B., {Ford}, K.~E.~S., {et~al.} 2024, arXiv e-prints, arXiv:2411.10590, \dodoi{10.48550/arXiv.2411.10590}

\bibitem[{{Davies} \& {Lin}(2020)}]{DaviesLin20}
{Davies}, M.~B., \& {Lin}, D. N.~C. 2020, \mnras, 498, 3452, \dodoi{10.1093/mnras/staa2590}

\bibitem[{{DeLaurentiis} {et~al.}(2023{\natexlab{a}}){DeLaurentiis}, {Epstein-Martin}, \& {Haiman}}]{Stan23}
{DeLaurentiis}, S., {Epstein-Martin}, M., \& {Haiman}, Z. 2023{\natexlab{a}}, \mnras, 523, 1126, \dodoi{10.1093/mnras/stad1412}

\bibitem[{{DeLaurentiis} {et~al.}(2023{\natexlab{b}}){DeLaurentiis}, {Epstein-Martin}, \& {Haiman}}]{DeLaurentiis23}
---. 2023{\natexlab{b}}, \mnras, 523, 1126, \dodoi{10.1093/mnras/stad1412}

\bibitem[{{Delfavero} {et~al.}(2024){Delfavero}, {Ford}, {McKernan}, {Cook}, {Nathaniel}, {Postiglione}, {Ray}, \& {O'Shaughnessy}}]{Delfavero24}
{Delfavero}, V., {Ford}, K.~E.~S., {McKernan}, B., {et~al.} 2024, arXiv e-prints, arXiv:2410.18815, \dodoi{10.48550/arXiv.2410.18815}

\bibitem[{{Di Matteo} {et~al.}(2005){Di Matteo}, {Springel}, \& {Hernquist}}]{DiMatteo05}
{Di Matteo}, T., {Springel}, V., \& {Hernquist}, L. 2005, \nat, 433, 604, \dodoi{10.1038/nature03335}

\bibitem[{{Dittmann} {et~al.}(2024){Dittmann}, {Dempsey}, \& {Li}}]{AlexD24}
{Dittmann}, A.~J., {Dempsey}, A.~M., \& {Li}, H. 2024, \apj, 964, 61, \dodoi{10.3847/1538-4357/ad23ce}

\bibitem[{{Dodici} \& {Tremaine}(2024)}]{DodiciTremaine24}
{Dodici}, M., \& {Tremaine}, S. 2024, arXiv e-prints, arXiv:2404.08138, \dodoi{10.48550/arXiv.2404.08138}

\bibitem[{{D'Orazio} \& {Loeb}(2020)}]{Dorazio20}
{D'Orazio}, D.~J., \& {Loeb}, A. 2020, \prd, 101, 083031, \dodoi{10.1103/PhysRevD.101.083031}

\bibitem[{{Fabj} {et~al.}(2020){Fabj}, {Nasim}, {Caban}, {Ford}, {McKernan}, \& {Bellovary}}]{Fabj20}
{Fabj}, G., {Nasim}, S.~S., {Caban}, F., {et~al.} 2020, \mnras, 499, 2608, \dodoi{10.1093/mnras/staa3004}

\bibitem[{{Fabj} \& {Samsing}(2024)}]{Fabj24}
{Fabj}, G., \& {Samsing}, J. 2024, \mnras, 535, 3630, \dodoi{10.1093/mnras/stae2499}

\bibitem[{{Farmer} {et~al.}(2019){Farmer}, {Renzo}, {de Mink}, {Marchant}, \& {Justham}}]{Farmer19}
{Farmer}, R., {Renzo}, M., {de Mink}, S.~E., {Marchant}, P., \& {Justham}, S. 2019, \apj, 887, 53, \dodoi{10.3847/1538-4357/ab518b}

\bibitem[{{Ford} \& {McKernan}(2022)}]{Ford22}
{Ford}, K.~E.~S., \& {McKernan}, B. 2022, \mnras, 517, 5827, \dodoi{10.1093/mnras/stac2861}

\bibitem[{{Ford} {et~al.}(2019){Ford}, {Bartos}, {McKernan}, {Haiman}, {Corsi}, {Keivani}, {Marka}, {Perna}, {Graham}, {Ross}, {Stern}, {Bellovary}, {Berti}, {O'Dowd}, {Lyra}, {MacLow}, \& {Marka}}]{FordWP}
{Ford}, K.~E.~S., {Bartos}, I., {McKernan}, B., {et~al.} 2019, \baas, 51, 247, \dodoi{10.48550/arXiv.1903.09529}

\bibitem[{{Fragione} {et~al.}(2019){Fragione}, {Grishin}, {Leigh}, {Perets}, \& {Perna}}]{Fragione19}
{Fragione}, G., {Grishin}, E., {Leigh}, N. W.~C., {Perets}, H.~B., \& {Perna}, R. 2019, \mnras, 488, 47, \dodoi{10.1093/mnras/stz1651}

\bibitem[{{Gabor} {et~al.}(2009){Gabor}, {Impey}, {Jahnke}, {Simmons}, {Trump}, {Koekemoer}, {Brusa}, {Cappelluti}, {Schinnerer}, {Smol{\v{c}}i{\'c}}, {Salvato}, {Rhodes}, {Mobasher}, {Capak}, {Massey}, {Leauthaud}, \& {Scoville}}]{Gabor09}
{Gabor}, J.~M., {Impey}, C.~D., {Jahnke}, K., {et~al.} 2009, \apj, 691, 705, \dodoi{10.1088/0004-637X/691/1/705}

\bibitem[{{Gangardt} {et~al.}(2024){Gangardt}, {Trani}, {Bonnerot}, \& {Gerosa}}]{pAGN24}
{Gangardt}, D., {Trani}, A.~A., {Bonnerot}, C., \& {Gerosa}, D. 2024, \mnras, 530, 3689, \dodoi{10.1093/mnras/stae1117}

\bibitem[{{Garg} {et~al.}(2022){Garg}, {Derdzinski}, {Zwick}, {Capelo}, \& {Mayer}}]{Mudit22}
{Garg}, M., {Derdzinski}, A., {Zwick}, L., {Capelo}, P.~R., \& {Mayer}, L. 2022, \mnras, 517, 1339, \dodoi{10.1093/mnras/stac2711}

\bibitem[{{Gayathri} {et~al.}(2023){Gayathri}, {Wysocki}, {Yang}, {Delfavero}, {O'Shaughnessy}, {Haiman}, {Tagawa}, \& {Bartos}}]{Gayathri23}
{Gayathri}, V., {Wysocki}, D., {Yang}, Y., {et~al.} 2023, \apjl, 945, L29, \dodoi{10.3847/2041-8213/acbfb8}

\bibitem[{{Gayathri} {et~al.}(2021{\natexlab{a}}){Gayathri}, {Yang}, {Tagawa}, {Haiman}, \& {Bartos}}]{Gayathri21}
{Gayathri}, V., {Yang}, Y., {Tagawa}, H., {Haiman}, Z., \& {Bartos}, I. 2021{\natexlab{a}}, \apjl, 920, L42, \dodoi{10.3847/2041-8213/ac2cc1}

\bibitem[{{Gayathri} {et~al.}(2021{\natexlab{b}}){Gayathri}, {Healy}, {Lange}, {O'Brien}, {Szczepanczyk}, {Bartos}, {Campanelli}, {Klimenko}, {Lousto}, \& {O'Shaughnessy}}]{GayathriH0}
{Gayathri}, V., {Healy}, J., {Lange}, J., {et~al.} 2021{\natexlab{b}}, \apjl, 908, L34, \dodoi{10.3847/2041-8213/abe388}

\bibitem[{{Gayathri} {et~al.}(2022){Gayathri}, {Healy}, {Lange}, {O'Brien}, {Szczepa{\'n}czyk}, {Bartos}, {Campanelli}, {Klimenko}, {Lousto}, \& {O'Shaughnessy}}]{Gayathri22}
---. 2022, Nature Astronomy, 6, 344, \dodoi{10.1038/s41550-021-01568-w}

\bibitem[{{Generozov} {et~al.}(2018){Generozov}, {Stone}, {Metzger}, \& {Ostriker}}]{Generozov18}
{Generozov}, A., {Stone}, N.~C., {Metzger}, B.~D., \& {Ostriker}, J.~P. 2018, \mnras, 478, 4030, \dodoi{10.1093/mnras/sty1262}

\bibitem[{{Genzel} {et~al.}(2003){Genzel}, {Sch{\"o}del}, {Ott}, {Eisenhauer}, {Hofmann}, {Lehnert}, {Eckart}, {Alexander}, {Sternberg}, {Lenzen}, {Cl{\'e}net}, {Lacombe}, {Rouan}, {Renzini}, \& {Tacconi-Garman}}]{Genzel03}
{Genzel}, R., {Sch{\"o}del}, R., {Ott}, T., {et~al.} 2003, \apj, 594, 812, \dodoi{10.1086/377127}

\bibitem[{{Gerosa} \& {Berti}(2019)}]{Gerosa19}
{Gerosa}, D., \& {Berti}, E. 2019, \prd, 100, 041301, \dodoi{10.1103/PhysRevD.100.041301}

\bibitem[{{Ghez} {et~al.}(2004){Ghez}, {Wright}, {Matthews}, {Thompson}, {Le Mignant}, {Tanner}, {Hornstein}, {Morris}, {Becklin}, \& {Soifer}}]{Ghez04}
{Ghez}, A.~M., {Wright}, S.~A., {Matthews}, K., {et~al.} 2004, \apjl, 601, L159, \dodoi{10.1086/382024}

\bibitem[{{Gnedin} {et~al.}(2014){Gnedin}, {Ostriker}, \& {Tremaine}}]{Gnedin14}
{Gnedin}, O.~Y., {Ostriker}, J.~P., \& {Tremaine}, S. 2014, \apj, 785, 71, \dodoi{10.1088/0004-637X/785/1/71}

\bibitem[{{Gond{\'a}n} \& {Kocsis}(2022)}]{BenceGW22}
{Gond{\'a}n}, L., \& {Kocsis}, B. 2022, \mnras, 515, 3299, \dodoi{10.1093/mnras/stac1985}

\bibitem[{{Gonz{\'a}lez-Mart{\'\i}n} {et~al.}(2009){Gonz{\'a}lez-Mart{\'\i}n}, {Masegosa}, {M{\'a}rquez}, {Guainazzi}, \& {Jim{\'e}nez-Bail{\'o}n}}]{Gonzalez09}
{Gonz{\'a}lez-Mart{\'\i}n}, O., {Masegosa}, J., {M{\'a}rquez}, I., {Guainazzi}, M., \& {Jim{\'e}nez-Bail{\'o}n}, E. 2009, \aap, 506, 1107, \dodoi{10.1051/0004-6361/200912288}

\bibitem[{{Goodman} \& {Tan}(2004)}]{Goodman04}
{Goodman}, J., \& {Tan}, J.~C. 2004, \apj, 608, 108, \dodoi{10.1086/386360}

\bibitem[{{Graham} {et~al.}(2020{\natexlab{a}}){Graham}, {Ford}, {McKernan}, {Ross}, {Stern}, {Burdge}, {Coughlin}, {Djorgovski}, {Drake}, {Duev}, {Kasliwal}, {Mahabal}, {van Velzen}, {Belecki}, {Bellm}, {Burruss}, {Cenko}, {Cunningham}, {Helou}, {Kulkarni}, {Masci}, {Prince}, {Reiley}, {Rodriguez}, {Rusholme}, {Smith}, \& {Soumagnac}}]{Graham20}
{Graham}, M.~J., {Ford}, K.~E.~S., {McKernan}, B., {et~al.} 2020{\natexlab{a}}, \prl, 124, 251102, \dodoi{10.1103/PhysRevLett.124.251102}

\bibitem[{{Graham} {et~al.}(2020{\natexlab{b}}){Graham}, {Ross}, {Stern}, {Drake}, {McKernan}, {Ford}, {Djorgovski}, {Mahabal}, {Glikman}, {Larson}, \& {Christensen}}]{Graham20a}
{Graham}, M.~J., {Ross}, N.~P., {Stern}, D., {et~al.} 2020{\natexlab{b}}, \mnras, 491, 4925, \dodoi{10.1093/mnras/stz3244}

\bibitem[{{Graham} {et~al.}(2023){Graham}, {McKernan}, {Ford}, {Stern}, {Djorgovski}, {Coughlin}, {Burdge}, {Bellm}, {Helou}, {Mahabal}, {Masci}, {Purdum}, {Rosnet}, \& {Rusholme}}]{Graham23}
{Graham}, M.~J., {McKernan}, B., {Ford}, K.~E.~S., {et~al.} 2023, \apj, 942, 99, \dodoi{10.3847/1538-4357/aca480}

\bibitem[{{Grishin} {et~al.}(2023){Grishin}, {Gilbaum}, \& {Stone}}]{Grishin23}
{Grishin}, E., {Gilbaum}, S., \& {Stone}, N.~C. 2023, arXiv e-prints, arXiv:2307.07546, \dodoi{10.48550/arXiv.2307.07546}

\bibitem[{{Gr{\"o}bner} {et~al.}(2020){Gr{\"o}bner}, {Ishibashi}, {Tiwari}, {Haney}, \& {Jetzer}}]{Grobner20}
{Gr{\"o}bner}, M., {Ishibashi}, W., {Tiwari}, S., {Haney}, M., \& {Jetzer}, P. 2020, \aap, 638, A119, \dodoi{10.1051/0004-6361/202037681}

\bibitem[{{Hailey} {et~al.}(2018){Hailey}, {Mori}, {Bauer}, {Berkowitz}, {Hong}, \& {Hord}}]{Hailey18}
{Hailey}, C.~J., {Mori}, K., {Bauer}, F.~E., {et~al.} 2018, \nat, 556, 70, \dodoi{10.1038/nature25029}

\bibitem[{{Han} {et~al.}(2024){Han}, {Yang}, {Tagawa}, {Jiang}, {Shen}, {Yun}, {Zhang}, \& {Zhong}}]{Han24}
{Han}, W.-B., {Yang}, S.-C., {Tagawa}, H., {et~al.} 2024, arXiv e-prints, arXiv:2401.01743, \dodoi{10.48550/arXiv.2401.01743}

\bibitem[{{Hankla} {et~al.}(2020){Hankla}, {Jiang}, \& {Armitage}}]{Hankla20}
{Hankla}, A.~M., {Jiang}, Y.-F., \& {Armitage}, P.~J. 2020, \apj, 902, 50, \dodoi{10.3847/1538-4357/abb4df}

\bibitem[{Harris {et~al.}(2020)Harris, Millman, van~der Walt, Gommers, Virtanen, Cournapeau, Wieser, Taylor, Berg, Smith, Kern, Picus, Hoyer, van Kerkwijk, Brett, Haldane, del R{\'{i}}o, Wiebe, Peterson, G{\'{e}}rard-Marchant, Sheppard, Reddy, Weckesser, Abbasi, Gohlke, \& Oliphant}]{harris2020array}
Harris, C.~R., Millman, K.~J., van~der Walt, S.~J., {et~al.} 2020, Nature, 585, 357, \dodoi{10.1038/s41586-020-2649-2}

\bibitem[{{Heggie} {et~al.}(1996){Heggie}, {Hut}, \& {McMillan}}]{Heggie96}
{Heggie}, D.~C., {Hut}, P., \& {McMillan}, S. L.~W. 1996, \apj, 467, 359, \dodoi{10.1086/177611}

\bibitem[{{Hern{\'a}ndez-Garc{\'\i}a} {et~al.}(2025){Hern{\'a}ndez-Garc{\'\i}a}, {Chakraborty}, {S{\'a}nchez-S{\'a}ez}, {Ricci}, {Cuadra}, {McKernan}, {Ford}, {Ar{\'e}valo}, {Rau}, {Arcodia}, {Kara}, {Liu}, {Merloni}, {Bruni}, {Goodwin}, {Arzoumanian}, {Assef}, {Baldini}, {Bayo}, {Bauer}, {Bernal}, {Brightman}, {Calistro Rivera}, {Gendreau}, {Homan}, {Krumpe}, {Lira}, {Mart{\'\i}nez-Aldama}, {Salvato}, \& {Sotomayor}}]{Hernandez25}
{Hern{\'a}ndez-Garc{\'\i}a}, L., {Chakraborty}, J., {S{\'a}nchez-S{\'a}ez}, P., {et~al.} 2025, Nature Astronomy, \dodoi{10.1038/s41550-025-02523-9}

\bibitem[{{Hills}(1975)}]{Hills75}
{Hills}, J.~G. 1975, \aj, 80, 809, \dodoi{10.1086/111815}

\bibitem[{{Hills}(1988)}]{Hills88}
---. 1988, \nat, 331, 687, \dodoi{10.1038/331687a0}

\bibitem[{{Ho}(2008)}]{Ho08}
{Ho}, L.~C. 2008, \araa, 46, 475, \dodoi{10.1146/annurev.astro.45.051806.110546}

\bibitem[{{Hopkins} {et~al.}(2024){Hopkins}, {Squire}, {Su}, {Steinwandel}, {Kremer}, {Shi}, {Grudic}, {Wellons}, {Faucher-Giguere}, {Angles-Alcazar}, {Murray}, \& {Quataert}}]{Hopkins24}
{Hopkins}, P.~F., {Squire}, J., {Su}, K.-Y., {et~al.} 2024, The Open Journal of Astrophysics, 7, 19, \dodoi{10.21105/astro.2310.04506}

\bibitem[{{Hopman} \& {Alexander}(2006{\natexlab{a}})}]{Hopman06a}
{Hopman}, C., \& {Alexander}, T. 2006{\natexlab{a}}, \apjl, 645, L133, \dodoi{10.1086/506273}

\bibitem[{{Hopman} \& {Alexander}(2006{\natexlab{b}})}]{Hopman06}
---. 2006{\natexlab{b}}, \apj, 645, 1152, \dodoi{10.1086/504400}

\bibitem[{{Hunt} \& {Malkan}(1999)}]{Malkan99}
{Hunt}, L.~K., \& {Malkan}, M.~A. 1999, \apj, 516, 660, \dodoi{10.1086/307150}

\bibitem[{Hunter(2007)}]{Hunter_2007}
Hunter, J.~D. 2007, Computing in Science \& Engineering, 9, 90, \dodoi{10.1109/MCSE.2007.55}

\bibitem[{{Inayoshi} {et~al.}(2017){Inayoshi}, {Tamanini}, {Caprini}, \& {Haiman}}]{Inayoshi17}
{Inayoshi}, K., {Tamanini}, N., {Caprini}, C., \& {Haiman}, Z. 2017, \prd, 96, 063014, \dodoi{10.1103/PhysRevD.96.063014}

\bibitem[{{Jermyn} {et~al.}(2022){Jermyn}, {Dittmann}, {McKernan}, {Ford}, \& {Cantiello}}]{Jermyn22}
{Jermyn}, A.~S., {Dittmann}, A.~J., {McKernan}, B., {Ford}, K.~E.~S., \& {Cantiello}, M. 2022, \apj, 929, 133, \dodoi{10.3847/1538-4357/ac5d40}

\bibitem[{{Jim{\'e}nez} \& {Masset}(2017)}]{JM17}
{Jim{\'e}nez}, M.~A., \& {Masset}, F.~S. 2017, \mnras, 471, 4917, \dodoi{10.1093/mnras/stx1946}

\bibitem[{{Kawamura} {et~al.}(2006){Kawamura}, {Nakamura}, {Ando}, {Seto}, {Tsubono}, {Numata}, {Takahashi}, {Nagano}, {Ishikawa}, {Musha}, {Ueda}, {Sato}, {Hosokawa}, {Agatsuma}, {Akutsu}, {Aoyanagi}, {Arai}, {Araya}, {Asada}, {Aso}, {Chiba}, {Ebisuzaki}, {Eriguchi}, {Fujimoto}, {Fukushima}, {Futamase}, {Ganzu}, {Harada}, {Hashimoto}, {Hayama}, {Hikida}, {Himemoto}, {Hirabayashi}, {Hiramatsu}, {Ichiki}, {Ikegami}, {Inoue}, {Ioka}, {Ishidoshiro}, {Itoh}, {Kamagasako}, {Kanda}, {Kawashima}, {Kirihara}, {Kiuchi}, {Kobayashi}, {Kohri}, {Kojima}, {Kokeyama}, {Kozai}, {Kudoh}, {Kunimori}, {Kuroda}, {Maeda}, {Matsuhara}, {Mino}, {Miyakawa}, {Miyoki}, {Mizusawa}, {Morisawa}, {Mukohyama}, {Naito}, {Nakagawa}, {Nakamura}, {Nakano}, {Nakao}, {Nishizawa}, {Niwa}, {Nozawa}, {Ohashi}, {Ohishi}, {Ohkawa}, {Okutomi}, {Oohara}, {Sago}, {Saijo}, {Sakagami}, {Sakata}, {Sasaki}, {Sato}, {Shibata}, {Shinkai}, {Somiya}, {Sotani}, {Sugiyama}, {Tagoshi}, {Takahashi}, {Takahashi}, {Takahashi}, {Takano}, {Tanaka}, {Taniguchi},
  {Taruya}, {Tashiro}, {Tokunari}, {Tsujikawa}, {Tsunesada}, {Yamamoto}, {Yamazaki}, {Yokoyama}, {Yoo}, {Yoshida}, \& {Yoshino}}]{Decigo06}
{Kawamura}, S., {Nakamura}, T., {Ando}, M., {et~al.} 2006, Classical and Quantum Gravity, 23, S125, \dodoi{10.1088/0264-9381/23/8/S17}

\bibitem[{{Kim} \& {Most}(2024)}]{KimMost24}
{Kim}, Y., \& {Most}, E.~R. 2024, arXiv e-prints, arXiv:2409.12359, \dodoi{10.48550/arXiv.2409.12359}

\bibitem[{{King} {et~al.}(2005){King}, {Lubow}, {Ogilvie}, \& {Pringle}}]{King05}
{King}, A.~R., {Lubow}, S.~H., {Ogilvie}, G.~I., \& {Pringle}, J.~E. 2005, \mnras, 363, 49, \dodoi{10.1111/j.1365-2966.2005.09378.x}

\bibitem[{{Knee} {et~al.}(2024){Knee}, {McIver}, {Naoz}, {Romero-Shaw}, {Hoang}, \& {Grishin}}]{Knee24}
{Knee}, A.~M., {McIver}, J., {Naoz}, S., {et~al.} 2024, \apjl, 971, L38, \dodoi{10.3847/2041-8213/ad6a10}

\bibitem[{{Kocsis} {et~al.}(2011){Kocsis}, {Yunes}, \& {Loeb}}]{Kocsis11}
{Kocsis}, B., {Yunes}, N., \& {Loeb}, A. 2011, \prd, 84, 024032, \dodoi{10.1103/PhysRevD.84.024032}

\bibitem[{{Kormendy} \& {Ho}(2013)}]{KormendyHo13}
{Kormendy}, J., \& {Ho}, L.~C. 2013, \araa, 51, 511, \dodoi{10.1146/annurev-astro-082708-101811}

\bibitem[{{Koudmani} {et~al.}(2024){Koudmani}, {Somerville}, {Sijacki}, {Bourne}, {Jiang}, \& {Profit}}]{Koudmani24}
{Koudmani}, S., {Somerville}, R.~S., {Sijacki}, D., {et~al.} 2024, \mnras, 532, 60, \dodoi{10.1093/mnras/stae1422}

\bibitem[{{Kritos} {et~al.}(2024{\natexlab{a}}){Kritos}, {Reali}, {Gerosa}, \& {Berti}}]{Kritos24}
{Kritos}, K., {Reali}, L., {Gerosa}, D., \& {Berti}, E. 2024{\natexlab{a}}, \prd, 110, 123017, \dodoi{10.1103/PhysRevD.110.123017}

\bibitem[{{Kritos} {et~al.}(2024{\natexlab{b}}){Kritos}, {Strokov}, {Baibhav}, \& {Berti}}]{KritosRapster24}
{Kritos}, K., {Strokov}, V., {Baibhav}, V., \& {Berti}, E. 2024{\natexlab{b}}, \prd, 110, 043023, \dodoi{10.1103/PhysRevD.110.043023}

\bibitem[{{Lazzati} {et~al.}(2022){Lazzati}, {Soares}, \& {Perna}}]{Lazzati22}
{Lazzati}, D., {Soares}, G., \& {Perna}, R. 2022, \apjl, 938, L18, \dodoi{10.3847/2041-8213/ac98ad}

\bibitem[{{Leigh} {et~al.}(2018){Leigh}, {Geller}, {McKernan}, {Ford}, {Mac Low}, {Bellovary}, {Haiman}, {Lyra}, {Samsing}, {O'Dowd}, {Kocsis}, \& {Endlich}}]{Leigh18}
{Leigh}, N.~W.~C., {Geller}, A.~M., {McKernan}, B., {et~al.} 2018, \mnras, 474, 5672, \dodoi{10.1093/mnras/stx3134}

\bibitem[{{Levin}(2007)}]{Levin07}
{Levin}, Y. 2007, \mnras, 374, 515, \dodoi{10.1111/j.1365-2966.2006.11155.x}

\bibitem[{{Li} {et~al.}(2022{\natexlab{a}}){Li}, {Lai}, \& {Rodet}}]{Jairu22}
{Li}, J., {Lai}, D., \& {Rodet}, L. 2022{\natexlab{a}}, \apj, 934, 154, \dodoi{10.3847/1538-4357/ac7c0d}

\bibitem[{{Li} \& {Lai}(2022)}]{Rixin22}
{Li}, R., \& {Lai}, D. 2022, \mnras, 517, 1602, \dodoi{10.1093/mnras/stac2577}

\bibitem[{{Li} {et~al.}(2022{\natexlab{b}}){Li}, {Chen}, {Lin}, \& {Wang}}]{YaPing2022}
{Li}, Y.-P., {Chen}, Y.-X., {Lin}, D. N.~C., \& {Wang}, Z. 2022{\natexlab{b}}, \apjl, 928, L1, \dodoi{10.3847/2041-8213/ac5b61}

\bibitem[{{Lousto} {et~al.}(2012){Lousto}, {Zlochower}, {Dotti}, \& {Volonteri}}]{Lousto12}
{Lousto}, C.~O., {Zlochower}, Y., {Dotti}, M., \& {Volonteri}, M. 2012, \prd, 85, 084015, \dodoi{10.1103/PhysRevD.85.084015}

\bibitem[{{Ma} {et~al.}(2025){Ma}, {Wang}, {Wu}, \& {Wang}}]{Ma25}
{Ma}, Z.-P., {Wang}, K., {Wu}, Q., \& {Wang}, J.-M. 2025, \prd, 111, 083033, \dodoi{10.1103/PhysRevD.111.083033}

\bibitem[{{MacLeod} \& {Lin}(2020)}]{MacLeod20}
{MacLeod}, M., \& {Lin}, D. N.~C. 2020, \apj, 889, 94, \dodoi{10.3847/1538-4357/ab64db}

\bibitem[{{Madau} \& {Dickinson}(2014)}]{madaudickinson}
{Madau}, P., \& {Dickinson}, M. 2014, \araa, 52, 415, \dodoi{10.1146/annurev-astro-081811-125615}

\bibitem[{{Maga{\~n}a Hernandez} \& {Palmese}(2024)}]{Ignacio24}
{Maga{\~n}a Hernandez}, I., \& {Palmese}, A. 2024, arXiv e-prints, arXiv:2407.02460, \dodoi{10.48550/arXiv.2407.02460}

\bibitem[{{Mandel} \& {Broekgaarden}(2022)}]{IlyaFloor22}
{Mandel}, I., \& {Broekgaarden}, F.~S. 2022, Living Reviews in Relativity, 25, 1, \dodoi{10.1007/s41114-021-00034-3}

\bibitem[{{McKernan} \& {Ford}(2023)}]{McKF24}
{McKernan}, B., \& {Ford}, K.~E.~S. 2023, arXiv e-prints, arXiv:2309.15213, \dodoi{10.48550/arXiv.2309.15213}

\bibitem[{{McKernan} {et~al.}(2022{\natexlab{a}}){McKernan}, {Ford}, {Callister}, {Farr}, {O'Shaughnessy}, {Smith}, {Thrane}, \& {Vajpeyi}}]{qX22}
{McKernan}, B., {Ford}, K.~E.~S., {Callister}, T., {et~al.} 2022{\natexlab{a}}, \mnras, 514, 3886, \dodoi{10.1093/mnras/stac1570}

\bibitem[{{McKernan} {et~al.}(2022{\natexlab{b}}){McKernan}, {Ford}, {Cantiello}, {Graham}, {Jermyn}, {Leigh}, {Ryu}, \& {Stern}}]{Starfall}
{McKernan}, B., {Ford}, K.~E.~S., {Cantiello}, M., {et~al.} 2022{\natexlab{b}}, \mnras, 514, 4102, \dodoi{10.1093/mnras/stac1310}

\bibitem[{{McKernan} {et~al.}(2024){McKernan}, {Ford}, {Cook}, {Delfavero}, {Nathaniel}, {Postiglione}, {Ray}, \& {O'Shaughnessy}}]{McFACTS}
{McKernan}, B., {Ford}, K.~E.~S., {Cook}, H.~E., {et~al.} 2024, arXiv e-prints, arXiv:2410.16515, \dodoi{10.48550/arXiv.2410.16515}

\bibitem[{{McKernan} {et~al.}(2014){McKernan}, {Ford}, {Kocsis}, {Lyra}, \& {Winter}}]{McK14}
{McKernan}, B., {Ford}, K.~E.~S., {Kocsis}, B., {Lyra}, W., \& {Winter}, L.~M. 2014, \mnras, 441, 900, \dodoi{10.1093/mnras/stu553}

\bibitem[{{McKernan} {et~al.}(2012){McKernan}, {Ford}, {Lyra}, \& {Perets}}]{McK12}
{McKernan}, B., {Ford}, K.~E.~S., {Lyra}, W., \& {Perets}, H.~B. 2012, \mnras, 425, 460, \dodoi{10.1111/j.1365-2966.2012.21486.x}

\bibitem[{{McKernan} {et~al.}(2020){McKernan}, {Ford}, \& {O'Shaughnessy}}]{McK20b}
{McKernan}, B., {Ford}, K.~E.~S., \& {O'Shaughnessy}, R. 2020, \mnras, 498, 4088, \dodoi{10.1093/mnras/staa2681}

\bibitem[{{McKernan} {et~al.}(2010){McKernan}, {Ford}, \& {Reynolds}}]{McK10}
{McKernan}, B., {Ford}, K.~E.~S., \& {Reynolds}, C.~S. 2010, \mnras, 407, 2399, \dodoi{10.1111/j.1365-2966.2010.17068.x}

\bibitem[{{McKernan} {et~al.}(2018){McKernan}, {Ford}, {Bellovary}, {Leigh}, {Haiman}, {Kocsis}, {Lyra}, {Mac Low}, {Metzger}, {O'Dowd}, {Endlich}, \& {Rosen}}]{McK18}
{McKernan}, B., {Ford}, K.~E.~S., {Bellovary}, J., {et~al.} 2018, \apj, 866, 66, \dodoi{10.3847/1538-4357/aadae5}

\bibitem[{{McKernan} {et~al.}(2019){McKernan}, {Ford}, {Bartos}, {Graham}, {Lyra}, {Marka}, {Marka}, {Ross}, {Stern}, \& {Yang}}]{McK19}
{McKernan}, B., {Ford}, K.~E.~S., {Bartos}, I., {et~al.} 2019, \apjl, 884, L50, \dodoi{10.3847/2041-8213/ab4886}

\bibitem[{{Milosavljevi{\'c}} \& {Merritt}(2003)}]{finalpc03}
{Milosavljevi{\'c}}, M., \& {Merritt}, D. 2003, \apj, 596, 860, \dodoi{10.1086/378086}

\bibitem[{{Miralda-Escud{\'e}} \& {Gould}(2000)}]{Miralda00}
{Miralda-Escud{\'e}}, J., \& {Gould}, A. 2000, \apj, 545, 847, \dodoi{10.1086/317837}

\bibitem[{{Mori} {et~al.}(2021){Mori}, {Hailey}, {Schutt}, {Mandel}, {Heuer}, {Grindlay}, {Hong}, {Ponti}, \& {Tomsick}}]{Mori21}
{Mori}, K., {Hailey}, C.~J., {Schutt}, T. Y.~E., {et~al.} 2021, \apj, 921, 148, \dodoi{10.3847/1538-4357/ac1da5}

\bibitem[{{Morris}(1993)}]{Morris93}
{Morris}, M. 1993, \apj, 408, 496, \dodoi{10.1086/172607}

\bibitem[{{Nasim} {et~al.}(2023){Nasim}, {Fabj}, {Caban}, {Secunda}, {Ford}, {McKernan}, {Bellovary}, {Leigh}, \& {Lyra}}]{Nasim22}
{Nasim}, S.~S., {Fabj}, G., {Caban}, F., {et~al.} 2023, \mnras, 522, 5393, \dodoi{10.1093/mnras/stad1295}

\bibitem[{{Natarajan} \& {Armitage}(1999)}]{NatarajanArmitage99}
{Natarajan}, P., \& {Armitage}, P.~J. 1999, \mnras, 309, 961, \dodoi{10.1046/j.1365-8711.1999.02917.x}

\bibitem[{{Nelson} \& {Gressel}(2010)}]{Nelson10}
{Nelson}, R.~P., \& {Gressel}, O. 2010, \mnras, 409, 639, \dodoi{10.1111/j.1365-2966.2010.17327.x}

\bibitem[{{Neumayer} {et~al.}(2020){Neumayer}, {Seth}, \& {B{\"o}ker}}]{Neumayer20}
{Neumayer}, N., {Seth}, A., \& {B{\"o}ker}, T. 2020, \aapr, 28, 4, \dodoi{10.1007/s00159-020-00125-0}

\bibitem[{{Olejak} {et~al.}(2024){Olejak}, {Klencki}, {Xu}, {Wang}, {Belczynski}, \& {Lasota}}]{Olejak24}
{Olejak}, A., {Klencki}, J., {Xu}, X.-T., {et~al.} 2024, \aap, 689, A305, \dodoi{10.1051/0004-6361/202450480}

\bibitem[{{Ostriker}(1999)}]{Ostriker99}
{Ostriker}, E.~C. 1999, \apj, 513, 252, \dodoi{10.1086/306858}

\bibitem[{{Ostriker}(1983)}]{Ostriker83}
{Ostriker}, J.~P. 1983, \apj, 273, 99, \dodoi{10.1086/161351}

\bibitem[{{Paardekooper} {et~al.}(2010){Paardekooper}, {Baruteau}, {Crida}, \& {Kley}}]{Paaardekooper10}
{Paardekooper}, S.~J., {Baruteau}, C., {Crida}, A., \& {Kley}, W. 2010, \mnras, 401, 1950, \dodoi{10.1111/j.1365-2966.2009.15782.x}

\bibitem[{{Padovani} {et~al.}(2017){Padovani}, {Alexander}, {Assef}, {De Marco}, {Giommi}, {Hickox}, {Richards}, {Smol{\v{c}}i{\'c}}, {Hatziminaoglou}, {Mainieri}, \& {Salvato}}]{Padovani17}
{Padovani}, P., {Alexander}, D.~M., {Assef}, R.~J., {et~al.} 2017, \aapr, 25, 2, \dodoi{10.1007/s00159-017-0102-9}

\bibitem[{{Palmese} {et~al.}(2023){Palmese}, {Bom}, {Mucesh}, \& {Hartley}}]{PalmeseH0}
{Palmese}, A., {Bom}, C.~R., {Mucesh}, S., \& {Hartley}, W.~G. 2023, \apj, 943, 56, \dodoi{10.3847/1538-4357/aca6e3}

\bibitem[{{Palmese} {et~al.}(2021){Palmese}, {Fishbach}, {Burke}, {Annis}, \& {Liu}}]{Palmese190521}
{Palmese}, A., {Fishbach}, M., {Burke}, C.~J., {Annis}, J., \& {Liu}, X. 2021, \apjl, 914, L34, \dodoi{10.3847/2041-8213/ac0883}

\bibitem[{{Papaloizou} \& {Larwood}(2000)}]{Papaloizou00}
{Papaloizou}, J.~C.~B., \& {Larwood}, J.~D. 2000, \mnras, 315, 823, \dodoi{10.1046/j.1365-8711.2000.03466.x}

\bibitem[{{Parkin} \& {Bicknell}(2013)}]{Parker13}
{Parkin}, E.~R., \& {Bicknell}, G.~V. 2013, \mnras, 435, 2281, \dodoi{10.1093/mnras/stt1450}

\bibitem[{{Peng} \& {Chen}(2021)}]{Peng21}
{Peng}, P., \& {Chen}, X. 2021, \mnras, 505, 1324, \dodoi{10.1093/mnras/stab1419}

\bibitem[{{Peters}(1964)}]{Peters64}
{Peters}, P.~C. 1964, Physical Review, 136, 1224, \dodoi{10.1103/PhysRev.136.B1224}

\bibitem[{{Postiglione} {et~al.}(2025){Postiglione}, {Ford}, {Best}, {McKernan}, \& {O'Dowd}}]{Postiglione25}
{Postiglione}, J., {Ford}, K.~E.~S., {Best}, H., {McKernan}, B., \& {O'Dowd}, M. 2025, arXiv e-prints, arXiv:2502.10591, \dodoi{10.48550/arXiv.2502.10591}

\bibitem[{{Prasad} {et~al.}(2024){Prasad}, {Wang}, {Perna}, {Ford}, \& {McKernan}}]{Prasad24}
{Prasad}, C., {Wang}, Y., {Perna}, R., {Ford}, K.~E.~S., \& {McKernan}, B. 2024, \mnras, 531, 1409, \dodoi{10.1093/mnras/stae1263}

\bibitem[{{Qian} {et~al.}(2024){Qian}, {Li}, \& {Lai}}]{Qian24}
{Qian}, K., {Li}, J., \& {Lai}, D. 2024, \apj, 962, 143, \dodoi{10.3847/1538-4357/ad1b53}

\bibitem[{{Rauch} \& {Tremaine}(1996)}]{RauchTremaine96}
{Rauch}, K.~P., \& {Tremaine}, S. 1996, \na, 1, 149, \dodoi{10.1016/S1384-1076(96)00012-7}

\bibitem[{{Reitze} {et~al.}(2019){Reitze}, {Adhikari}, {Ballmer}, {Barish}, {Barsotti}, {Billingsley}, {Brown}, {Chen}, {Coyne}, {Eisenstein}, {Evans}, {Fritschel}, {Hall}, {Lazzarini}, {Lovelace}, {Read}, {Sathyaprakash}, {Shoemaker}, {Smith}, {Torrie}, {Vitale}, {Weiss}, {Wipf}, \& {Zucker}}]{CE19}
{Reitze}, D., {Adhikari}, R.~X., {Ballmer}, S., {et~al.} 2019, in Bulletin of the American Astronomical Society, Vol.~51, 35, \dodoi{10.48550/arXiv.1907.04833}

\bibitem[{{Renzo} \& {Smith}(2024)}]{Renzo24}
{Renzo}, M., \& {Smith}, N. 2024, arXiv e-prints, arXiv:2407.16113, \dodoi{10.48550/arXiv.2407.16113}

\bibitem[{{Reynolds}(2021)}]{Reynolds21}
{Reynolds}, C.~S. 2021, \araa, 59, 117, \dodoi{10.1146/annurev-astro-112420-035022}

\bibitem[{{Rodriguez} {et~al.}(2019){Rodriguez}, {Zevin}, {Amaro-Seoane}, {Chatterjee}, {Kremer}, {Rasio}, \& {Ye}}]{Rodriguez19}
{Rodriguez}, C.~L., {Zevin}, M., {Amaro-Seoane}, P., {et~al.} 2019, \prd, 100, 043027, \dodoi{10.1103/PhysRevD.100.043027}

\bibitem[{{Rodriguez} {et~al.}(2022){Rodriguez}, {Weatherford}, {Coughlin}, {Amaro-Seoane}, {Breivik}, {Chatterjee}, {Fragione}, {K{\i}ro{\u{g}}lu}, {Kremer}, {Rui}, {Ye}, {Zevin}, \& {Rasio}}]{CMC22}
{Rodriguez}, C.~L., {Weatherford}, N.~C., {Coughlin}, S.~C., {et~al.} 2022, \apjs, 258, 22, \dodoi{10.3847/1538-4365/ac2edf}

\bibitem[{{Rodr{\'\i}guez-Ram{\'\i}rez} {et~al.}(2025){Rodr{\'\i}guez-Ram{\'\i}rez}, {Nemmen}, \& {Bom}}]{Nemmen25}
{Rodr{\'\i}guez-Ram{\'\i}rez}, J.~C., {Nemmen}, R., \& {Bom}, C.~R. 2025, \prd, 111, 083020, \dodoi{10.1103/PhysRevD.111.083020}

\bibitem[{{Romero-Shaw} {et~al.}(2022){Romero-Shaw}, {Lasky}, \& {Thrane}}]{Isobel22}
{Romero-Shaw}, I., {Lasky}, P.~D., \& {Thrane}, E. 2022, \apj, 940, 171, \dodoi{10.3847/1538-4357/ac9798}

\bibitem[{{Romero-Shaw} {et~al.}(2020){Romero-Shaw}, {Lasky}, {Thrane}, \& {Calder{\'o}n Bustillo}}]{RomeroShaw20}
{Romero-Shaw}, I., {Lasky}, P.~D., {Thrane}, E., \& {Calder{\'o}n Bustillo}, J. 2020, \apjl, 903, L5, \dodoi{10.3847/2041-8213/abbe26}

\bibitem[{{Ross} {et~al.}(2020){Ross}, {Graham}, {Calderone}, {Ford}, {McKernan}, \& {Stern}}]{Ross20}
{Ross}, N.~P., {Graham}, M.~J., {Calderone}, G., {et~al.} 2020, \mnras, 498, 2339, \dodoi{10.1093/mnras/staa2415}

\bibitem[{{Rowan} {et~al.}(2023){Rowan}, {Boekholt}, {Kocsis}, \& {Haiman}}]{Rowan23}
{Rowan}, C., {Boekholt}, T., {Kocsis}, B., \& {Haiman}, Z. 2023, \mnras, 524, 2770, \dodoi{10.1093/mnras/stad1926}

\bibitem[{{Rowan} {et~al.}(2025{\natexlab{a}}){Rowan}, {Whitehead}, {Fabj}, {Kirkeberg}, {Pessah}, \& {Kocsis}}]{Rowan25}
{Rowan}, C., {Whitehead}, H., {Fabj}, G., {et~al.} 2025{\natexlab{a}}, arXiv e-prints, arXiv:2505.23739, \dodoi{10.48550/arXiv.2505.23739}

\bibitem[{{Rowan} {et~al.}(2025{\natexlab{b}}){Rowan}, {Whitehead}, {Fabj}, {Saini}, {Kocsis}, {Pessah}, \& {Samsing}}]{Rowan25a}
---. 2025{\natexlab{b}}, \mnras, 539, 1501, \dodoi{10.1093/mnras/staf547}

\bibitem[{{Rozner} {et~al.}(2023){Rozner}, {Generozov}, \& {Perets}}]{Rozner23}
{Rozner}, M., {Generozov}, A., \& {Perets}, H.~B. 2023, \mnras, 521, 866, \dodoi{10.1093/mnras/stad603}

\bibitem[{{Ryu} {et~al.}(2024){Ryu}, {McKernan}, {Ford}, {Cantiello}, {Graham}, {Stern}, \& {Leigh}}]{Ryu24}
{Ryu}, T., {McKernan}, B., {Ford}, K.~E.~S., {et~al.} 2024, \mnras, 527, 8103, \dodoi{10.1093/mnras/stad3487}

\bibitem[{{Samsing} {et~al.}(2022){Samsing}, {Bartos}, {D'Orazio}, {Haiman}, {Kocsis}, {Leigh}, {Liu}, {Pessah}, \& {Tagawa}}]{Samsing22}
{Samsing}, J., {Bartos}, I., {D'Orazio}, D.~J., {et~al.} 2022, \nat, 603, 237, \dodoi{10.1038/s41586-021-04333-1}

\bibitem[{{S{\'a}nchez-S{\'a}ez} {et~al.}(2024){S{\'a}nchez-S{\'a}ez}, {Hern{\'a}ndez-Garc{\'\i}a}, {Bernal}, {Bayo}, {Calistro Rivera}, {Bauer}, {Ricci}, {Merloni}, {Graham}, {Cartier}, {Ar{\'e}valo}, {Assef}, {Concas}, {Homan}, {Krumpe}, {Lira}, {Malyali}, {Mart{\'\i}nez-Aldama}, {Mu{\~n}oz Arancibia}, {Rau}, {Bruni}, {F{\"o}rster}, {Pavez-Herrera}, {Tub{\'\i}n-Arenas}, \& {Brightman}}]{Sanchez24}
{S{\'a}nchez-S{\'a}ez}, P., {Hern{\'a}ndez-Garc{\'\i}a}, L., {Bernal}, S., {et~al.} 2024, \aap, 688, A157, \dodoi{10.1051/0004-6361/202347957}

\bibitem[{{S{\'a}nchez-Salcedo} \& {Chametla}(2014)}]{Sanchez14}
{S{\'a}nchez-Salcedo}, F.~J., \& {Chametla}, R.~O. 2014, \apj, 794, 167, \dodoi{10.1088/0004-637X/794/2/167}

\bibitem[{{Sanders} {et~al.}(1988){Sanders}, {Soifer}, {Elias}, {Madore}, {Matthews}, {Neugebauer}, \& {Scoville}}]{Sanders88}
{Sanders}, D.~B., {Soifer}, B.~T., {Elias}, J.~H., {et~al.} 1988, \apj, 325, 74, \dodoi{10.1086/165983}

\bibitem[{{Santini} {et~al.}(2023){Santini}, {Gerosa}, {Cotesta}, \& {Berti}}]{Santini23}
{Santini}, A., {Gerosa}, D., {Cotesta}, R., \& {Berti}, E. 2023, arXiv e-prints, arXiv:2308.12998, \dodoi{10.48550/arXiv.2308.12998}

\bibitem[{{Secunda} {et~al.}(2019){Secunda}, {Bellovary}, {Mac Low}, {Ford}, {McKernan}, {Leigh}, {Lyra}, \& {S{\'a}ndor}}]{Secunda19}
{Secunda}, A., {Bellovary}, J., {Mac Low}, M.-M., {et~al.} 2019, \apj, 878, 85, \dodoi{10.3847/1538-4357/ab20ca}

\bibitem[{{Secunda} {et~al.}(2020){Secunda}, {Bellovary}, {Mac Low}, {Ford}, {McKernan}, {Leigh}, {Lyra}, {S{\'a}ndor}, \& {Adorno}}]{Secunda20}
---. 2020, \apj, 903, 133, \dodoi{10.3847/1538-4357/abbc1d}

\bibitem[{{Shakura} \& {Sunyaev}(1973)}]{SS73}
{Shakura}, N.~I., \& {Sunyaev}, R.~A. 1973, \aap, 24, 337

\bibitem[{{Sijacki} {et~al.}(2007){Sijacki}, {Springel}, {Di Matteo}, \& {Hernquist}}]{Sijacki07}
{Sijacki}, D., {Springel}, V., {Di Matteo}, T., \& {Hernquist}, L. 2007, \mnras, 380, 877, \dodoi{10.1111/j.1365-2966.2007.12153.x}

\bibitem[{{Silpa} {et~al.}(2022){Silpa}, {Kharb}, {Harrison}, {Girdhar}, {Mukherjee}, {Mainieri}, \& {Jarvis}}]{Silpa22}
{Silpa}, S., {Kharb}, P., {Harrison}, C.~M., {et~al.} 2022, \mnras, 513, 4208, \dodoi{10.1093/mnras/stac1044}

\bibitem[{{Sirko} \& {Goodman}(2003)}]{SG03}
{Sirko}, E., \& {Goodman}, J. 2003, \mnras, 341, 501, \dodoi{10.1046/j.1365-8711.2003.06431.x}

\bibitem[{{Soltan}(1982)}]{Soltan82}
{Soltan}, A. 1982, \mnras, 200, 115, \dodoi{10.1093/mnras/200.1.115}

\bibitem[{{Speri} {et~al.}(2023){Speri}, {Antonelli}, {Sberna}, {Babak}, {Barausse}, {Gair}, \& {Katz}}]{Speri23}
{Speri}, L., {Antonelli}, A., {Sberna}, L., {et~al.} 2023, Physical Review X, 13, 021035, \dodoi{10.1103/PhysRevX.13.021035}

\bibitem[{{Stegmann} {et~al.}(2025){Stegmann}, {Gerosa}, {Romero-Shaw}, {Fumagalli}, {Tagawa}, \& {Zwick}}]{Stegmann25}
{Stegmann}, J., {Gerosa}, D., {Romero-Shaw}, I., {et~al.} 2025, arXiv e-prints, arXiv:2505.13589, \dodoi{10.48550/arXiv.2505.13589}

\bibitem[{{Stern} {et~al.}(2018){Stern}, {McKernan}, {Graham}, {Ford}, {Ross}, {Meisner}, {Assef}, {Balokovi{\'c}}, {Brightman}, {Dey}, {Drake}, {Djorgovski}, {Eisenhardt}, \& {Jun}}]{Stern18}
{Stern}, D., {McKernan}, B., {Graham}, M.~J., {et~al.} 2018, \apj, 864, 27, \dodoi{10.3847/1538-4357/aac726}

\bibitem[{{Stone} {et~al.}(2017){Stone}, {Metzger}, \& {Haiman}}]{Stone17}
{Stone}, N.~C., {Metzger}, B.~D., \& {Haiman}, Z. 2017, \mnras, 464, 946, \dodoi{10.1093/mnras/stw2260}

\bibitem[{{Storchi-Bergmann} \& {Schnorr-M{\"u}ller}(2019)}]{Storchi19}
{Storchi-Bergmann}, T., \& {Schnorr-M{\"u}ller}, A. 2019, Nature Astronomy, 3, 48, \dodoi{10.1038/s41550-018-0611-0}

\bibitem[{{Syer} {et~al.}(1991){Syer}, {Clarke}, \& {Rees}}]{Syer91}
{Syer}, D., {Clarke}, C.~J., \& {Rees}, M.~J. 1991, \mnras, 250, 505, \dodoi{10.1093/mnras/250.3.505}

\bibitem[{{Tagawa} {et~al.}(2020{\natexlab{a}}){Tagawa}, {Haiman}, {Bartos}, \& {Kocsis}}]{Hiromichi20spin}
{Tagawa}, H., {Haiman}, Z., {Bartos}, I., \& {Kocsis}, B. 2020{\natexlab{a}}, \apj, 899, 26, \dodoi{10.3847/1538-4357/aba2cc}

\bibitem[{{Tagawa} {et~al.}(2020{\natexlab{b}}){Tagawa}, {Haiman}, \& {Kocsis}}]{Hiromichi20}
{Tagawa}, H., {Haiman}, Z., \& {Kocsis}, B. 2020{\natexlab{b}}, \apj, 898, 25, \dodoi{10.3847/1538-4357/ab9b8c}

\bibitem[{{Tagawa} {et~al.}(2023{\natexlab{a}}){Tagawa}, {Kimura}, \& {Haiman}}]{Hiromichineutrinos23}
{Tagawa}, H., {Kimura}, S.~S., \& {Haiman}, Z. 2023{\natexlab{a}}, \apj, 955, 23, \dodoi{10.3847/1538-4357/ace71d}

\bibitem[{{Tagawa} {et~al.}(2023{\natexlab{b}}){Tagawa}, {Kimura}, {Haiman}, {Perna}, \& {Bartos}}]{Hiromichi23}
{Tagawa}, H., {Kimura}, S.~S., {Haiman}, Z., {Perna}, R., \& {Bartos}, I. 2023{\natexlab{b}}, \apjl, 946, L3, \dodoi{10.3847/2041-8213/acc103}

\bibitem[{{Tagawa} {et~al.}(2024){Tagawa}, {Kimura}, {Haiman}, {Perna}, \& {Bartos}}]{Hiromichi24}
---. 2024, \apj, 966, 21, \dodoi{10.3847/1538-4357/ad2e0b}

\bibitem[{{Tagawa} {et~al.}(2021){Tagawa}, {Kocsis}, {Haiman}, {Bartos}, {Omukai}, \& {Samsing}}]{Hiromichi21ecc}
{Tagawa}, H., {Kocsis}, B., {Haiman}, Z., {et~al.} 2021, \apjl, 907, L20, \dodoi{10.3847/2041-8213/abd4d3}

\bibitem[{{Tanaka} \& {Ward}(2004)}]{Tanaka04}
{Tanaka}, H., \& {Ward}, W.~R. 2004, \apj, 602, 388, \dodoi{10.1086/380992}

\bibitem[{{The LIGO Scientific Collaboration} {et~al.}(2021){The LIGO Scientific Collaboration}, {the Virgo Collaboration}, {the KAGRA Collaboration}, \& {Abbott}}]{o3b}
{The LIGO Scientific Collaboration}, {the Virgo Collaboration}, {the KAGRA Collaboration}, \& {Abbott}, R. e.~a. 2021, arXiv e-prints, arXiv:2111.03606, \dodoi{10.48550/arXiv.2111.03606}

\bibitem[{{Thompson} {et~al.}(2005){Thompson}, {Quataert}, \& {Murray}}]{TQM05}
{Thompson}, T.~A., {Quataert}, E., \& {Murray}, N. 2005, \apj, 630, 167, \dodoi{10.1086/431923}

\bibitem[{{Trani} \& {Di Cintio}(2025)}]{Trani25}
{Trani}, A.~A., \& {Di Cintio}, P. 2025, arXiv e-prints, arXiv:2506.02173, \dodoi{10.48550/arXiv.2506.02173}

\bibitem[{{Tremmel} {et~al.}(2017){Tremmel}, {Karcher}, {Governato}, {Volonteri}, {Quinn}, {Pontzen}, {Anderson}, \& {Bellovary}}]{Tremmel17}
{Tremmel}, M., {Karcher}, M., {Governato}, F., {et~al.} 2017, \mnras, 470, 1121, \dodoi{10.1093/mnras/stx1160}

\bibitem[{{Ueda} {et~al.}(2014){Ueda}, {Akiyama}, {Hasinger}, {Miyaji}, \& {Watson}}]{Ueda14}
{Ueda}, Y., {Akiyama}, M., {Hasinger}, G., {Miyaji}, T., \& {Watson}, M.~G. 2014, \apj, 786, 104, \dodoi{10.1088/0004-637X/786/2/104}

\bibitem[{{Vaccaro} {et~al.}(2024){Vaccaro}, {Mapelli}, {P{\'e}rigois}, {Barone}, {Artale}, {Dall'Amico}, {Iorio}, \& {Torniamenti}}]{Vaccaro24}
{Vaccaro}, M.~P., {Mapelli}, M., {P{\'e}rigois}, C., {et~al.} 2024, \aap, 685, A51, \dodoi{10.1051/0004-6361/202348509}

\bibitem[{{Vajpeyi} {et~al.}(2022){Vajpeyi}, {Thrane}, {Smith}, {McKernan}, \& {Saavik Ford}}]{Vajpeyi22}
{Vajpeyi}, A., {Thrane}, E., {Smith}, R., {McKernan}, B., \& {Saavik Ford}, K.~E. 2022, \apj, 931, 82, \dodoi{10.3847/1538-4357/ac6180}

\bibitem[{{Varma} {et~al.}(2019){Varma}, {Gerosa}, {Stein}, {H{\'e}bert}, \& {Zhang}}]{SurfinBH}
{Varma}, V., {Gerosa}, D., {Stein}, L.~C., {H{\'e}bert}, F., \& {Zhang}, H. 2019, \prl, 122, 011101, \dodoi{10.1103/PhysRevLett.122.011101}

\bibitem[{{Veronesi} {et~al.}(2023){Veronesi}, {Rossi}, \& {van Velzen}}]{Veronesi23}
{Veronesi}, N., {Rossi}, E.~M., \& {van Velzen}, S. 2023, \mnras, 526, 6031, \dodoi{10.1093/mnras/stad3157}

\bibitem[{{Veronesi} {et~al.}(2022){Veronesi}, {Rossi}, {van Velzen}, \& {Buscicchio}}]{Veronesi22}
{Veronesi}, N., {Rossi}, E.~M., {van Velzen}, S., \& {Buscicchio}, R. 2022, \mnras, 514, 2092, \dodoi{10.1093/mnras/stac1346}

\bibitem[{{Veronesi} {et~al.}(2025){Veronesi}, {van Velzen}, {Rossi}, \& {Storey-Fisher}}]{Veronesi25a}
{Veronesi}, N., {van Velzen}, S., {Rossi}, E.~M., \& {Storey-Fisher}, K. 2025, \mnras, 536, 375, \dodoi{10.1093/mnras/stae2575}

\bibitem[{{Vijaykumar} {et~al.}(2023){Vijaykumar}, {Tiwari}, {Kapadia}, {Arun}, \& {Ajith}}]{Aditya23}
{Vijaykumar}, A., {Tiwari}, A., {Kapadia}, S.~J., {Arun}, K.~G., \& {Ajith}, P. 2023, \apj, 954, 105, \dodoi{10.3847/1538-4357/acd77d}

\bibitem[{Virtanen {et~al.}(2020)Virtanen, Gommers, Oliphant, Haberland, Reddy, Cournapeau, Burovski, Peterson, Weckesser, Bright, {van der Walt}, Brett, Wilson, Millman, Mayorov, Nelson, Jones, Kern, Larson, Carey, Polat, Feng, Moore, {VanderPlas}, Laxalde, Perktold, Cimrman, Henriksen, Quintero, Harris, Archibald, Ribeiro, Pedregosa, {van Mulbregt}, \& {SciPy 1.0 Contributors}}]{2020SciPy-NMeth}
Virtanen, P., Gommers, R., Oliphant, T.~E., {et~al.} 2020, Nature Methods, 17, 261, \dodoi{10.1038/s41592-019-0686-2}

\bibitem[{{Wang} {et~al.}(2024{\natexlab{a}}){Wang}, {Zhu}, \& {Lin}}]{Wang24}
{Wang}, Y., {Zhu}, Z., \& {Lin}, D. N.~C. 2024{\natexlab{a}}, \mnras, 528, 4958, \dodoi{10.1093/mnras/stae321}

\bibitem[{{Wang} {et~al.}(2024{\natexlab{b}}){Wang}, {Zhu}, \& {Lin}}]{WZL24}
---. 2024{\natexlab{b}}, \mnras, 528, 4958, \dodoi{10.1093/mnras/stae321}

\bibitem[{{Wang} {et~al.}(2021){Wang}, {McKernan}, {Ford}, {Perna}, {Leigh}, \& {Mac Low}}]{Yihan21}
{Wang}, Y.-H., {McKernan}, B., {Ford}, S., {et~al.} 2021, \apjl, 923, L23, \dodoi{10.3847/2041-8213/ac400a}

\bibitem[{{Whitehead} {et~al.}(2024){Whitehead}, {Rowan}, {Boekholt}, \& {Kocsis}}]{Whitehead24}
{Whitehead}, H., {Rowan}, C., {Boekholt}, T., \& {Kocsis}, B. 2024, \mnras, 531, 4656, \dodoi{10.1093/mnras/stae1430}

\bibitem[{{Whitehead} {et~al.}(2025){Whitehead}, {Rowan}, \& {Kocsis}}]{Whitehead25}
{Whitehead}, H., {Rowan}, C., \& {Kocsis}, B. 2025, arXiv e-prints, arXiv:2505.23899, \dodoi{10.48550/arXiv.2505.23899}

\bibitem[{{Wong} {et~al.}(2019){Wong}, {Baibhav}, \& {Berti}}]{Kaze19}
{Wong}, K. W.~K., {Baibhav}, V., \& {Berti}, E. 2019, \mnras, 488, 5665, \dodoi{10.1093/mnras/stz2077}

\bibitem[{{Xue} {et~al.}(2025){Xue}, {Tagawa}, {Haiman}, \& {Bartos}}]{Xue25}
{Xue}, L., {Tagawa}, H., {Haiman}, Z., \& {Bartos}, I. 2025, arXiv e-prints, arXiv:2504.19570, \dodoi{10.48550/arXiv.2504.19570}

\bibitem[{{Yamada} {et~al.}(2024){Yamada}, {Kawamuro}, {Mizumoto}, {Ricci}, {Ogawa}, {Noda}, {Ueda}, {Enoto}, {Kokubo}, {Minezaki}, {Sameshima}, {Horiuchi}, \& {Mizukoshi}}]{Yamada24}
{Yamada}, S., {Kawamuro}, T., {Mizumoto}, M., {et~al.} 2024, \apjs, 274, 8, \dodoi{10.3847/1538-4365/ad5961}

\bibitem[{{Yang} {et~al.}(2021){Yang}, {Estrada-Carpenter}, {Papovich}, {Vito}, {Walsh}, {Yao}, \& {Yuan}}]{Yang21a}
{Yang}, G., {Estrada-Carpenter}, V., {Papovich}, C., {et~al.} 2021, \apj, 921, 170, \dodoi{10.3847/1538-4357/ac2233}

\bibitem[{{Yang} {et~al.}(2023){Yang}, {Caputi}, {Papovich}, {Arrabal Haro}, {Bagley}, {Behroozi}, {Bell}, {Bisigello}, {Buat}, {Burgarella}, {Cheng}, {Cleri}, {Dav{\'e}}, {Dickinson}, {Elbaz}, {Ferguson}, {Finkelstein}, {Grogin}, {Hathi}, {Hirschmann}, {Holwerda}, {Huertas-Company}, {Hutchison}, {Iani}, {Kartaltepe}, {Kirkpatrick}, {Kocevski}, {Koekemoer}, {Kokorev}, {Larson}, {Lucas}, {P{\'e}rez-Gonz{\'a}lez}, {Rinaldi}, {Shen}, {Trump}, {de la Vega}, {Yung}, \& {Zavala}}]{Yang23MIRI}
{Yang}, G., {Caputi}, K.~I., {Papovich}, C., {et~al.} 2023, \apjl, 950, L5, \dodoi{10.3847/2041-8213/acd639}

\bibitem[{{Yang} {et~al.}(2019{\natexlab{a}}){Yang}, {Bartos}, {Haiman}, {Kocsis}, {M{\'a}rka}, {Stone}, \& {M{\'a}rka}}]{Yang19ApJ}
{Yang}, Y., {Bartos}, I., {Haiman}, Z., {et~al.} 2019{\natexlab{a}}, \apj, 876, 122, \dodoi{10.3847/1538-4357/ab16e3}

\bibitem[{{Yang} {et~al.}(2019{\natexlab{b}}){Yang}, {Bartos}, {Gayathri}, {Ford}, {Haiman}, {Klimenko}, {Kocsis}, {M{\'a}rka}, {M{\'a}rka}, {McKernan}, \& {O'Shaughnessy}}]{Yang19}
{Yang}, Y., {Bartos}, I., {Gayathri}, V., {et~al.} 2019{\natexlab{b}}, \prl, 123, 181101, \dodoi{10.1103/PhysRevLett.123.181101}

\bibitem[{{Zhang} {et~al.}(2025){Zhang}, {Wang}, {Yuan}, {Tagawa}, {Wei}, {Li}, \& {Cai}}]{241125}
{Zhang}, S.-R., {Wang}, Y., {Yuan}, Y.-F., {et~al.} 2025, arXiv e-prints, arXiv:2505.10395, \dodoi{10.48550/arXiv.2505.10395}

\bibitem[{{Zhu} \& {Chen}(2025)}]{Zhu25}
{Zhu}, L.-G., \& {Chen}, X. 2025, arXiv e-prints, arXiv:2505.02924, \dodoi{10.48550/arXiv.2505.02924}

\end{thebibliography}
\bibliographystyle{aasjournal}



\end{document}